\documentclass{aa}
\usepackage{graphicx,txfonts,float,footmisc,twoopt}
\usepackage{natbib} 
\usepackage{xcolor}
\usepackage[normalem]{ulem}
\usepackage{amsmath}	
\usepackage{amssymb}
\usepackage[version=3]{mhchem}
\usepackage{hyperref}
\newcommand{\Hy}{\mathrm{H}}
\newcommand{\Hp}{{\mathrm{H}^+}}
\newcommand{\Hm}{{\mathrm{H}^-}}
\newcommand{\He}{\mathrm{He}}
\newcommand{\dd}{\mathrm{d}}

\begin{document} 

\title{Impact of H\,{\sc i} cooling and study of accretion disks in AGB wind-companion smoothed particle hydrodynamic simulations}
\titlerunning{H\,{\sc i} cooling and accretion disks in AGB binaries}

\author{J. Malfait \inst{1} \and  L. Siess \inst{2}  \and M. Esseldeurs \inst{1} \and F. De Ceuster \inst{1} \and { S. H. J. Wallstr\"om} \inst{1} \and A. de Koter \inst{3,1} \and L. Decin \inst{1}}

\institute{ Institute of Astronomy, KU Leuven, Celestijnenlaan 200D, 3001 Leuven, Belgium \and Institut d'Astronomie et d'Astrophysique, Universit\'e Libre de Bruxelles (ULB), CP 226, 1050 Brussels, Belgium \and University of Amsterdam, Anton Pannekoek Institute for Astronomy, Amsterdam, Amsterdam, 1098 XH, The Netherlands)
}

\date{
Accepted, 20/08/2024}

\abstract
	{High-resolution observations reveal that the outflows of evolved low- and intermediate-mass stars harbour complex morphological structures that are linked to the presence of one or multiple companions. {Hydrodynamical simulations provide a way to study the impact of a companion on the shaping of the AGB (Asymptotic Giant Branch star) outflow.} }
        {Using smoothed particle hydrodynamic (SPH) simulations of a mass losing AGB {star} with a binary companion, we study the impact of including  H\,{\sc i} atomic line cooling on the flow morphology and on the properties of the accretion disks that form around the companion. }
        {We use the \textsc{Phantom} code to perform high-resolution 3D SPH simulations of the interaction of a solar-mass companion with the outflow of an AGB star using different wind velocities and eccentricities. We compare the models properties computed with and without the inclusion of H\,{\sc i} cooling.} 
	{The inclusion of H\,{\sc i} cooling produces a large decrease in the temperature up to one order of magnitude in the region closely surrounding the companion star. As a consequence, morphological irregularities and relatively energetic (bipolar) outflows that were obtained without H\,{\sc i} cooling no longer appear. In case of an eccentric orbit and a low wind velocity, the morphologies are still highly asymmetric, but the same structures recur at every orbital period, making the morphology more regular. Flared accretion disks, with a (sub)-Keplerian velocity profile, are found to form around the companion in all our models with H\,{\sc i} cooling  provided the accretion radius is small enough. The disks have radial sizes ranging from about 0.4 to 0.9\,au and masses around $10^{-7}$--$10^{-8} \, {\rm M_{\odot}}$. For the considered wind velocities, mass accretion onto the companion is up to a factor of 2 higher than predicted by the standard Bondi Hoyle Littleton rate, and ranges between $\sim 4 $ to $21 \%$ of the AGB wind mass loss rate.  The lower the wind velocity at the location of the companion, the larger and the more massive the disk, and the higher the mass accretion efficiency. In eccentric systems, the disk size, disk mass and mass accretion efficiency vary depending on the orbital phase.}
        {H\,{\sc i} cooling is an essential ingredient to properly model the medium around the companion where density-enhanced wind structures form and it favours the formation of an accretion disk.}

\maketitle

\authorrunning{Malfait et al.}

\section{Introduction}
\label{ch:intro}

When low- and intermediate-mass stars reach the Asymptotic Giant Branch (AGB) evolutionary phase, they shed their outer layers through a strong, pulsation-enhanced, dust-driven stellar wind, with mass-loss rates of $\dot{M} \sim 10^{-8}$--$10^{-5} \, {\rm M_\odot\, yr^{-1}}$ \citep{Hofner2018,Decin2021}. These outflows span a large range in density and temperature, making them interesting chemical laboratories, that contribute to $\sim 85 \%$ of the gas and $\sim 35 \%$ of the dust enrichment of the inter-stellar-medium (ISM) \citep{Tielens2005}.
For decades these outflows where believed to be spherically symmetric, and where modelled in a 1-dimensional approach. 
However, high-resolution observations reveal that asymmetrical structures such as spirals, arcs, disks, and bipolar outflows are embedded in the winds of AGB stars \citep{Mauron2006,Maercker2012,Decin2020,Danilovich2024}.
{The environment of post-AGB stars and planetary nebulae (PNe), that descend from these AGB circumstellar envelopes (CSEs), are also known to show complex morphologies, including e.g. bipolar nebulae and shells \citep{Balick2002} and consecutive rings and arcs \citep{RamosLarios2016}. }
Currently the main theories for the formation of non-spherical planetary nebulae involve interactions with one or more companions \citep{Demarco2009,DeMarco2013,Jones2017,Akashi2017,Garcia2018}.
Likewise, the complex morphologies of post-AGB stars are linked to the presence of a binary system with a circumbinary disk and to a companion with circumstellar accretion disk and bipolar jets, though the exact formation mechanism is unknown \citep{VanWinckel2003,Bollen2022,Corporaal2023,Verhamme2024,DePrins2024}. These post-AGB binaries are often found in highly eccentric orbits with periods of $10^{2}$--$10^{4} \, {\rm days}$ \citep{Oomen2018,Jorissen2019,Escorza2019}. This is in strong disagreement with population synthesis results, that predict closer, circularised orbits due to tides, or wider, possibly eccentric orbits \citep{Toonen2012}.
Similarly, it is now commonly believed that the observed structures in AGB outflows are the result of the impact of companions that are hidden within the dense wind. This hypotheses is supported by binary population statistics \citep{Moe2017,Offner2023}{, that states that about $50\%$ of solar mass stars has a binary companion, with this percentage increasing with stellar mass. Further, several} recent observational studies of AGB stars manage to identify the indirect presence of a companion, for example through its impact on molecular abundances \citep[][Montargès et al. in prep]{Ramstedt2014,Kervella2016,Homan2020(2),Decin2020,Danilovich2024,Planquart2024a}.
Finally, high resolution 3-dimensional (3D) hydrodynamic simulations of binary AGB stars reveal how complex structures naturally emerge from the presence of a companion. {The earliest of such studies \citep{TheunsI,TheunsII,MastrodemosI,MastrodemosII} already showed that Archimedes-like spiral structures, elongated morphologies, and accretion disks can naturally form, depending on the orbital and wind parameters of the modelled system. Later higher resolution simulations analysed in more detail the formation of different structures, including spirals, arc- and ring-shaped morphologies, equatorial density enhancements, and accretion disks, and studied the impact of these structures on the mass transfer and orbital evolution \citep[see e.g.][]{Kim2012B,Kim2019,Chen2017,Chen2020,Saladino2018,Saladino2019A,ElMellah2020,Malfait2021,Maes2021,Aydi2022}.
The structures that arise in the AGB wind due to binary interaction, might play an important role in the shaping of the CSE at later evolutionary stages. For example, \cite{Lora2023} use 3D radiation-hydrodynamic models to mimic the transition from the AGB to the proto-PN phase by artificially ejecting jets and a spherically symmetric fast wind, after a spiral structure has already been shaped in the AGB wind by a companion star. With this method, they are able to recover the morphologies of proto-PNe and PNe with ring-like structures in their haloes.}

In previous work by \cite{Malfait2021} (which we will refer to as Paper I), 3D smoothed particle hydrodynamic (SPH) simulations where performed with the \textsc{Phantom} code\footnote{\hyperlink{https://github.com/danieljprice/phantom}{https://github.com/danieljprice/phantom}} \citep{phantom}, for a grid of $9$ AGB binary-wind models, with different orbital eccentricities and wind velocities. Their main conclusions were as follows. For high wind velocities (initial velocity $v_{\rm ini} = 20\, {\rm{km\,s^{-1}}}$), the limited gravitational attraction of the companion on the wind particles results in a 2-edged spiral structure originating behind the companion. This stable wind structure shapes the global morphology into a regular Archimedes spiral in the orbital plane, with concentric arcs in the meridional plane. For lower wind velocities, more wind material is compressed around the companion into a high-pressure bubble, that is delimited by a bow shock spiral originating in front of the companion. In case of an intermediate wind velocity ($v_{\rm ini} = 10\, {\rm{km\,s^{-1}}}$), this high-pressure region and bow shock are stable, resulting in a regular spiral with meridional bicentric rings. When the wind velocity is very low ($v_{\rm ini} = 5 \, {\rm{km\,s^{-1}}}$), periodic instabilities occur around the high-pressure region, in some cases followed by relatively energetic outflows with radial velocities up to $~30 \, {\rm km\,s^{-1} }$. This results in irregular morphologies in both the orbital and meridional plane view.
In case the orbit is eccentric, the phase-dependency of the wind-companion interaction strength introduces additional complexities. The wind structures close to the stars vary throughout the orbital phases, and instabilities are more likely to occur, such that highly asymmetric wind morphologies result.
The instabilities that occurred were associated with hot wind material with gas temperatures reaching $T \geq 10^{5} \, \rm{K}$ that accumulated around the companion. 

In this work, we include H\,{\sc i} atomic line cooling in our SPH model, following \cite{Spitzer1978} (see Section~\ref{ch:method}). This cooling has been considered in previous simulations by {\cite{MastrodemosI,MastrodemosII} and \cite{Chen2017}} 
to provide an efficient cooling especially impacting the material in the close vicinity of  the companion. We select $6$ of the simulations analysed in Paper I, and study the effect of H\,{\sc i} cooling on the wind structures (Section~\ref{ch:impactCooling}).
Due to the presence of cooling, accretion disks form around the solar-mass companion. We perform higher-resolution simulations focusing on these accretion disks, and study in detail their morphology and properties, how they change for different input wind velocities and in eccentric orbits, and how efficiently mass is accreted through the disks (Section~\ref{ch:accrDisks}). Our conclusions are presented in Section~\ref{ch:conclusion}.

\section{Method}
\label{ch:method}
\subsection{Input Physics}
\label{ch:NumMethod}

The hydrodynamic simulations are performed with the smoothed particle hydrodynamics (SPH) code \textsc{Phantom} \citep{phantom}.
The numerical setup of our models is the same as described in paper I and \citet{Maes2021}, but with additional cooling in the form of H\,{\sc i} atomic line cooling. The wind consists of SPH gas particles, that follow the adiabatic equation of state (EOS)
\begin{equation}
        \label{eq:EOS}
	P = (\gamma -1) \rho u 
\end{equation}
where $P$ is the pressure, $\rho$ the gas density, $u$ the internal energy, and $\gamma$ the polytropic index. 
The gas temperature $T$ is obtained using Eq.~\ref{eq:EOS} and
the ideal gas law
\begin{equation}
	P = \frac{\rho k_\mathrm{B}T}{\mu m_\mathrm{H}} 
\end{equation}
where $k_\mathrm{B}$ is Boltzmann's constant, $m_\mathrm{H}$ the mass of a hydrogen atom, and $\mu$ is the mean molecular weight.
The polytropic index $\gamma$ and mean molecular weight $\mu$ are set to a constant value throughout the model, with $\gamma = 1.2$ a typical value to account for the temperature profile in AGB winds \citep{Millar2004,Maes2021}, and $\mu = 2.38$ or $1.26$ depending on whether we consider a molecular or atomic wind, respectively (see also Table~\ref{ta:binarySetup}). {Note, however, that the temperature profile is not imposed throughout the simulations, and greatly depends on the cooling and heating mechanisms.}

In this work, H\,{\sc i} cooling is included in the setup through an implicit integration method that is acting on substeps. The cooling term is calculated by use of the relation
\begin{equation}
\Lambda_\mathrm{cool\_HI} = 7.3 \times 10^{-19} {\rm g \, cm^{3}} \times \frac{n_e \, n_\mathrm{H} \mathrm{e}^{-118\,400 {\rm \, K} / T}}{\rho}\ \mathrm{erg\,g}^{-1}\,\mathrm{s}^{-1},
\end{equation}
presented in \citet{Spitzer1978}, where $n_e$ and $n_\mathrm{H}$ are the electron and total hydrogen number density, respectively (for details see Appendix~\ref{AP:elNumbDens}).
This cooling is mostly effective at the high temperatures reached inside shocks or close to an accreting companion, and becomes dominant at temperatures above $\sim 8000$~K \citep{Bowen1988}.
This term is included in the energy equation, along with adiabatic work done by expansion and compression, and shock heating $\Lambda_{\rm shock}$ \citep{phantom}, as
\begin{equation}
    \label{Eq:energy}
    \frac{\mathrm{d} u}{\text{d} t} =  -\frac{P}{\rho} (\vec{\nabla} \cdot \textit{\textbf{v}}) + \Lambda_{\rm shock} - \Lambda_\mathrm{cool\_HI}.
\end{equation}

The binary companion is modelled by a sink particle, with an accretion radius $R_{\rm{s,accr}}$, such that gas particles and their {mass and} momentum are indiscriminately accreted {by the companion sink particle} if they fall within $0.8 \, R_{\rm{s,accr}}$, and additional checks on gravitational binding and angular momentum are performed in the boundary region $0.8 \, R_{\rm{s,accr}} < r < R_{\rm{s,accr}}$, as described in sect 2.8.2 of {\cite{phantom}}.
To model the wind, SPH particles are distributed on spheres around the primary sink particle and assigned an initial velocity $v_{\rm ini}$ \cite[for details, see][]{Siess2022,Esseldeurs2023}. To drive the wind, the free-wind approximation is used, in which the gravity of the AGB star is artificially balanced by the radiation force \cite[introduced by][]{TheunsI}. Although this approximation does not take into account the complex driving mechanism of dust-driven winds, it avoids additional assumptions on e.g.\,pulsations, dust, and luminosity, while still retrieving reasonable results in the low mass-loss rate regime \citep{Esseldeurs2023}.

\subsection{Simulation grid setup}
\label{ch:SetupMethod}

A grid of six models is calculated to analyse how H\,{\sc i} cooling impacts the wind-companion interaction and resulting wind morphologies. The setup parameters are presented in Table~\ref{ta:binarySetup}. As in paper I, the primary star is an AGB star with mass $M_{\rm p} = 1.5\,{\rm M_{\odot}}$, accretion radius $R_{\rm p,accr}= 1.2 \, \rm{au}= 258\, \rm R_{\odot}$, and effective temperature $T_{\rm p} = 3000$ K. The AGB star is part of a binary system with a solar-mass companion $M_{\rm s} = 1\, \rm M_{\odot}$ located at a semi-major axis of $a = 6\, \rm{au}$, giving the binary an orbital period of $P_{\rm orb} = 9.3 \, \rm{yrs}$, and orbital velocities $v_{\rm orb,p} = 7.69 \, {\rm km \, s^{-1}}$ and $ v_{\rm orb,c} = 11.54\, {\rm km \, s^{-1}}$. The stars are deep in their potential well, with Roche radii of $2.487 \, {\rm au}$ and $2.066 \, {\rm au}$ for the primary and secondary, respectively.

{The simulations are run for $10$ orbital periods, corresponding to $t_{\rm max} = 93 \, {\rm yrs} $. 
The grid of binary models is presented in Table ~\ref{inputTable}, and is a selection of the models analysed in Paper I, but with H\,{\sc i} cooling included. To study the effect of orbital eccentricity, we model systems with eccentricities $e = 0$ and $e= 0.5$. Further, we consider systems with input wind velocities $v_{\rm ini} = 5, 10$ and $20 \, {\rm km \, s^{-1}}$ to model different wind-companion interaction regimes and wind morphologies \citep{ElMellah2020,Malfait2021,Maes2021}.
The mass-loss rate of the AGB star is set to $\dot{M} = 10^{-7}\,{\rm M_{\odot} \, {\rm yr}^{-1}}$. For wind velocities of $\sim 20 \, {\rm km \, s^{-1}}$ observations \citep{Ramstedt2020} suggest higher mass-loss rates, but by setting the same $\dot{M}$ for all simulations, we isolate the effect of changing the wind velocity, without {introducing an additional variable parameter (we come back to this in Sect.~\ref{ImpactMLR})}. {Moreover, we note that a better treatment of radiative acceleration (not assuming a free-wind) is needed to optimally model the effect of the mass-loss rate on the resulting wind structures \citep{Esseldeurs2023}.}

Compared to paper I, we slightly increased the boundary radius, that sets the distance from the AGB sink particle above which particles are removed from the simulation, from $200$ to $250 \, {\rm au}$, and we slightly decreased the accretion radius of the companion $R_{\rm s,accr}= 0.05 \, \rm{au}$ to $R_{\rm s,accr}= 0.04 \, \rm{au}= 8.6\,{\rm R_{\odot}}$. By decreasing $R_{\rm s,accr}$ we increase the computation time, but we enhance our ability to numerically resolve an accretion disk around the companion. At the end of the computation, the simulations have reached a quasi-steady state and contain $\sim 10^6$ particles (the precise number depends on the input wind velocity and the number of particles that are accreted and thereby removed from the simulation).

\begin{table}
	\caption{Initial binary models setup.}
	\begin{center}
		\begin{tabular}{lll}
			\hline 
			\hline
			Parameter     & Unit      &  Initial value \\
			\hline 
			$M_{\rm p}$        & [${\rm M_{\odot}}$] &  $1.5$\\
			$R_{\rm p,accr}$   & [${\rm R_{\odot}}$] &  $258$  \\
			$T_{\rm p}$		&	[K] &  $3000$ \\
			$M_{\rm s}$      &   [${\rm M_{\odot}}$] & $1$ \\
			$R_{{\rm s,accr}}$ & [${\rm R_{\odot}}$] &  $8.6^{a}, 2.15^{b}$  \\
			$a$                & [au] &  $6$ \\
   			$\dot{M}$		&	[${\rm{M_\odot\,yr^{-1}}}$] & $10^{-7}$ \\
			$t_{\rm max}$    &   [yrs] &  $93$  \\
			$R_{\rm bound}$   &  [au] &  $250^{a}, 20^{b}$  \\
                $\mu$              &  &  $2.38^{a}, 1.26^{b}$ \\
                $\gamma$           &  & $1.2$ \\
			\hline
		\end{tabular}
	\end{center}
	{\textbf{Notes.} \footnotesize{Superscript ${^a}$ indicates values that are only used in Sect.~\ref{ch:impactCooling} for the analysis of the impact of H\,{\sc i} cooling, and superscript ${^b}$ for the accretion disk models discussed in Sect.~\ref{ch:accrDisks}. $M_{\rm p}$, $M_{\rm s}$, $R_{\rm p,accr}$, and $R_{\rm s,accr}$ are the initial masses and accretion radii of the primary (AGB) and secondary (companion) star, $\dot{M}$ is the mass-loss rate of the AGB star, and $T_{\rm p}$ its effective temperature, $a$ is the semi-major axis of the system, and $R_{\rm bound}$ is the boundary distance from the AGB star beyond which gas particles are killed. The models run for $t_{\rm max} = 93\, {\rm yrs}$ which corresponds to $10$ orbital periods. Finally, $\mu$ is the mean molecular weight, and $\gamma$ the adiabatic index of the wind, both set constant.
	}}
        \label{ta:binarySetup}

\end{table}

\begin{table}
	\caption{Binary model characteristics.}
	\begin{center}
		\begin{tabular}{ccc}
			\hline
			\hline
			&&\\[-2ex]
			Model name &  $v_{\rm ini}$ [${\rm{km\,s^{-1}}}$] & $e$ \\
			&&\\[-2ex]
			\hline
			v05e00   & $5$  & $0.00$  \\
			v05e50   & $5$  & $0.50$  \\
			
			v10e00   & $10$ & $0.00$ \\
			v10e50   & $10$ & $0.50$ \\
			
			v20e00   & $20$ & $0.00$ \\
			v20e50   & $20$ & $0.50$ \\
			\hline
		\end{tabular}
	\end{center}
	{\textbf{Notes.} \footnotesize{Binary models with their characteristic input values. The model names are set in such a way that the characteristics can be deduced from it, with `v...' denoting the input wind velocity in ${\rm{km\,s^{-1}}}$ and `e...' the value of the eccentricity multiplied by a factor $100$.}}
	\label{inputTable}
\end{table}

\section{Impact {of} H\,{\sc i} cooling on binary simulations}
\label{ch:impactCooling}

In Paper I, cooling was regulated by the equation of state of an ideal gas with a polytropic index $\gamma = 1.2$, without additional cooling mechanisms. This led to the formation of a high-pressure region around the companion star, where the SPH particles reached temperatures above $10^5 \, {\rm K}$ in combination with high velocities. This gave rise to periodic instabilities and energetic outflows both in the direction of the orbital plane and towards the poles. We here include a cooling mechanism that reduces the temperatures in this hot region around the companion, and study the effect on the wind shaping.

The main effects of including H\,{\sc i} atomic line cooling are the following. (i) The disappearance of periodic instabilities that occurred due to heated-up material around the companion, and that were giving rise for example to a square-like structure in model v05e00 of paper I. (ii) Instead, {material around the companion is cooled down by about an order of magnitude,} and is able to accumulate and form an accretion disk, which we study in detail in Sect.~\ref{ch:accrDisks} with higher-resolution models. (iii) The bipolar outflows with velocities up to $\sim 30 \, \rm{km \, s^{-1}}$ that developed in the low AGB wind velocities simulations of paper I are no longer present in our improved models. (iv) Even with H\,{\sc i} cooling and formation of stable accretion disks, complex, very asymmetric wind morphologies emerge in the eccentric systems. (v) Cooling also decreases the wind velocity in the enhanced density wind structures, which affects the outflow morphology. These effects are discussed in more detail below.
Following the same format as Paper I, we analyse the wind structures that are created close to the stars, and the larger-scale wind morphologies that result from them. 
{Although the addition of H\,{\sc i} cooling yields a more realistic temperature distribution, in most of our simulations, the resulting temperatures in the inner wind region are still too high for efficient dust nucleation to take place. Therefore additional cooling mechanisms, such as {molecular or atomic fine structure} line cooling, are required to model a self-consistent dust-driven wind.}

\begin{figure*}
	\centering
	\includegraphics[width = \textwidth]{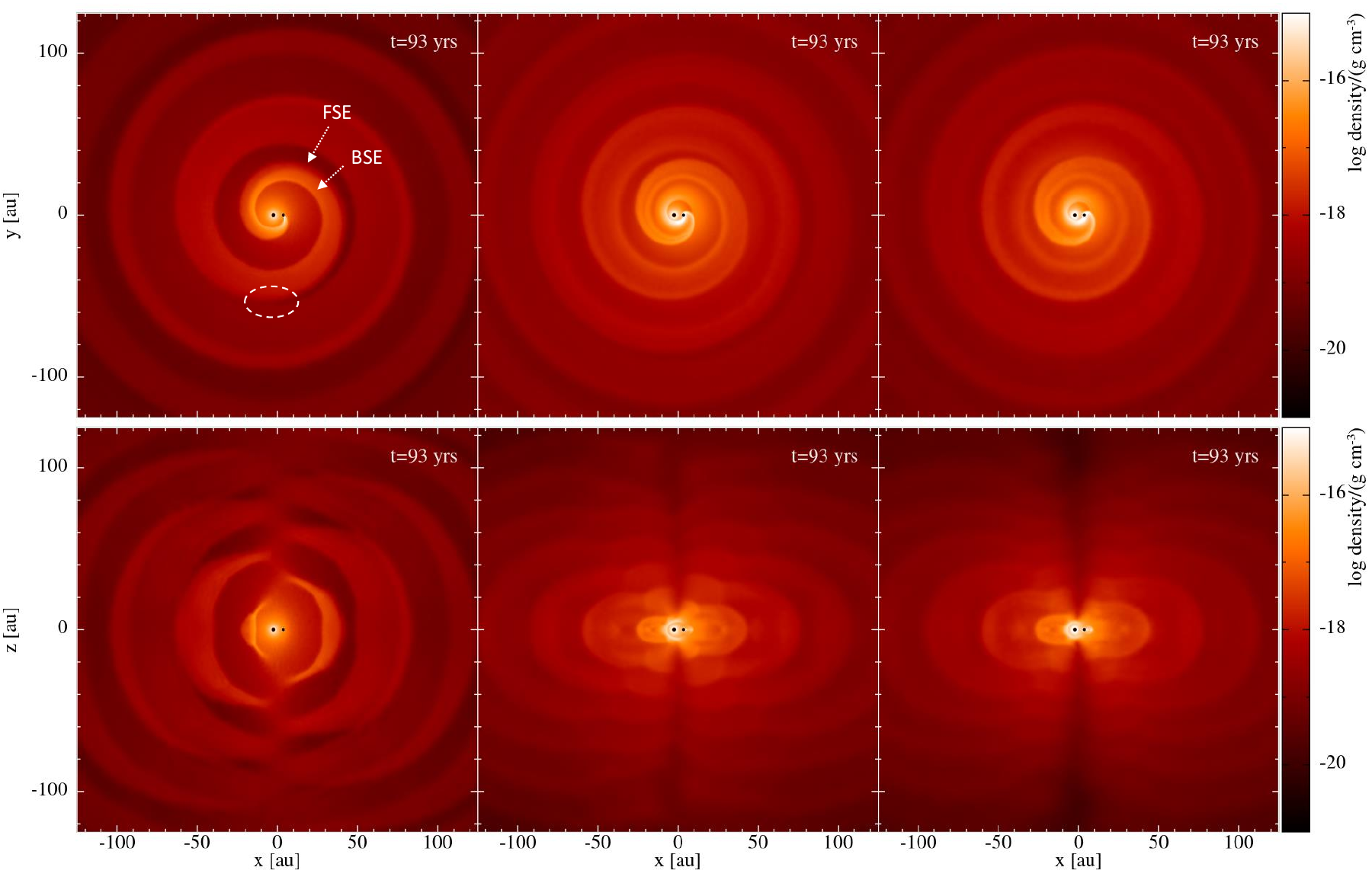}
	\caption{Density profile in a slice through (top row) and perpendicular to (bottom row) the orbital plane of binary models v20e00 (left), v10e00 (middle), and v05e00 (right) at the end of the simulation ($t = 93 \, {\rm yrs}$), with log = (log$_{10}$). The AGB and companion star are annotated as the left and right dot, respectively, not to scale. In the upper left panel, the `backward spiral edge' (BSE) and `frontward spiral edge' (FSE) and the location where they merge into one spiral structure are indicated. {A version without H\,{\sc i} cooling is presented in Fig.~1 in Paper I.}}
	\label{Fig:z125e00}
\end{figure*}

\subsection{Wind structures in circular binary systems, with H\,{\sc i} cooling}

For the three non-eccentric systems v05e00, v10e00 and v20e00 (with initial wind velocities $v_{\rm ini} = 5, 10, \, {\rm and} \, 20 \, {\rm{km\,s^{-1}}}$) the density distribution in the orbital and meridional plane is shown in the top and bottom row in Fig.~\ref{Fig:z125e00}, respectively. {The same plots for the models without cooling are presented in Fig.~1 in Paper I.}
{In models v20e00 (left column) and v10e00 (middle column), the global wind structures presented in these figures appear not significantly altered by the inclusion of H\,{\sc i} cooling. In the low wind velocity model v05e00 (left column), however, the wind morphology is strongly impacted by the H\,{\sc i} cooling. }

\subsubsection{model v20e00}
\begin{figure*}
	\centering
	\includegraphics[width = \textwidth]{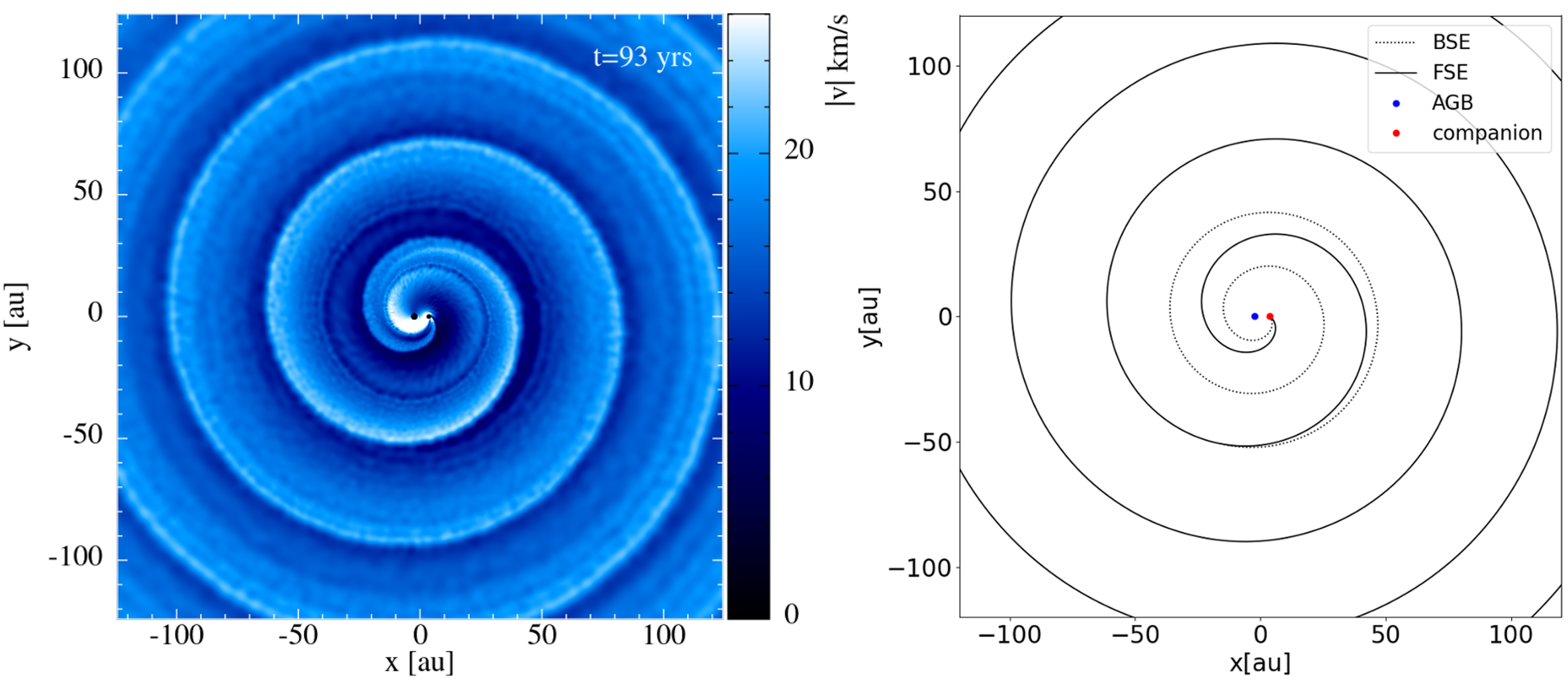}
	\caption{Velocity profile in a slice through the orbital plane (left) and analytical Archimedean spiral fit of the merging `backward spiral edge' (BSE) and `frontward spiral edge' (FSE) of model v20e00 (right). {A version without H\,{\sc i} cooling is presented in Fig.~2 and Fig.~3 of Paper I. }}
	\label{theorSpiral}
\end{figure*}

{In model v20e00 (Fig.~\ref{Fig:z125e00}, left column), the passage of the companion in the AGB wind creates a spiral structure in the wake behind the companion, best seen in the orbital plane. }
The gravitational slingshot effect creates a velocity dispersion leading to the formation of a radially slower inner `backward spiral edge' (BSE) and faster moving outward `frontward spiral edge' (FSE), for a more detailed description, see Paper I and \cite{Maes2021}. 
After $\sim1.25$ orbital period, the FSE merges with the BSE that originated one orbital period earlier (indicated at $x\approx0$, $y\approx-50\, {\rm au}$ in the upper left panel of Fig.~\ref{Fig:z125e00}), and one Archimedean spiral structure results. The spirals can be approximated by a mathematical Archimedean spiral described by equations (6)-(9) in Paper I, with approximate radial velocity of the wind particles within the 2 spiral edges of $v_{r,{\rm BSE}} = 11\,{\rm{km\,s^{-1}}}$ and $v_{r,{\rm FSE}} = 19.5 \,{\rm{km\,s^{-1}}}$. Fig.~\ref{theorSpiral} shows the orbital plane velocity distribution (left) and the analytical Archimedean spiral (right) for comparison. This figure illustrates again how the faster FSE catches up with the slower BSE around $x=0 \, {\rm au}, y = -50 \, {\rm au}$, and after this interaction, the FSE dominates the global orbital plane spiral structure. 

{A difference with model v20e00 in Paper I without H\,{\sc i} cooling} is the much lower (radial) velocity of the FSE and of the material within the two-edged gravity wake behind the companion, which results in a smaller opening angle of the two-edged wake. {This can be seen in Fig.~\ref{Fig:v20e00TPu} that compares the pressure, temperature, internal energy and velocity distribution in the inner wind region of the model with (top row) and without (bottom row) H\,{\sc i} cooling.} 
{Fig.~\ref{Fig:v20e00TPu} reveals that} with H\,{\sc i} cooling the internal energy $u$ and temperature $T$ of the wind particles decrease within the wake, which in turn lowers the pressure $P$ (Eq.~\ref{eq:EOS}) and velocity $v$. 
More precisely, the temperature closely around the companion drops by an order of magnitude from $\sim 10^{5}$ to $\sim 10^{4}\, {\rm K}$. Further away from the companion, after an orbital phase of $\pi/2$ (around $x = 0$, $y<0$ in Fig.~\ref{Fig:v20e00TPu}) the temperature within the gravity wake has decreased from $\sim 3\times 10^{4}$ to $\sim 10^{4} \, {\rm K}$. This cooling by a factor $\sim3$ remains present further out in the wind, up to large radii, as can be seen in Fig.~\ref{Fig:v20e00Tfull}. 
As a result, the radial velocity of the FSE ($v_{r,{\rm FSE}} = 19.5 \,{\rm{km\,s^{-1}}}$) is about $\sim 4.5 \,{\rm{km\,s^{-1}}}$ lower than the approximated velocity in paper I ($v_{r,{\rm FSE,paperI}} = 24 \,{\rm{km\,s^{-1}}}$).
The consequences of this slower FSE are the following. At first, because the velocity of the BSE has not decreased, the width of the gravity wake, so the radial distance between the BSE and FSE, is smaller. Secondly, the FSE catches up with the BSE later, and in this model about half an orbital period later. At last, the inter-arm distance between two windings of the resulting Archimedean spiral is shorter, as this distance is determined by the radial distance the spiral material travels during one orbital period. 
In the meridional plane (Fig.~\ref{Fig:z125e00}, bottom left), the 3D spiral of model v20e00 is displayed in the form of concentric arcs. The same three effects of the lower radial velocity of the material within the spiral as described for the orbital plane spiral can be identified. The banana-shaped arcs are again delineated by the BSE and FSE (respectively the edge closest to and furthest away from the stars). 
{Due to the inclusion of H\,{\sc i} cooling (i) the width of the arcs is smaller, (ii) the FSE catches up with a previous BSE later}, and (iii) the inter-arc distance after overlap is smaller, making the entire morphology appear more compressed.

\subsubsection{model v10e00}
\label{ch:v10e00}
In models v10e00 and v05e00 (middle and right column of Fig.~\ref{Fig:z125e00}), the gravitational impact of the companion on the wind material is stronger due to the lower wind velocity, such that more material can accumulate around the companion.
Similarly to model v20e00, an Archimedean spiral structure results in the orbital plane, but the wind structure close to the companion differs. 

Fig.~\ref{fig:v10e00zoom} displays the density, velocity, temperature and pressure distribution in the orbital plane close to the stars of model v10e00. Instead of one 2-edged spiral, there is one dense inner flow behind the companion that wraps tightly around the AGB star, and that terminates as it collides with a bow shock spiral (around $x\approx 9 \, {\rm au}, y \approx 1 \, {\rm au}$) that originates in front of the moving companion. As can be seen in the density plot in Fig.~\ref{Fig:z125e00} (middle column, top row), the bow shock spiral itself is delineated again by a slower BSE and faster FSE, and after almost $1$ orbit the FSE catches up with the previous BSE (around $x=35 \, {\rm au}$, $y=20\, {\rm au}$ in these plots) such that one Archimedean spiral results (for a more detailed description, see section 3.1.2 in Paper I).  

{The density and pressure distribution in Fig.~\ref{fig:v10e00zoom} show the dense inner spiral flow behind the companion, and the less dense bow shock spiral (and can be compared to Fig.~4 in Paper I without H\,{\sc i} cooling).}
The first, {main impact of the cooling} is in the material closely surrounding the companion (around $x=3.6 \, {\rm au}, y = 0 \, {\rm au}$). In this region, the gas temperature has dropped by more than an order of magnitude from $T \gtrsim 10^{5} \, {\rm K}$ to $T \lesssim 10^{4} \, {\rm K}$, such that the bow shock particles also have significantly lower temperatures. As a result of this decreased temperature, more wind material can accumulate closely around the companion, and is gravitationally captured into an accretion disk. The high density results in a high pressure, such that this circumstellar disk is also clearly visible in the pressure distribution plot (see Sect.~\ref{ch:accrDisks}).
A second difference caused by the cooling is that, analogous to model v20e00, the decrease in internal energy results in a lower pressure, and thereby a lower radial velocity of the material within the high-density Archimedean spiral and FSE. 
This makes the orbital plane inter-arm and meridional plane inter-arc distance smaller, such that the global morphology is more compressed {(which can be seen from comparison between middle row in Fig.~\ref{Fig:z125e00} and Fig.~1 in Paper I)}.

\begin{figure}
	\centering
	\includegraphics[width = 0.49 \textwidth]{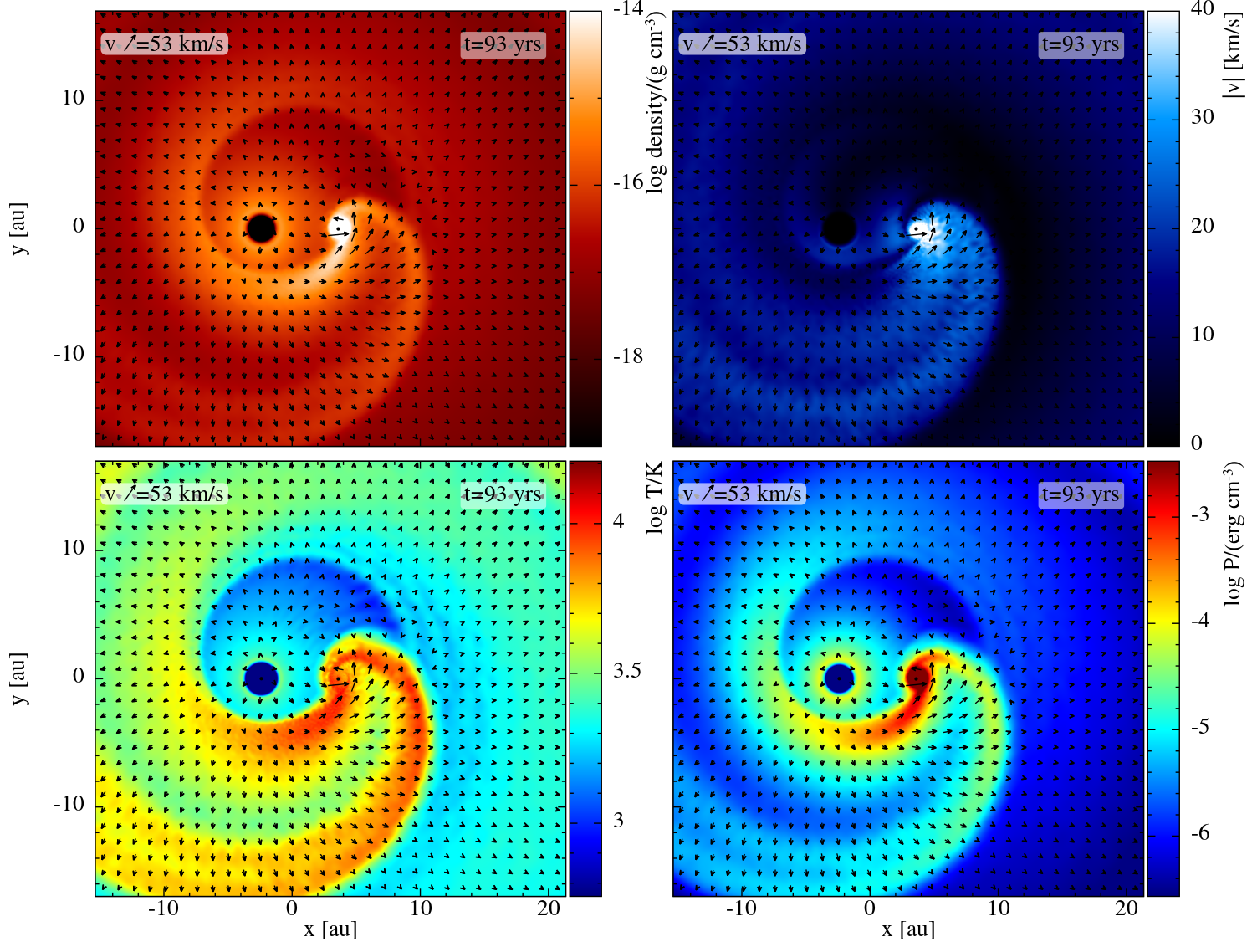}
	\caption{Density, velocity, temperature, and pressure profile in the orbital plane of model v10e00, superimposed with the velocity vector profile which is shown as black arrows. {A version without H\,{\sc i} cooling is presented in Fig.~4 of Paper I. }}
	\label{fig:v10e00zoom}
\end{figure}

\subsubsection{model v05e00}

{Without H\,{\sc i} cooling}, the low AGB wind velocity in model v05e00 resulted in an unstable region around the companion. This led to irregular wind structures with a square orbital plane spiral and irregular meridional morphology (presented in Section 3.1.3. in Paper I). 
With H\,{\sc i} cooling, the morphology of model v05e00 (shown on the right column of Fig.~\ref{Fig:z125e00}) is very similar to the one of model v10e00, with a regular Archimedean spiral in the orbital plane, and regular bicentric arcs in the meridional plane.

Fig.~\ref{v05e00zoom} confirms that the density, velocity, temperature and pressure distribution in the orbital plane close to the stars is almost the same as for model v10e00 (Fig.~\ref{fig:v10e00zoom}). {Fig.~\ref{v05e00densT2ts}} shows the density and temperature distribution around the companion at two different time-steps within one orbital period. Because the material around the companion is now efficiently cooled down to $T \lesssim 10 ^{4} \, {\rm K}$, the previous instabilities with irregular varying wind structures {(visible in Fig.~5 of Paper I)}, no longer appear. Instead there is a stable bow shock and accretion disk (studied in more detail in Sect.~\ref{ch:accrDisks}).

\subsection{Eccentric models}
\label{ch:eccentricModelsHIcooling}
In case the binaries are in eccentric orbits, the wind-companion interaction becomes phase-dependent: the companion spends more time around apastron passage, while the gravitational interaction with the wind is stronger when it resides in the higher density region near the shorter periastron passage.
In the circular orbit models, the wind structure created close to the companion (such as the two-edged spiral or bow shock) does not change. The phase-dependency in the eccentric models implies that the type of structure around the stars varies throughout one orbital period. These varying inner structures then travel radially outward with the wind, making the global morphology asymmetric.

Fig.~\ref{e50fullanotations} shows the density distribution of the binary models with eccentricity $e=0.5$ (top row: v20e05, middle row: v10e50, bottom row: v05e50) in three different perpendicular slices (left: orbital plane $xy$, middle: edge-on $xz$, right: edge-on $yz$). This figure illustrates the asymmetry, diversity and irregularity in three dimensions of the AGB outflow of eccentric binaries. {Compared to the same models without cooling (which can be found in Paper I in Fig. 6 (v20e50, right column), Fig.~10 (v10e50, second row) and Fig.~13 (v05e50, third row), the current density morphologies harbour similar features and are slightly more regular, but still asymmetric, as we explain in the following paragraphs in more detail.}

\begin{figure*}
	\centering
	\includegraphics[width =\textwidth]{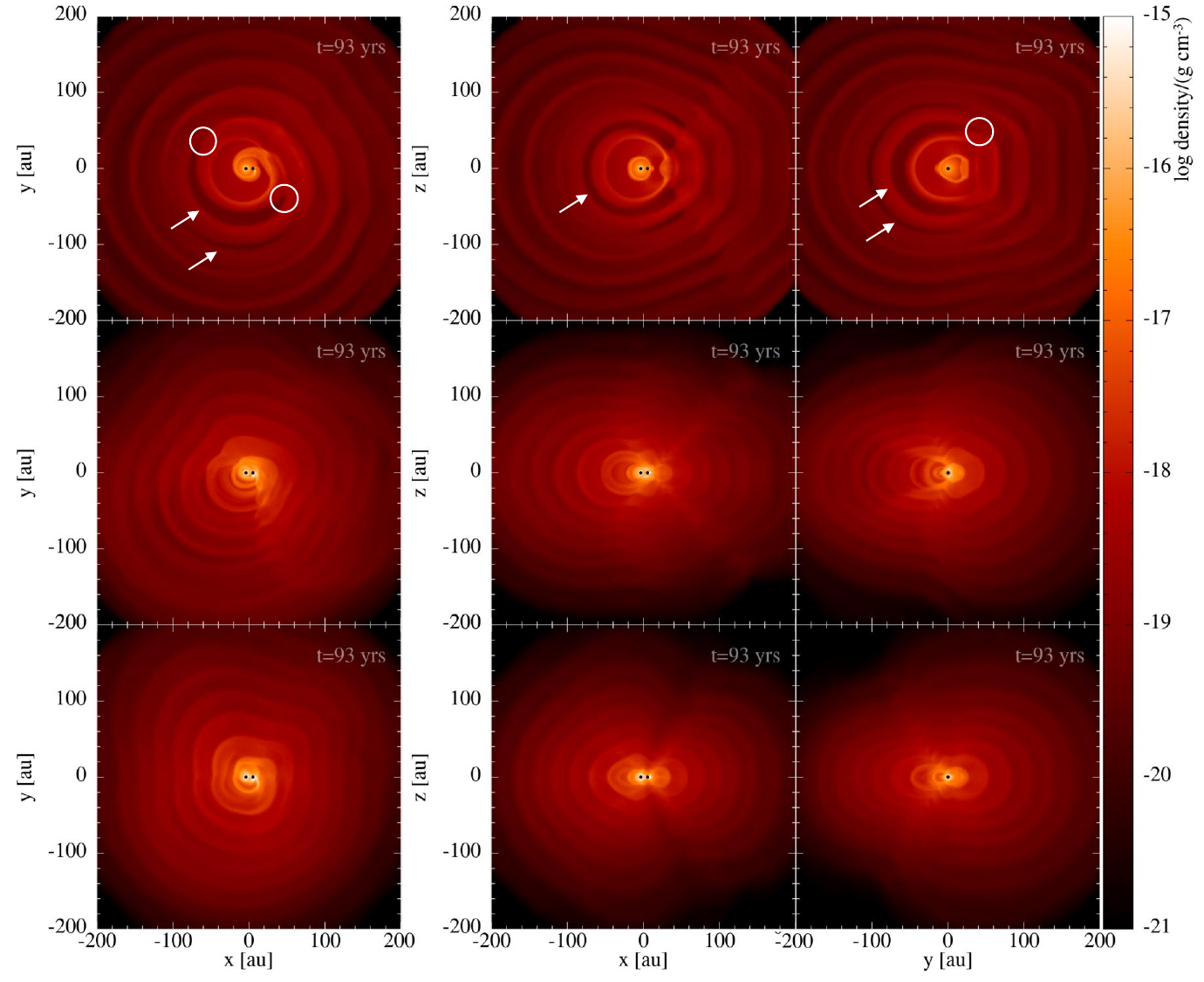}
	\caption{
		Density distribution in a slice through the orbital plane (left, $x$-$y$ plane) and in two perpendicular meridional plane slices (middle, $x$-$z$ plane, and right, $y$-$z$ plane) of the three eccentric binary models v20e50 (top), v10e50 (middle) and v05e50 (bottom). The white arrows indicate low-density gaps (darker regions in figure), and the white circles annotate some of the bifurcations (at edges of low-density gaps). {A version without H\,{\sc i} cooling is presented in Fig.~6 (v20e50, right column), Fig.~10 (v10e50, second row) and Fig.~13 (v05e50, third row) of Paper I. }}
	\label{e50fullanotations}
\end{figure*}

\subsubsection{v20e50}
\label{ch:HIcooling_v20e50}
Apart from overall asymmetry, two other features characteristic of eccentric systems are low-density gaps and bifurcations \citep{Kim2019,Malfait2021}. Bifurcations are locations where two spirals seem to split up or come together, and they arise at the edges of the low-density regions. These features appear in model v20e50, as can be seen in Fig.~\ref{e50fullanotations}, where the darker-coloured low-density gaps are annotated by white arrows, and the bifurcations are encircled. 
These characteristic patterns are created through the phase-dependent wind-companion interaction, which makes the inner wind structure vary between a 2-edged spiral and bow shock, as is explained in detail in Section 3.2.1. of Paper I. The same mechanism is shaping the wind in model v20e50, with the main difference that because of the cooling, the companion is surrounded by an accretion disk throughout all orbital phases, as discussed in more detail in Sect.~\ref{ch:accrDisks}.
Fig.~\ref{v20e50zoom} displays the density, velocity, temperature and pressure distribution close to the stars at 4 consecutive orbital phases. 
In the top row, the stars just passed periastron passage, and the companion is surrounded by an accretion disk and a bow shock. { This is similar to the simulation without H\,{\sc i} cooling (presented in Fig.~9 of Paper I), except that now there is an accretion disk instead of a `high-energy' bubble, such that the radial extent of the bow shock surrounding this region is now also much smaller. } Towards apastron passage (second row), the increase of orbital separation induces a decrease in the orbital velocity of the companion. The radially outward moving bow shock tends to overtake the companion, which results in a larger opening angle, more towards the front of the companion.
After apastron passage (third row) the companion orbits radially inward again with an increasing orbital velocity. This creates a kink in the structure behind the companion, as can be seen in time-step 2 and 3 of Fig.~\ref{v20e50zoom}. 
The decreasing wind-companion interaction around apastron induces again the creation of a 2-edged spiral structure without bow shock (third and fourth row). The companion keeps its accretion disk, which remains present throughout every orbital phase (see Fig.~\ref{fig:v20e50} in Sect.~\ref{ch:accrDisks}).

\subsubsection{v10e50 \& v05e50}
The characteristic low-density gaps and bifurcations do not form in the models with lower wind velocity. Because of the stronger wind-companion interaction intensity, the eccentricity causes asymmetric wind structures that deviate stronger from the regular spiral-arc morphology of non-eccentric binaries. These morphologies are displayed in the density distribution in the middle and bottom row of Fig.~\ref{e50fullanotations}, for model v10e50 and v05e50 respectively.
Contrary to the simulations without cooling (displayed in Figs. 10 and 13 of paper I), the wind morphology is much more regular, and self-similar (recurrent density structures at different radii, originating at specific orbital phases). 
Additionally, the simulations without H\,{\sc i} cooling harboured bipolar outflows, which are now no longer present, as can be seen in the edge-on velocity plots in Fig.~\ref{Fig:v05e50vel_edgeOn}.

Figs.~\ref{fig:v10e50zoom} and ~\ref{fig:v05e50zoom} display the density and temperature distribution in the orbital plane slice close to the stars at two different orbital phases. 
Because of the low wind velocity and the eccentric nature of the orbit, the wind structures around the companion do not only strongly vary because of the phase-dependent wind-companion interaction, but they also interact and collide with other high density structures that originated one orbital period earlier and were very tightly wound around the AGB star.
{The inclusion of cooling prevents the formation of hot, escaping `high-energy' bubbles after apastron passage that were present in paper I (in Figs. 12 and 14).}
However, there are similarities in the shaping of the wind structures. For example, when the star is orbiting from periastron to apastron passage (e.g.\,at $t=80.74 \, {\rm yrs}$), there is always a bow shock, but in the model with cooling, the opening angle is smaller.
For a better visualisation of how these structures vary over each orbital period, and how the bow shock interacts with other high density structures, we refer to {online movie 1} (see Appendix~\ref{movies}) showing the evolution of the density distribution in the orbital plane slice of system v10e50 in Fig.~\ref{fig:v10e50zoom} in a co-rotating coordinate frame.

{The temperature plots in Figs.~\ref{fig:v10e50zoom} and ~\ref{fig:v05e50zoom} confirm that the H\,{\sc i} cooling is efficient in the region closely around the companion sink particle, that used to have temperatures $T > 10^5 \, {\rm K}$ (as shown in Figs. 11 and 14 in paper I).}

\begin{figure}
	\centering
	\includegraphics[width = 0.49 \textwidth]{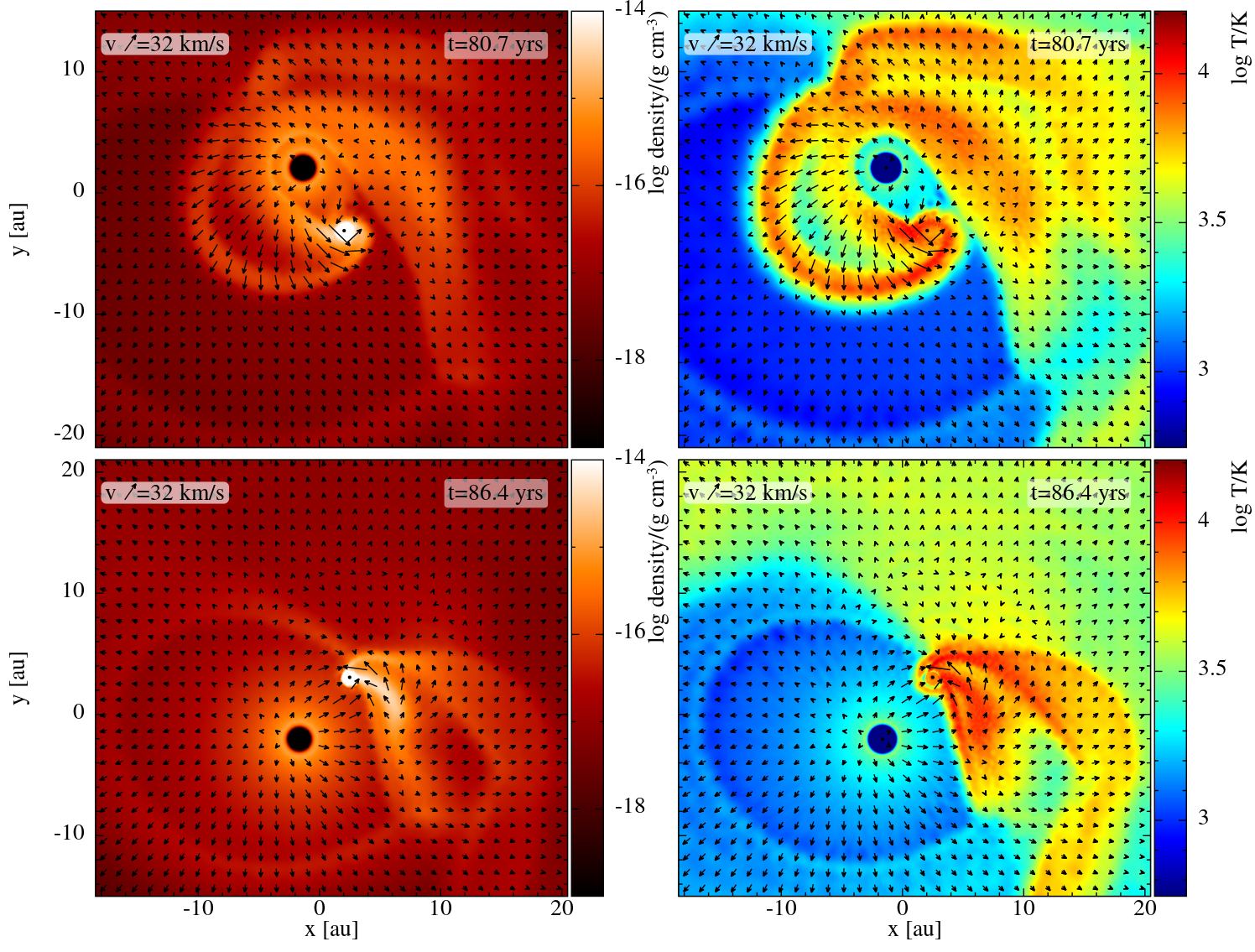}
	\caption{Density and temperature distribution in a slice through the orbital plane of model v10e50 at two consecutive orbital phases. {A version without H\,{\sc i} cooling is presented in Fig.~12 of Paper I. } The time-evolution of this density distribution in a co-rotating frame is available as {online movie 1} (see Appendix~\ref{movies}). }
	\label{fig:v10e50zoom}
\end{figure}

\begin{figure}
	\centering
	\includegraphics[width = 0.49 \textwidth]{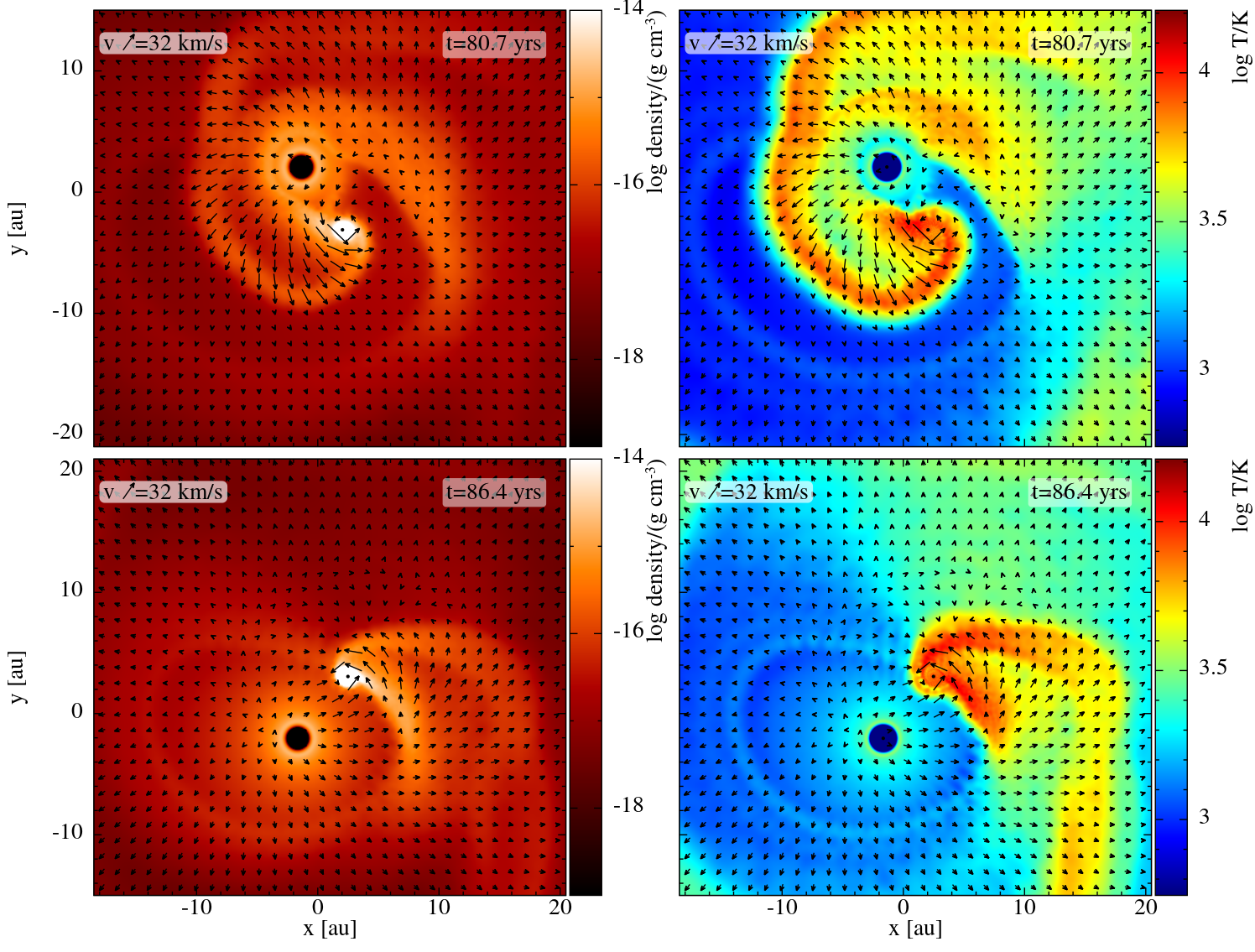}
	\caption{Density and temperature distribution in a slice through the orbital plane of model v05e50 at two consecutive orbital phases.}
	\label{fig:v05e50zoom}
\end{figure}

\section{Accretion disk around the companion}
\label{ch:accrDisks}

Early 3D SPH simulation studies by \cite{TheunsI} showed that gravitationally compressed gas around the companion star can form an accretion disk, if the temperature and pressure of the gas surrounding the companion are not too high, so the gas remains bound to the companion star. This is the case in their isothermal EOS ($\gamma = 1$), and not in the adiabatic EOS ($\gamma = 1.5$) model. They predict that the formation of a disk will be determined by the efficiency of cooling.
{In their isothermal models, an inclined accretion disk with a radius of about $1 \, {\rm au}$ forms, with two streams of matter inflow. The inclination of the disk is believed to be due to numerical round-off errors.
With the inclusion of more realistic cooling and heating prescriptions, including H\,{\sc i} cooling \citep{Spitzer1978} and rotational H$_2$O cooling \citep{Neufeld1993}, \cite{MastrodemosI} also find that permanent accretion disks form in their models, with radii around $0.6 \, {\rm au}$.
The first higher-resolution 3D simulations of disk formation in AGB binaries was conducted by \cite{HuarteEspinosa2013}. They investigated the formation of wind-capture accretion disks in binary systems with orbital separations $a \le 10 \, {\rm au}$ using an adaptive mesh refinement (AMR) code. They describe that accretion disks are able to form because the wind material that will be accreted is not flowing directly towards the companion, but is streaming towards an earlier position, slightly behind the current position of the moving companion. They define the distance between this off-centre location and the current position of the companion as the accretion line impact parameter 
\begin{equation}
    b = 2\, \frac{M_{\rm s}^2 (M_{\rm s}+M_{\rm p})^3 }{M_{\rm p}^5} \, \frac{\displaystyle\left(\frac{1}{a}\right)^{5/2}}{\left(\displaystyle\frac{1}{a} + \displaystyle \frac{1}{r_{\rm w}} \right)^{7/2}},
    \label{Eq:b}
\end{equation}
with $r_{\rm w} = \frac{G M_{\rm p}} {(M_{\rm p} + M_{\rm s}) \,  v_{\infty}^2}$.
The authors conclude that the accretion radius needs to be smaller than $b$, to resolve the formation of the disk.
Further, \cite{Saladino2018} performed SPH simulations with a cooling rate following \cite{Bowen1988} and \cite{Schure2009}, and find that, in case the terminal wind velocity is smaller than the orbital one ($v_\infty<v_{\rm orb}$), an accretion disk forms around the companion, but not when $v_\infty \gtrsim v_{\rm orb}$. Accretion disks were also modelled in the AMR simulations of \cite{ElMellah2020}. They agree that disks are more likely to form in case of lower wind speed, and show that the presence of a disk increases the mass accretion efficiency onto the companion, which can lead to a shrinkage of the orbit if the primary is a few times more massive than the companion star. 
Finally, \cite{Lee2022} modelled wind accretion on a $0.6\, {\rm M_\odot}$ white dwarf orbiting a giant star at a separation of $2 \, {\rm au}$, with a similar cooling prescription as \cite{Saladino2018}. They found that an accretion disk always forms, regardless of the wind velocity, as long as the accretion radius is sufficiently small and the resolution high enough. They differentiate between two types of disks, depending on whether the wind velocity at the location of the companion in a 1D single star wind is larger or smaller than the orbital velocity of the companion. They found that for a slow wind, the disks have a typical radius of $\sim0.15 \, {\rm au}$ and an asymmetric density distribution with two inflowing spiral features. Up to $> 20 \%$ of the mass lost by the AGB star is accreted through the disk onto the companion. For a fast wind, this mass accretion efficiency is lower and consistent with BHL theory. Further, the disks are spatially smaller for increasing wind velocity. }

To study in more detail the accretion disks in our simulations, we recomputed our models with a higher resolution and a smaller accretion radius. We decrease the boundary radius $R_{\rm bound}$ (distance from the AGB sink particle where SPH particles are removed from the simulation) to $20 \, \rm{au}$ to compensate for the large increase in computation time. We change the value of the (constant) mean molecular weight to $\mu = 1.26$, as we expect the hot accretion region to consist mainly of atomic hydrogen, instead of molecular hydrogen ($\mu = 2.38$) in the larger-scale wind. The effect of changing $\mu$ is discussed in Section~\ref{ch:impactMu}.

\subsection{Resolution study}
We perform a resolution study to find the optimal particle resolution and accretion radius to resolve the accretion disk.
We first change the number of particles launched on each sphere around the AGB star (set by \texttt{iwind\_res}, see Table~\ref{resolutionTable} and \cite{Siess2022} for more details), which is equivalent to reducing the mass of the SPH particles.
We start our resolution study with model v10e00 for which an accretion disk was present in the simulations in Section~\ref{ch:impactCooling}, with \texttt{iwind\_res}$ = 5$. Fig.~\ref{fig:v10racc04} shows a slice of the density distribution in the orbital and meridional plane within $2 \, {\rm au}$ from the companion sink particle after $6$ orbital periods with four different resolutions (\texttt{iwind\_res}$ = 5, 7, 8, 9$, corresponding to $\sim  10^{5}, 3\times 10^{5}, 5 \times 10^{5}{\rm, and} \, 7\times 10^{5}$ particles, respectively, {with the increase from lowest to highest resolution approaching that of a factor $2$ in each direction in 3D grid-based codes}). {Although this test covers a limited difference in number of particles, due to computational limitations, it} shows that increasing the resolution is important to resolve the density structures in and closely around the accretion disk. All characteristic disk features (that are discussed in Section~\ref{ch:subsDiskMorph}) are resolved from \texttt{iwind\_res}$=8$ onward.
Next we investigate the effect of decreasing the accretion radius in our models. Following \cite{HuarteEspinosa2013}, to resolve the accretion disk, $R_{\rm{s,accr}}$ must be lower than the accretion line impact parameter $b$ (Eq.~\ref{Eq:b}). This requirement is already fulfilled in our models with $R_{\rm{s,accr}} = 0.04 \, {\rm au}$, since for models v20e00, v10e00, and v05e00, $b = 0.08$, 0.46 and 0.82~au, respectively. 
However, with this setup, model v20e00 does not reveal the presence of an accretion disk even when changing the resolution, as illustrated in Fig.~\ref{v20e00ra04}, comparing the simulations with \texttt{iwind\_res} = 5 and 8 after $6$ orbital periods. But if we decrease the accretion radius to $R_{\rm{s,accr}} = 0.01 \, {\rm au} = 2.15\, R_\odot$, an accretion disk appears at all the explored resolutions. This can be seen in Fig.~\ref{fig:v20racc01} which displays the density distribution for this model with \texttt{iwind\_res} = 6, 7, 8.  Therefore, we find that the criterion given by \cite{HuarteEspinosa2013} of $r_{\rm {s,accr}} < b$ is indeed a necessary, but not a sufficient condition to model the formation of an accretion disk.
Our results are in agreement with the conclusion of \cite{Lee2022} that an accretion disk always forms as long as $R_{\rm{s,accr}}$ is set small enough.
For this model, the higher the resolution, the more confined and the higher the maximum density in the disk becomes. This trend is less obvious in model v10e00 (see Fig.~\ref{fig:v10racc04}). For model v20e00, the disk might become more confined and denser for higher resolution, but due to computational limits it is not feasible to calculate simulations with $R_{\rm{s,accr}} = 0.01 \, {\rm{au}}$ and \texttt{iwind\_res} $> 8$. 
We therefore conclude that the best setup for the analysis of accretion disks in our models is an accretion radius of $R_{\rm{s,accr}} = 0.01 \, {\rm{au}} \approx2.15 R_\odot$, to ensure that the accretion disk is always present, and a resolution \texttt{iwind\_res} set to $8$, to resolve the disk properly, while keeping the computational cost reasonable. 
The simulations are calculated over $10$ orbital periods, such that we can study the evolution over time of the disk mass, size and mass accretion. With these settings, one model takes approximately four weeks to complete.

\subsection{Impact of mean molecular weight choice}
\label{ch:impactMu}
As explained before, we changed the mean molecular weight from $\mu = 2.38$ to $\mu = 1.26$ to model the hot accretion region around the companion in an optimal way, considering that $\mu$ is not yet self-consistently calculated in our model.
{This, however, has implications on the internal energy $u$ (Eq.~\ref{Eq:energy}), the density distribution and velocity of the wind. }
To investigate this, we look at the analytical 1D profiles produced by our \textsc{Phantom} models. We conclude that decreasing $\mu$ does not significantly change the analytic 1D density, temperature and pressure profile. However, the internal energy $u$ increases by a factor $\sim 2$, and the radial wind velocity undergoes a stronger acceleration. Table~\ref{ta:impactMuOnVel} gives the wind velocity of the 1D models at $r = a = 6 \, {\rm au}$ and the terminal velocity $v_\infty$ for the different input velocities, for $\mu = 2.38$ and $\mu = 1.26$.
Changing $\mu$ from $2.38$ to $1.26$ makes the 1D wind velocity $\sim 1-3 \, {\rm km \, s^{-1}}$ higher at the location of our companion, and adds $\sim 2-4 \, {\rm km \, s^{-1}}$ to the terminal velocity, with the largest difference for the lowest input wind velocity. 
The wind velocity at the location of the companion is one of the prime parameters determining the strength and effects of the wind-companion interaction, so this increase in wind velocity will impact the wind structures and the accretion disk in our simulations. This emphasises the importance of calculating $\mu$ self-consistently in future models, so that all regions of the wind can be simultaneously modelled in a physically correct way.

The change in $\mu$ in our models leads to a shift in the velocity parameter with respect to the models in Section~\ref{ch:impactCooling}. From Table~\ref{ta:impactMuOnVel} we notice that the velocity profile in model v05e00 with $\mu=1.26$ resembles well the one in model v10e00 with $\mu = 2.38$, and the velocity profile in model v10e00 with $\mu=1.26$ lies in-between models v10e00 and v20e00 with $\mu=2.38$.

The density structures surrounding the accretion disks in the orbital plane in the models v05e00, v10e00 and v20e00 with $\mu=1.26$ are displayed in Fig.~\ref{fig:ADz10}. As expected from the similar wind velocity profile, the density profile of model v05e00$_{\mu 1.26}$ resembles well the one of model v10e00$_{\mu 2.38}$ (Fig.~\ref{fig:v10e00zoom}). The companion is surrounded by a bow shock spiral in front of the circumstellar disk, and a second flow attached behind the companion.
The increased wind velocity due to the decreased $\mu$ in model v10e00 results in a weaker wind-companion interaction, such that the companion in model v10e00$_{\mu 1.26}$ is surrounded by a weaker bow shock, with a smaller opening angle, compared to model v10e00$_{\mu 2.38}$. 
The velocity shift in model v20e00 is not that large, such that the companion and accretion disk in model v20e00$_{\mu 1.26}$ are in front of a 2-edged spiral that is very similar to model v20e00$_{\mu 2.38}$ (Fig.~\ref{Fig:z125e00}).

\begin{table}[H]
    \caption{Impact of $\mu$ on the wind velocity}
    \begin{center}
    \begin{tabular}{lcc}
    \hline
    \hline
    \centering
       $\mu = 2.38$ &$v_{\rm w}(6 \,{\rm au})$ & $v_\infty$ \\
      \hline
    v05e00&$9.5$&$11.7$  \\
    v10e00&$12.7$&$14.6$  \\
    v20e00&$21.4$&$22.7$  \\
      \hline
    \end{tabular}
    \begin{tabular}{lcc}
    \hline
    \hline
    \centering
        $\mu = 1.26$& $v_{\rm w}(6 \,{\rm au})$
       &$v_\infty$ \\
      \hline
     v05e00& $12.4$&$15.5$  \\
    v10e00& $14.8$&$17.7$  \\
    v20e00& $22.6$&$24.8$  \\      \hline
    \end{tabular}
    
    \end{center}
    {\footnotesize{\textbf{Notes.}} 1D analytical wind velocity at location of companion $v_{\rm w}(6 \,{\rm au})$ and terminal wind velocity $v_\infty$, both in $\rm km \, s^{-1}$, for $\mu = 2.38$ (left), and $\mu = 1.26$ (right).} 
    \label{ta:impactMuOnVel}
\end{table}

\begin{figure*}
    \centering
    \includegraphics[width = \textwidth]{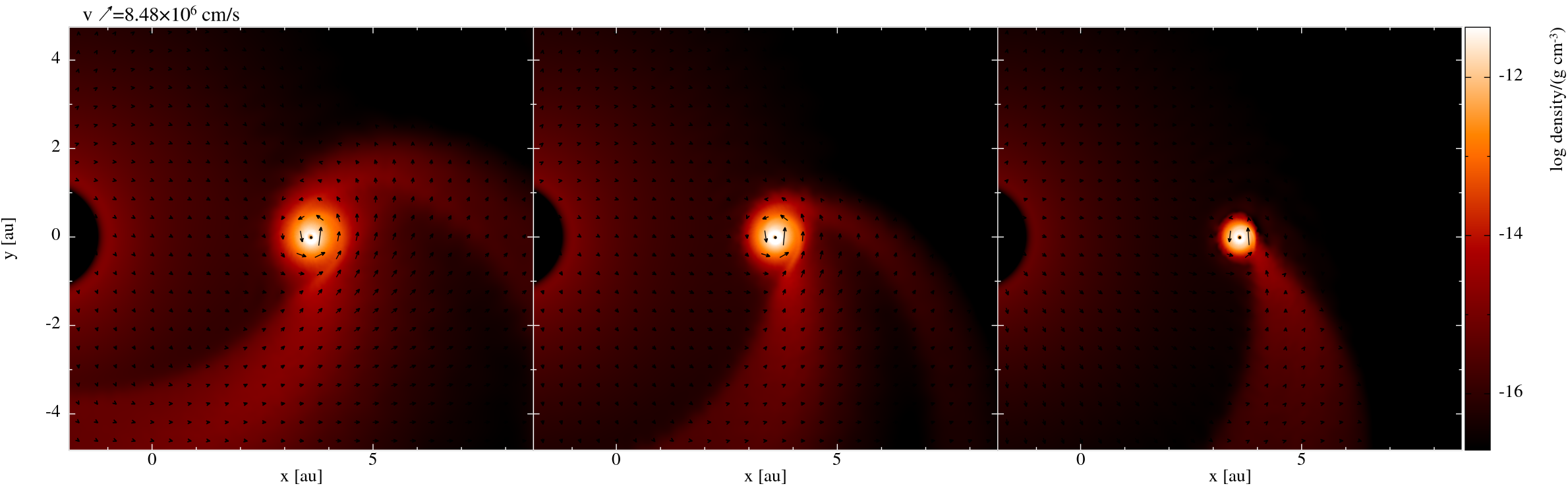}
    \caption{Density distribution in a slice through the orbital plane of models v05e00 (left), v10e00 (middle), and v20e00 (right) with $\mu = 1.26$, $R_{\rm{s,accr}} = 0.01 \, {\rm{au}}$ and \texttt{iwind\_res} $= 8$.}
    \label{fig:ADz10}
\end{figure*}

\subsection{Disk morphology}
\label{ch:subsDiskMorph}
Figs.~\ref{fig:ADzoomv05e00}, ~\ref{fig:ADzoomv10e00} and ~\ref{fig:ADzoomv20e00} display the density distribution, overplotted with the velocity vector field (in the reference frame of the stationary center-of-mass), of the accretion disk in the orbital plane and in 2 perpendicular slices for the circular orbit models v05e00, v10e00 and v20e00, respectively, after $10$ orbital periods. The orbital plane plots show that the disks consist of high-density material that is rotating counter-clockwise around the companion (prograde with respect to the orbital motion). The perpendicular plane slices reveal the 3D flared shape of the accretion disks, that are aligned with and quasi-symmetric with respect to the orbital plane.
As seen in Section~\ref{ch:impactCooling}, for low wind velocity, the relatively strong interaction of the companion with the wind produces a bow shock spiral. This bow shock protects the disk from exposure to the wind and thereby from ram pressure stripping, such that it can grow more easily \citep{Lee2022}, as seen in model v05e00 (see left plot in Fig.~\ref{fig:ADz10}). In model v20e00 there is no bow shock, and in model v10e00 there is the onset of a protecting bow shock (Fig.~\ref{fig:ADz10}). As a consequence, in Figs.~\ref{fig:ADzoomv05e00}, ~\ref{fig:ADzoomv10e00} and ~\ref{fig:ADzoomv20e00}, the disk size is larger for model v05e00, a bit smaller in model v10e00, and significantly smaller for model v20e00.
Additionally, in models v05e00 and v10e00, a thin, high-density, spiral-arm structure, that extends counter-clockwise from the flow behind the companion into the accretion disk, is resolved. Material accumulates through this structure behind the companion into the accretion disk (see velocity vectors in Fig.~\ref{fig:ADz10}). As shown in Fig.~\ref{fig:v05e00ADTemp}, that displays the temperature profile in and around the accretion disk for model v05e00, this structure has high temperatures close to $10^4\, {\rm K}$ in the orbital plane. 
Similarly, there is a second high-density structure connecting the accretion disk and the bow shock in front of the companion, that is more difficult to resolve in the density plots.
The two structures continue inside the accretion disk and are annotated in Fig.~\ref{fig:spiralsv05e00} to guide the eye. 

The density distribution in the edge-on planes in Figs.~\ref{fig:ADzoomv05e00}, ~\ref{fig:ADzoomv10e00} and~\ref{fig:ADzoomv20e00} provides insight in the 3D morphology of the disks, which is not axisymmetric. 
In models v05e00 and v10e00, the disk has a large flaring angle and reaches large scale heights on the bow shock side ($y>0$ in upper right plot of Figs.~\ref{fig:ADzoomv05e00} and ~\ref{fig:ADzoomv10e00}), whereas on the opposite side ($y<0$ in same plot), the disk is less puffy and more confined towards the orbital plane. Moreover, the disk side towards the AGB star ($x<3.6 \, {\rm au}$ in the bottom plot) appears to have a roughly constant flaring angle at increasing distance from the companion, whereas the flaring angle on the side away from the AGB star ($x> 3.6 \, {\rm au}$) seems to increase with radius (best visible at very small radii, close to the location of the companion at $x=3.6 \, {\rm au}$). 
The smaller disk in model v20e00 (Fig.~\ref{fig:ADzoomv20e00}) is more symmetric, as is expected from the absence of a bow shock, which affects the outer region of the 3D disk morphology in the lower velocity models.
These asymmetries are analysed in more detail in Section~\ref{ch:asymmetriesDisks}.

\begin{figure}
    \includegraphics[width = 0.49\textwidth]{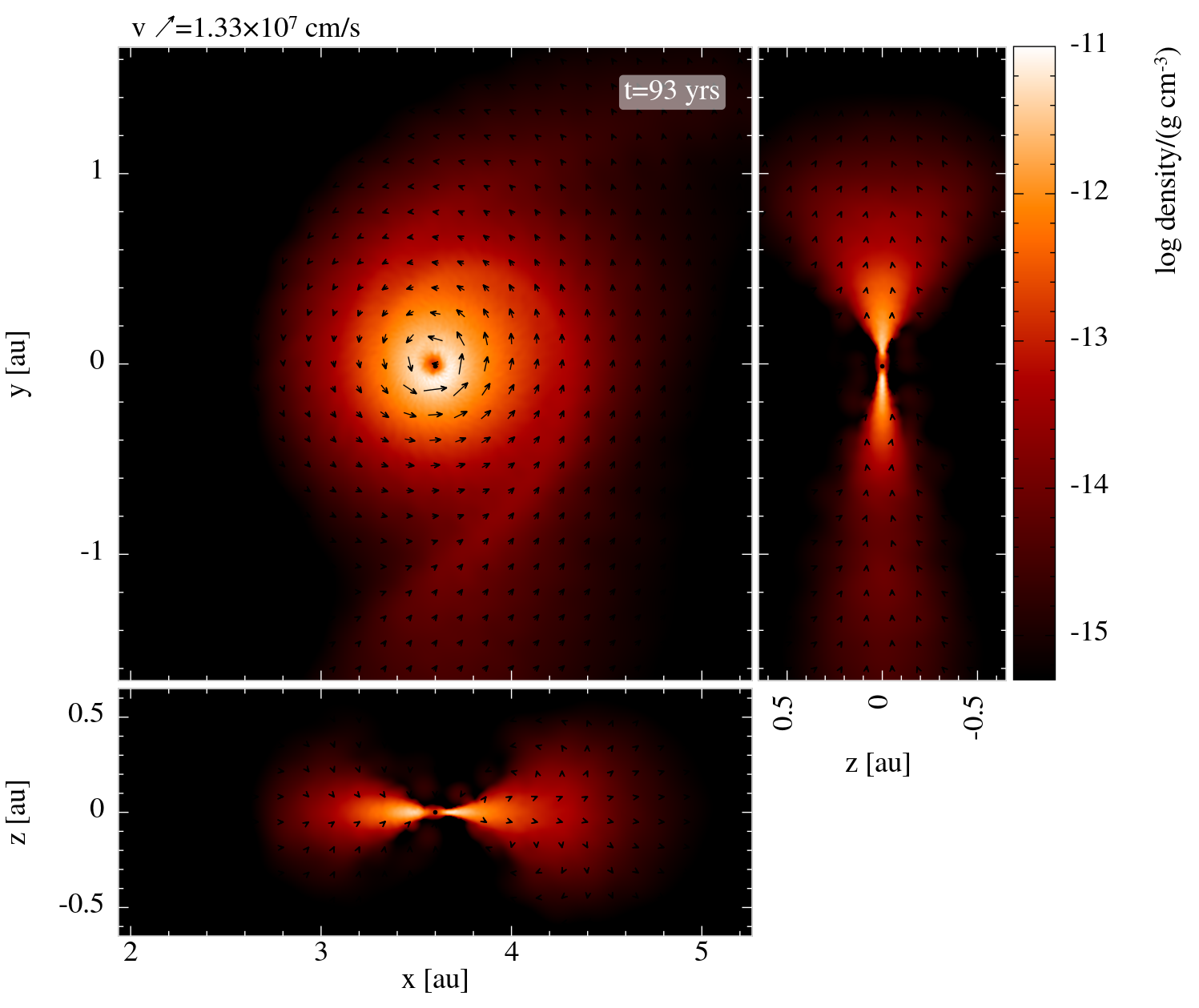}
    \caption{Density distribution in the accretion disk around the companion in model v05e00 (with $R_{\rm{s,accr}} = 0.01 \, {\rm au}$ and \texttt{iwind\_res} $= 8$) in a slice through the orbital plane (upper left), and in two perpendicular slices (through $y=0$ in lower panel, through $x=3.6$ in upper right panel). The AGB sink particle is located at $y=0, z=0, x<0$. Velocity vectors are annotated by black arrows on top of the density distribution.}
    \label{fig:ADzoomv05e00}
\end{figure}

\begin{figure}
    \includegraphics[width = 0.49\textwidth]{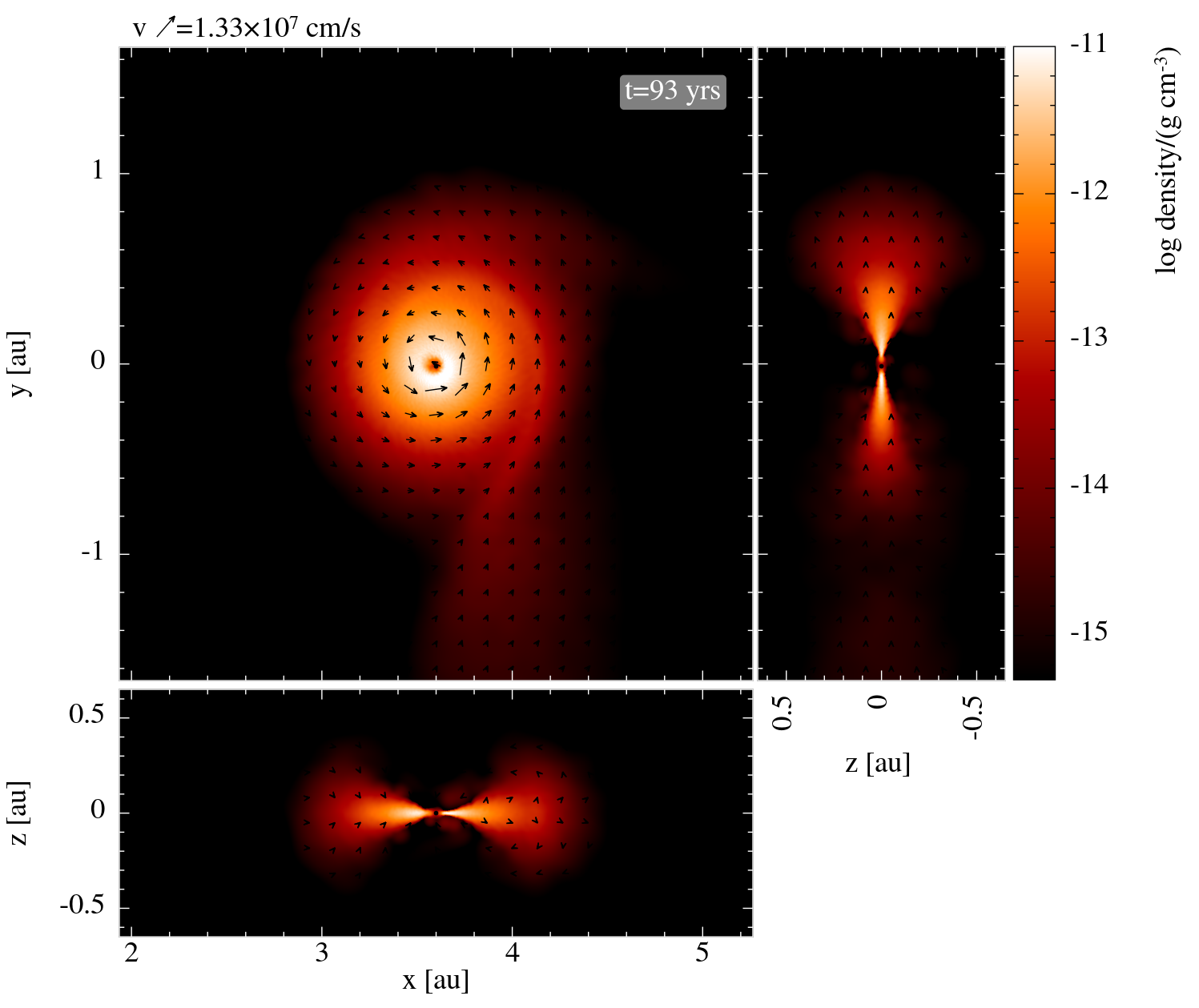}
    \caption{Same as Fig.~\ref{fig:ADzoomv05e00}, but for model v10e00.}
    \label{fig:ADzoomv10e00}
\end{figure}

\begin{figure}
    \includegraphics[width = 0.49\textwidth]{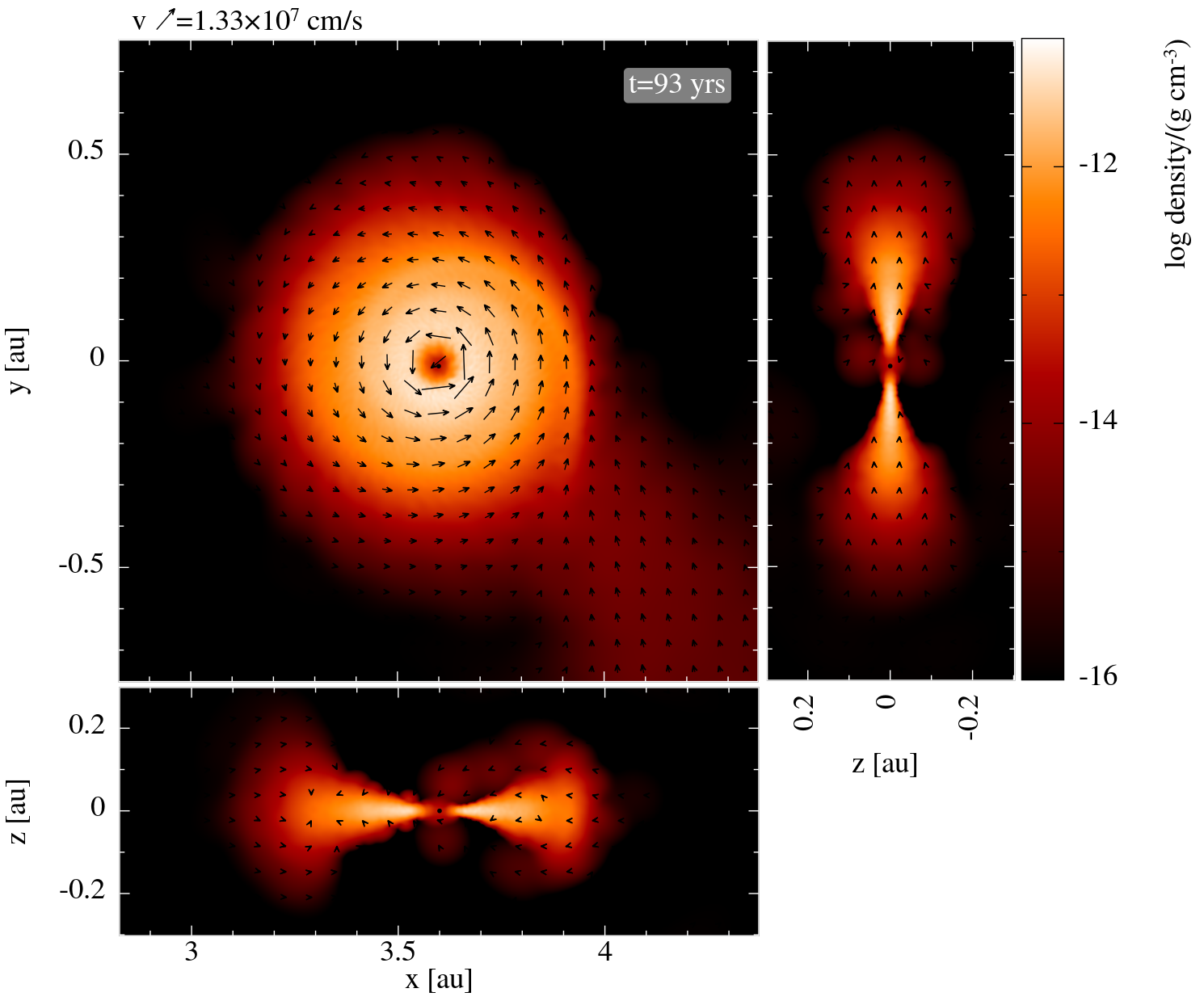}
    \caption{Same as Fig.~\ref{fig:ADzoomv05e00}, but for model v20e00. Note that the spatial limits are smaller compared to the accretion disks in models v05e00 and v10e00.}
    \label{fig:ADzoomv20e00}
\end{figure}

\begin{figure}
    \centering
    \includegraphics[width = 0.49\textwidth]{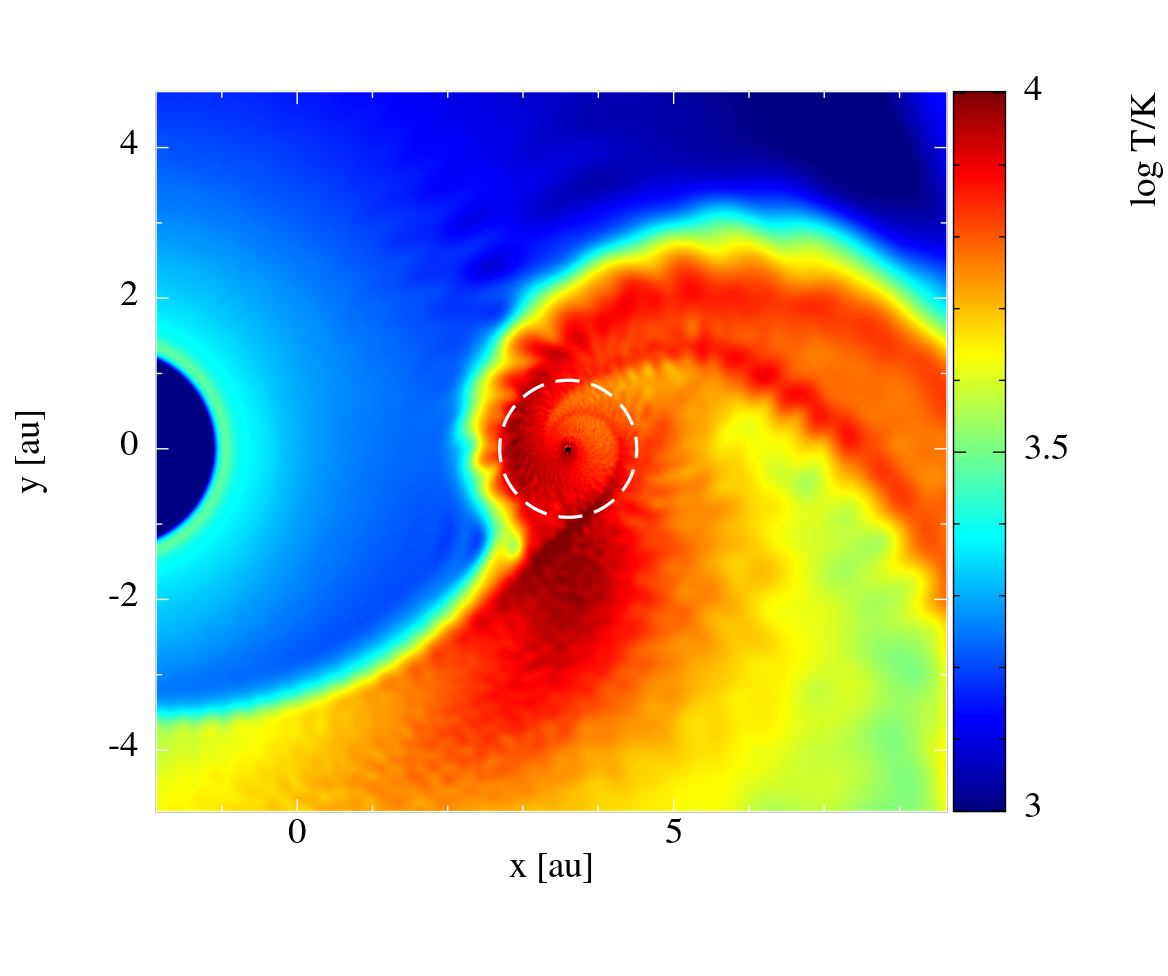}
    \caption{Temperature distribution in a slice through the orbital plane focusing on the accretion disk in model v05e00, of which the estimated radius is annotated in white.}
    \label{fig:v05e00ADTemp}
\end{figure}

\subsection{Quantitative estimates of radius, mass and scale height}
\label{ch:quantitativeAnalysisDisks}

To estimate the radius $r$, density scale height $H(r)$, and mass $M_\mathrm{disk}$ of the accretion disks, we adopt a similar method as \citet{Lee2022}. We integrate the disk structure over radii $r_i$, with a step $\Delta r = 0.01 \, {\rm au}$, starting from $R_{\rm{s,accr}} = 0.01 \, {\rm au}$ (with origin of the coordinate system at the location of the companion sink particle as in Fig~\ref{fig:thetaDir}). 
For each point ($r_i$, $\theta_j$, $z=0$) in the reference frame of the companion, the density scale height $H(r_i,\theta_j)$ can be calculated on a vertical line through ($r_i$,$\theta_j$).
A theoretical prediction for the vertical density that holds for a thin disk where the gas is isothermal in the $z$-direction, is 
\begin{equation}
    \rho_{\rm theor}(r,z) = \rho_{\rm{max}}(r) \, e^{-\frac{1}{2} \left(\frac{z}{H}\right)^2} 
    \label{eq:RhoEquation}
\end{equation}
with $\rho_{\rm{max}}$ the maximum density on the vertical line \citep[generally in the orbital plane, at $z=0$,][]{Chiang1997}. 
On a vertical line through a point $(r_i,\theta_i)$, we define two scale heights as
\begin{equation}
     \rho(z=H) = \frac{\rho_{\rm{max}}(r_i,\theta_j)}{\sqrt{e}} \approx 0.60 \, \rho_{\rm{max}}(r_i,\theta_j)   
    \label{eq:SH}
\end{equation}
and
\begin{equation}
    \rho(z = 2H = \Tilde{H}) = \frac{\rho_{\rm{max}}(r_i,\theta_j)}{e^2} \approx 0.14 \, \rho_{\rm{max}}(r_i,\theta_j) .
    \label{eq:2SH}
\end{equation}
We estimate the scale height $H(r_i,\theta_j)$ in our simulations as the mean of the absolute values of the positive and negative $z$-values for which the density drops by a factor $1/\sqrt{e}$. 
Because our disks are not axisymmetric, the density scale height at a given radius, $H(r_i)$, is then calculated as the median of $10$ $H(r_i,\theta_j)$ values for equally distributed $\theta_j$ between $0$ and $2\pi$. By calculating the median, and not mean value, outlier values are ignored (when other high-density structures are present at certain $\theta_j$). 
At each radius $r_i$, the added mass w.r.t. the previous step, $M(r_i) - M(r_{i-1})$, is calculated as the sum of the mass of all SPH particles for which $r_{i-1}<r<r_i$ and $|z|<\Tilde{H}(r_i)$. Note that we use a disk height of $\Tilde{H}(r_i)$, as this provides a more realistic estimate for the total mass in the disk, and has a negligible effect on our estimate of the disk radius. 
We assume that the disk edge is reached if the relative added mass w.r.t. the previously calculated mass satisfies the following criterion: 
\begin{equation}
    {\rm{if }}\, \, \frac{M(r_i)-M(r_{i-1})}{M(r_{i-1})} \le 0.3 \, \frac{\Delta r}{\rm{au}} \ \implies r_{\rm{disk}}\le r_i .
    \label{eq:diskRadius}
\end{equation}
The factor $0.3$ in the criterion is the smallest constant towards which the relative added mass consistently keeps decreasing for all models. 

The resulting estimates for the disk radius $r_{\rm disk}$, total disk mass $M_{\rm disk}$, and scale height estimates at radius $r_{\rm disk}$ for the circular orbit models are given in Table~\ref{ta:diskPropCircModels}. 
We find that the lower the wind velocity, the larger the disk radius, scale height and mass estimates, with the most extended disk having a radius $r_{\rm disk}= 0.91 \, {\rm au}$ and a total mass estimated to $M_{\rm disk}\approx 10^{-7} \, M_\odot$.
This confirms that with decreasing AGB wind velocity, more material can be captured by the companion in the accretion disk, which results in a more massive and extended disk. 
Note that $M_{\rm{disk}}/ M_{\rm{s}} \ll 0.1$, so no gravitational instabilities and formation of moons or planets is expected \citep{Armitage2020}.
Fig.~\ref{fig:MdiskEv} shows the evolution of the total disk mass throughout $10$ orbital periods. The roughly constant disk mass near the end of the simulation indicates that the mass accretion rate into the disk equals the mass accretion rate onto the companion sink particle, which is discussed in more detail in Section~\ref{ch:massAccr}.

\begin{table}[H]
    \caption{Disk properties for circular models}
    \begin{center}
    \begin{tabular}{lcccc}
    \hline
    \hline
    \centering
       & $r_{\rm disk}$ [au] & $M_{\rm disk} \, [M_\odot]$ & $H_{\rm{max}}$ [au] & $\Tilde{H}_{\rm{max}}$ [au]\\
    \hline
      v05e00  & $0.91$ & $1.05 \times 10^{-7}$  & $0.20$ & $0.58$ \\
      v10e00 & $0.70$& $9.04 \times 10^{-8}$ & $0.14$ & $0.32$\\
      v20e00 &  $0.41$ & $3.45 \times 10^{-8}$  & $0.06$ & $0.13$\\
      \hline
    \end{tabular}
    \end{center}

    {\footnotesize{\textbf{Notes.} Properties of the disk in circular orbit models as calculated from the analysis described in Section~\ref{ch:quantitativeAnalysisDisks}, at $t = 93 \, {\rm yrs}$.}}
    \label{ta:diskPropCircModels}
\end{table}

\begin{figure}
    \centering
    \includegraphics[width = 0.5\textwidth]{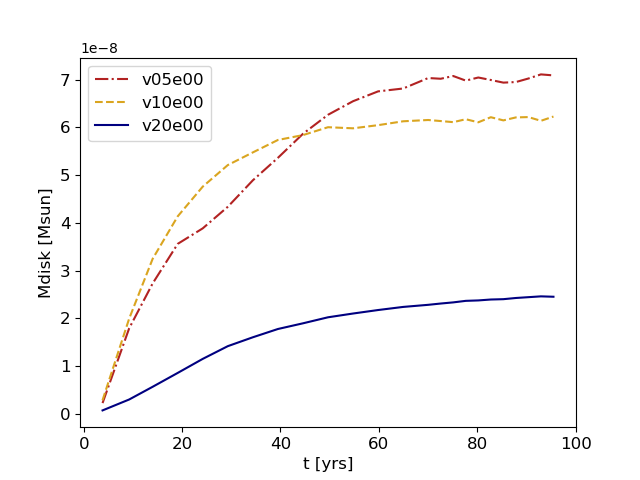}
    \caption{Evolution of the accretion disk mass for the circular orbit binaries.}
    \label{fig:MdiskEv}
\end{figure}

To get a more quantitative view of the disk shape, Fig.~\ref{fig:e00SH_plot} displays the radial profile of the scale height $H(r)$, the aspect ratio $H(r)/r$, and the midplane density $\rho_{\rm max}$ within the disk (averaged over $\theta$). The density scale height and aspect ratio profiles are similar for all models.
The aspect ratio $H(r)/r$ increases with $r$, indicating that the disk is flaring.
At the inner rim of the disk, for $r \lesssim 0.07 \, {\rm au}$, the aspect ratio shows a strong decrease. This indicates that if the disk would be optically thick, the inner rim might self-shadow the disk. However, the midplane density $\rho_{\rm max}$ is small in this region.

\begin{figure}
    \centering
    \includegraphics[width = 0.4 \textwidth]{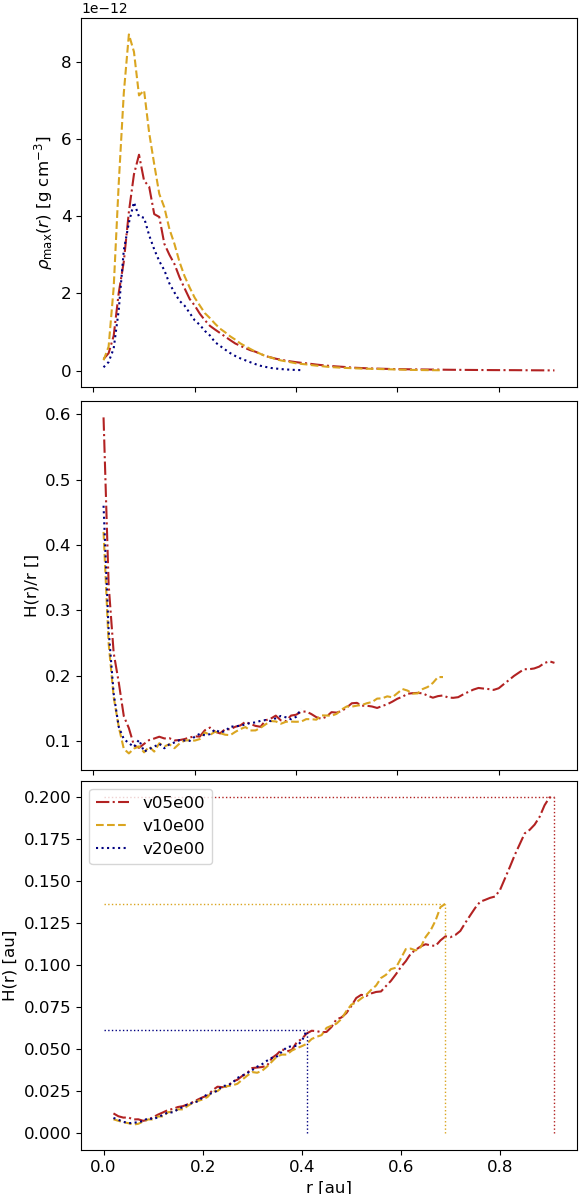}
    \caption{Midplane density $\rho_{\rm max}$, aspect ratio $H(r)/r$, and median density scale height $H(r)$ (Eq.~\ref{eq:SH}) as a function of disk radius for the circular orbit models. The dotted lines in the bottom plot indicate the scale height at the estimated outer disk radius (Eq.~\ref{eq:diskRadius}).
    }
    \label{fig:e00SH_plot}
\end{figure}

\subsection{Disk velocity distribution}
Figs.~\ref{fig:vr_e00} and~\ref{fig:vt_e00} display the radial and tangential velocity field, respectively, of the accretion disks in the orbital plane. In contrast to previous figures, the coordinate frame is now the reference frame of the companion, such that the companion sink particle (annotated as a black dot) is located at $(0,0,0)$ and has a zero-velocity $(0,0,0)$, and the AGB star is located at the left side on the $y=0$ axis. The estimated radius of the disk (calculated as described in Section~\ref{ch:quantitativeAnalysisDisks}) is annotated as a dotted circle. 
In the tangential velocity field plots (Fig.~\ref{fig:vt_e00}) it can be seen that the material orbiting around the companion is accelerating as it gets closer to the companion. The radial velocity profiles in Fig.~\ref{fig:vr_e00} are more complex, with material approaching and moving away from the companion on different sides of the disk. {This indicates that the particles are on eccentric orbits around the companion, with apastron (periastron) located where $v_r$ switches from positive to negative (negative to positive) in the counterclockwise orbit.} Interpreting these velocity fields {in more detail} is not intuitively straight-forward, taking into account that these are the velocities in the co-rotating frame of the moving companion and accretion disk. Note however that the tangential velocities are much larger than the radial velocity, and will therefore dominate the velocity field.
The angular average of the absolute value of the radial velocity as a function of the radial coordinate $r$ is given in Fig.~\ref{fig:mean_abs_vr}, and shows that $|v_r| \ll |v_t|$.

To understand the general behaviour of particles in this velocity field, Fig.~\ref{fig:vt_vr_vKepl} shows the tangential $v_t$ and radial velocity distribution $v_r$ as a function of the radial coordinate $r$, averaged over all $\theta$-directions in the orbital plane. The tangential velocity follows very closely a Keplerian rotation in the inner disk up to $r \approx 0.2 \, {\rm au}$ after which the rotation becomes sub-Keplerian. The mean (outward-directed, not absolute value) radial velocity is for most radii close to, but below zero, indicating that material slowly spirals inward while it is rapidly rotating around the companion. Once particles get very close to the companion ($r< 0.05 \, {\rm au}$), within a few accretion radii ($R_{\rm{s,accr}} = 0.01 \, {\rm au}$), they are rapidly accreted.

The two spiral-like features that were identified as enhanced density structures in model v05e00, appear in the radial velocity distribution in Fig.~\ref{fig:vr_e00} as material with lower radial velocity compared to closely surrounding material (so blueshifted in the figure). The spirals are drawn on top of the radial velocity distribution of model v05e00 in Fig.~\ref{fig:spiralsv05e00_vr} to guide the eye.
This indicates that in these spiral-like structures, material tends to move inward towards the companion sink particle (compared relatively to the surrounding material). The first spiral is visible on the $x>0$ side (white-coloured structure in the red velocity area). It initiates behind and goes towards the front of the companion, where it spirals in towards the companion sink particle ($y>0$, dark-blue in the blue velocity area). The second spiral-like structure initiates in the bow shock in front of the companion (dark blue structure in the blue velocity area at $y>0$).
In model v20e00, no density enhanced spiral-like feature was resolved in the density distribution (Fig.~\ref{fig:ADzoomv20e00}), but the radial velocity distribution in Fig.~\ref{fig:vr_e00} (right plot) shows that there is a spiral structure with radially inflowing material. {This structure originates from the flow behind the companion, in which material is pulled towards the disk with negative radial velocities up to $-40 \, {\rm km \, s^{-1}}$, and is visible as a thin blue spiral arising at the lower-right hand side of the disk.}

\begin{figure*}
    \centering
    \includegraphics[width = \textwidth]{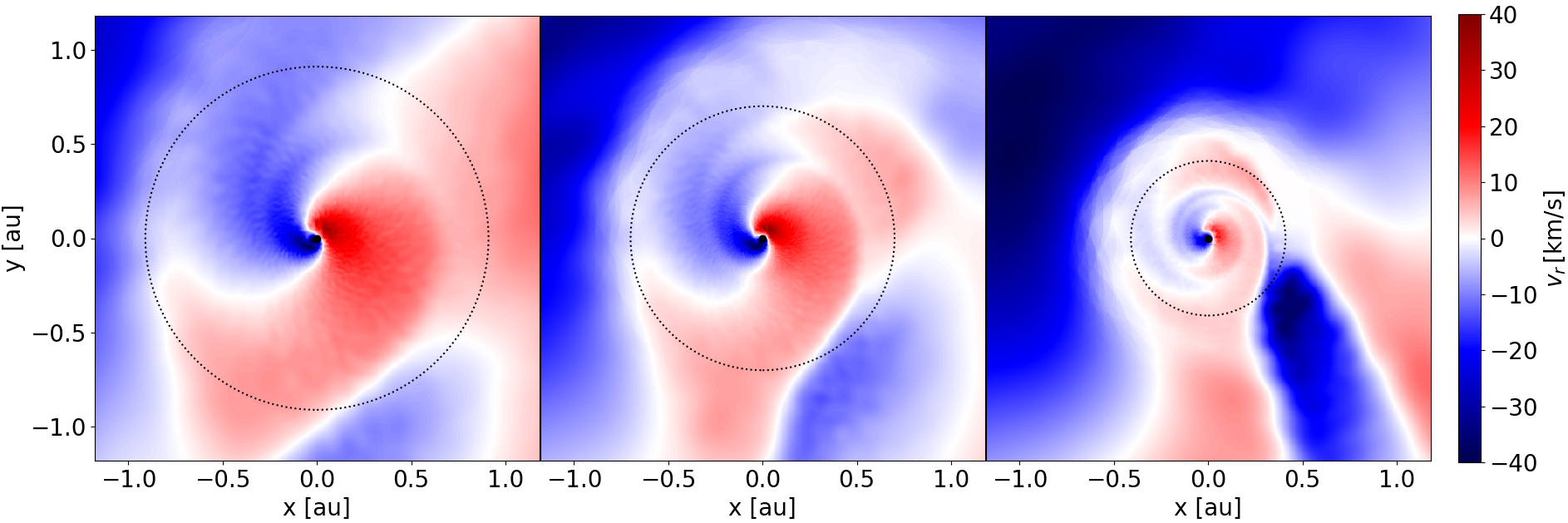}
    \caption{Radial velocity distribution ($v_r$) in accretion disks in models v05e00 (left), v10e00 (middle) and v20e00 (right). The coordinate frame is such that the companion sink particle (annotated as a black dot) is located at $(0,0)$ and has velocity $(0,0,0)$, and the AGB star is located at the left side on the $y=0$ axis.  The estimated radius of the circumstellar disks, as calculated in Section~\ref{ch:quantitativeAnalysisDisks}, are annotated as a dotted circle.}
    \label{fig:vr_e00}
\end{figure*}

\subsection{Asymmetries within the accretion disk}
\label{ch:asymmetriesDisks}
From the density plots of Figs.~\ref{fig:ADzoomv05e00} and~\ref{fig:ADzoomv10e00}, discussed in Section~\ref{ch:subsDiskMorph}, we concluded that the bow-shock surrounded disk in the low-velocity models do not have an axisymmetric density distribution.
To investigate the asymmetries in more detail, the scale height $H(r)$ (Eq.~\ref{eq:SH}), aspect ratio $H(r)/r$, and midplane density $\rho_{\rm max}$ are calculated separately in $4$ different directions within the disk, $\theta \in [-\pi/4,\pi/4], [\pi/4, 3\pi/4], [3\pi/4, 5\pi/4], [5\pi/4,7\pi/4]$.
The resulting radial profiles are shown in Figs.~\ref{fig:v05e00_diffTh_SH_ar},~\ref{fig:v10e00_diffTh_SH_ar} and~\ref{fig:v20e00_diffTh_SH_ar} for models v05e00, v10e00 and v20e00, respectively.
The profiles for model v20e00 are very similar for the different $\theta$ regions, confirming the approximately axisymmetric morphology depicted in Fig.~\ref{fig:ADzoomv20e00}. 
For models v05e00 and v10e00, the strongest asymmetry is in the quadrant $\theta \sim \pi/2$, that corresponds to the front side of the moving companion ($y>0$ side, see \ref{fig:thetaDir}). In this region,  the scale height, and aspect ratio  are larger compared to the median values in the full disk (compare yellow dashed to black solid line in Figs.~\ref{fig:v05e00_diffTh_SH_ar},~\ref{fig:v10e00_diffTh_SH_ar}) because of the onset of the bow shock. The quadrant centred around $\theta \sim 3\pi /2$, corresponding to the region behind the companion ($y<0$ side, see \ref{fig:thetaDir}), has the smallest radial and vertical extent.
We additionally note that the disk side towards the AGB star ($\theta \sim \pi$) appears to have a more constant flaring angle, compared to the stronger increasing flaring angle in the other quadrants. This is reflected in the relatively low and slowly increasing aspect ratio at radii from $r\approx 0.1$ to $0.3 \, {\rm au}$.

\begin{figure}
    \centering
    \includegraphics[width = 0.4 \textwidth]{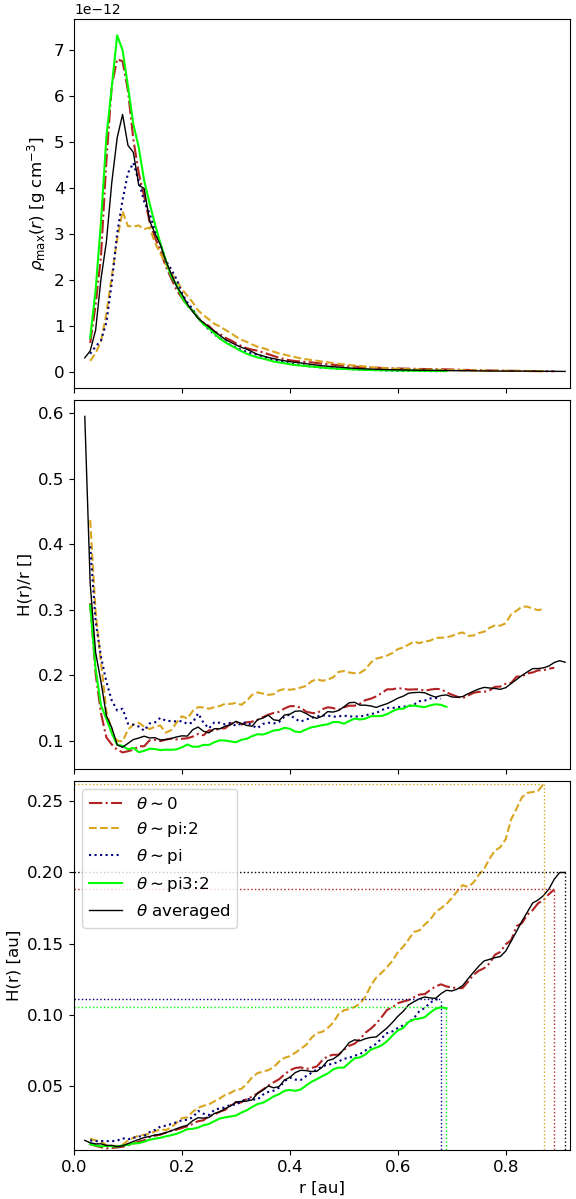}
    \caption{Midplane density $\rho_{\rm max}$, aspect ratio $H(r)/r$, and density scale height $H(r)$ (Eq.~\ref{eq:SH}), for model v05e00, at 4 different $\theta$ quadrants centred around $\theta = 0, \pi/2, \pi$ and $3\pi/2$, and averaged over all $\theta$ (full, black line), giving insight into the asymmetric nature of the accretion disk. $\theta = 0$ is the side away from AGB star, as annotated in Fig.~\ref{fig:thetaDir}.}
    \label{fig:v05e00_diffTh_SH_ar}
\end{figure}

\subsection{Accretion disks in eccentric orbits}
\label{ch:accrDisksEcc}

In the eccentric orbit models v05e50, v10e50 and v20e50, the wind-companion interaction is phase-dependent. Therefore the disk morphology and properties change throughout one orbital period, ranging from a larger bow-shock surrounded type, to a smaller symmetric disk type, as well as to in-between phases and more asymmetric shapes. 

To see how the accretion disk properties change, Figs.~\ref{fig:v05e50},~\ref{fig:v10e50}, and~\ref{fig:v20e50} visualise the disk at $4$ timesteps within one orbital period for models v05e50, v10e50, and v20e50, respectively. These $4$ timesteps are selected to optimally represent the different disk morphologies that appear during each orbital period. The figures contain (i) the orbital plane density distribution for $r \lesssim 10 \, {\rm au}$ (top row), (ii) the orbital plane density distribution of the disk in the reference frame of the companion at coordinate $(0,0,0)$, with the AGB located on the $x<0, y=0$ axis (as indicate by the arrow, middle row), and (iii) the radial velocity within the disk in the orbital plane, again in the reference frame of the companion (bottom row). 
Animations of the density distribution during the full last 4 orbital periods in a co-rotating frame are also available as {online movie 2 and 3} for model v05e50 and v20e50, respectively (see Appendix~\ref{movies}).

Additionally, estimates of the disk radius $r_{\rm disk}$, mass $M_{\rm disk}$ and scale heights at these $4$ different phases are given in Tables~\ref{ta:diskPropv05e50}, ~\ref{ta:diskPropv10e50} and~\ref{ta:diskPropv20e50} for models v05e50, v10e50 and v20e50, respectively (properties calculated as described in Section~\ref{ch:quantitativeAnalysisDisks}). 

Models v05e50 and v10e50 behave in a very similar way. Their accretion disk and surrounding structures are influenced by the phase-dependent orbital velocity and distance to the AGB (determining the density of the surrounding wind), and the high-density structures the companion encounters on its orbit. The complex interplay between these effects results in the different disk shapes and velocity profiles visualised in Figs.~\ref{fig:v05e50}, ~\ref{fig:v10e50} and {online movie 2} (see Appendix~\ref{movies}). After apastron passage, when the companion approaches the AGB, pronounced spiral density enhancements are present in the disk (see snapshot for $t = 85.5 \, {\rm yrs}$), while around periastron passage, when the orbital velocity is maximal, the disk has a strong asymmetry in the density distribution {and velocity field} (see $t = 88.3 \, {\rm yrs}$). 
Compared to the circular orbit, the disks contain more mass, but are spatially slightly smaller. This is not the case for model v20e50, for which the disk is less massive (Fig.~\ref{fig:v20e50}). 
This could be attributed to the fact that in the low-velocity models, the disk encounters a slowly outward moving high-density structure around apastron that can feed the disk (see $t = 85.5$ and $93 \, {\rm yrs}$ in the top row of Fig.~\ref{fig:v05e50}). The accretion disk surrounded companion seems to collide with the high-density stream in front of its orbital movement direction. 
In model v20e50, the companion does not encounter such enhanced density structures in the wind, but the structures around the companion do vary from a 2-edged spiral (e.g.\,at $t = 86.3 \, {\rm yrs}$) to bow-shock (e.g.\,at $t = 90.6 \, {\rm yrs}$, see top row of Fig.~\ref{fig:v20e50} and {online movie 3}, see Appendix~\ref{movies}). This also affects the accretion disk shape, as can be seen in the lower two rows of this Figure. In this model, the eccentric motion leads to a slightly more extended accretion disk that is less massive than in the circular case (compare Table~\ref{ta:diskPropv20e50} to Table~\ref{ta:diskPropCircModels}). 
{For all models, the radial velocity field in the vicinity of the companion (bottom row of Figs.~\ref{fig:v05e50}, \ref{fig:v10e50}, \ref{fig:v20e50}) strongly varies over the orbital phases. This shows that the orbit of the particles within the disk, and the direction from which material is accreted onto the disk, can change rapidly.}

\begin{figure*}
    \centering
    \includegraphics[width = \textwidth]{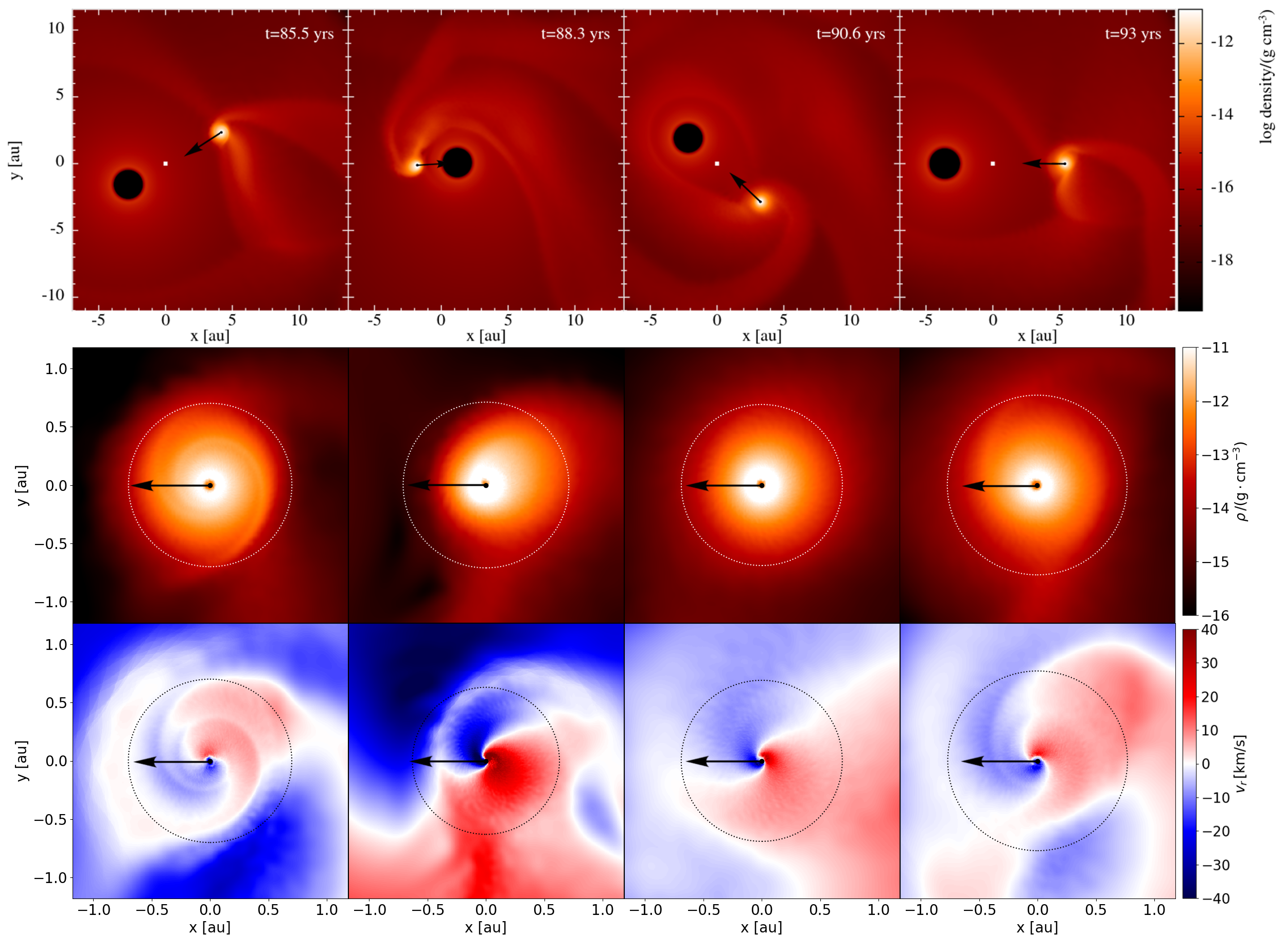}
    \caption{Top: density distribution in a slice through the orbital plane of model v05e50 at $4$ consecutive orbital phases within one orbital period. The center-of-mass is annotated by a white square. The black arrow indicates the direction from the companion to the location of the AGB. {Online movie 2} shows the time-evolution of this density distribution during the final $4$ orbital periods of the simulation, in a co-rotating frame (see Appendix~\ref{movies}).
    Middle: more zoomed-in density distribution in a slice through the orbital plane at the same $4$ consecutive orbital phases, but the coordinate frame is rotated and translated such that the companion sink particle (annotated as a black dot) is located at $(0,0)$, and the AGB star on the $x<0$ side of the $y=0$ axis.
    Bottom: radial velocity distribution of the accretion disk in model v05e50 at the same timesteps, and in the same coordinate frame as in the middle row. The companion has a zero-velocity $(0,0,0)$. 
    The estimated radius of the circumstellar disk as calculated in Section~\ref{ch:quantitativeAnalysisDisks} is annotated as a dotted circle in middle and bottom row.}
    \label{fig:v05e50}
\end{figure*}

\begin{table}[H]
    \caption{Disk properties v05e50 at different orbital phases}
    \begin{center}
    \begin{tabular}{lccccc}
    \hline
    \hline
    \centering
       $t$ [yrs] & $\phi$ & $r_{\rm disk}$ [au] & $M_{\rm disk} \, [M_\odot]$ & $H_{\rm{max}}$ [au] & $\Tilde{H}_{\rm{max}}$ [au]\\
        \hline
      85.5  & $0.16\, \pi$  & $0.70$ & $2.10 \times 10^{-7}$ & $0.13$ & $0.30$\\
      88.3  & $1.02\, \pi$  & $0.71$ & $2.07 \times 10^{-7}$ & $0.18$ & $0.34$\\
      90.6  & $1.77\, \pi$  & $0.69$ & $1.69 \times 10^{-7}$ & $0.13$ & $0.31$\\
      93.0  & $2.00\, \pi$  & $0.77$ & $1.65 \times 10^{-7}$ & $0.16$  & $0.38$\\
    \end{tabular}
    \end{center}
    {\footnotesize{\textbf{Notes.} Disk radius $r_{\rm disk}$, mass $M_{\rm disk}$, and scale heights $H_{\rm max}$ and $\Tilde{H}_{\rm max}$ as calculated from the analysis described in Section~\ref{ch:quantitativeAnalysisDisks}, at 4 consecutive orbital phases $\phi$ for model v05e50 (corresponding to the plots in Fig.~\ref{fig:v05e50}) .}}
    \label{ta:diskPropv05e50}
\end{table}

\begin{table}[H]
    \caption{Disk properties v20e50 at different orbital phases}
    \begin{center}
    \begin{tabular}{lccccc}
    \hline
    \hline
    \centering
       $t$ [yrs] & $\phi$ & $r_{\rm disk}$ [au] & $M_{\rm disk} \, [M_\odot]$ & $H_{\rm{max}}$ [au] & $\Tilde{H}_{\rm{max}}$ [au] \\
     \hline
      86.3& $0.26\, \pi$  & $0.49$ & $2.69 \times 10^{-8}$  & $0.09$ & $0.19$\\
      88.8& $1.33\, \pi$  & $0.42$ & $2.40 \times 10^{-8}$  & $0.06$ & $0.16$\\
      90.6& $1.77\, \pi$  & $0.47$ & $2.82 \times 10^{-8}$  & $0.08$ & $0.17$\\
      93.0& $2.00\, \pi$  & $0.40$ & $2.94 \times 10^{-8}$  & $0.06$ & $0.13$\\
    \end{tabular}
    \end{center}
    {\footnotesize{\textbf{Notes.} Same as Table~\ref{ta:diskPropv05e50}, but for model v20e50 (corresponding to the plots in Fig.~\ref{fig:v20e50}).}}
    \label{ta:diskPropv20e50}
\end{table}

\begin{figure*}
    \centering
    \includegraphics[width = \textwidth]{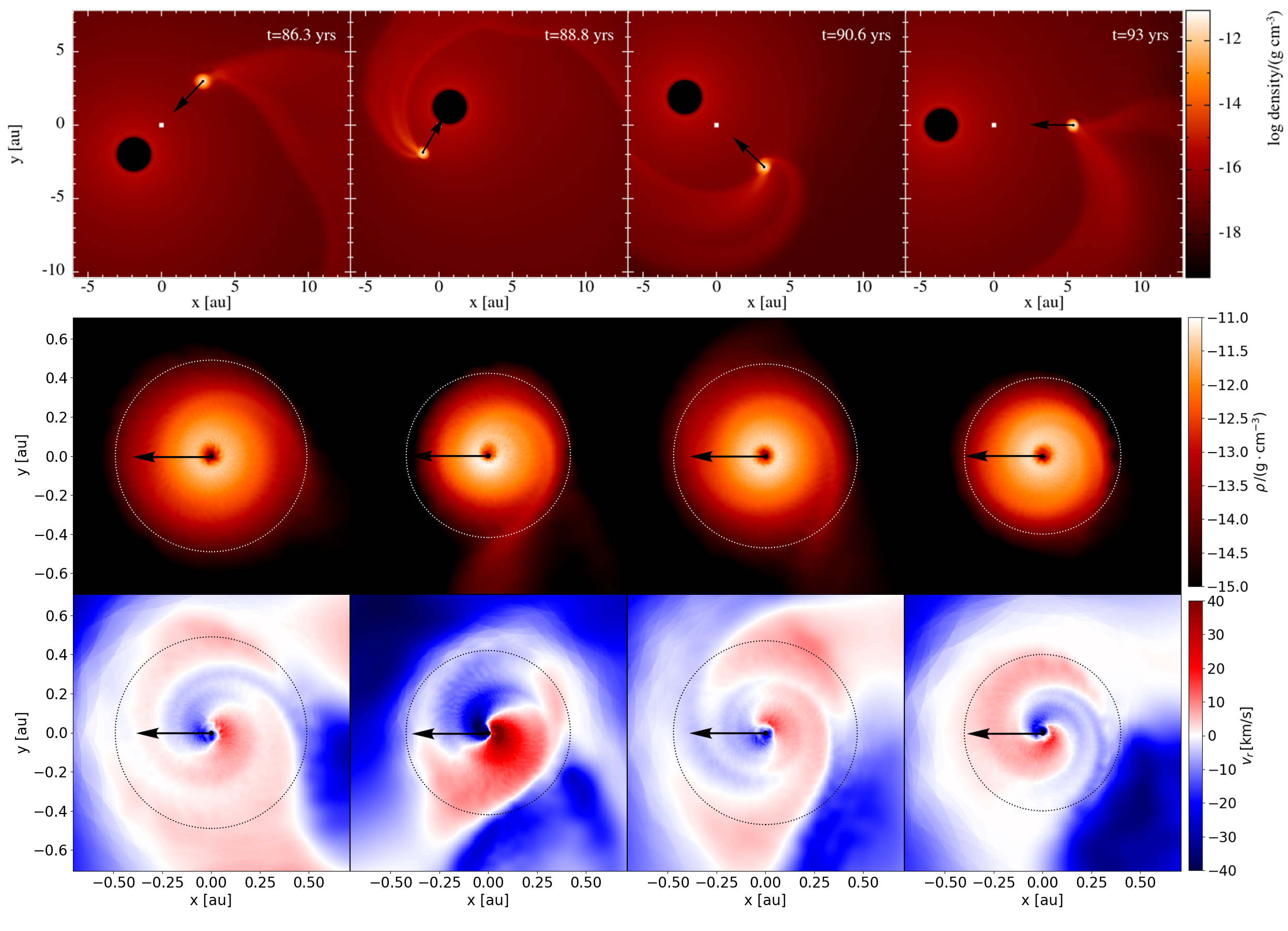}
    \caption{Same as Fig.~\ref{fig:v05e50}, but for model v20e50. {Online movie 3} shows the time-evolution of the orbital plane density distribution during the final $4$ orbital periods of the simulation, in a co-rotating frame (see Appendix~\ref{movies}).}
    \label{fig:v20e50}
\end{figure*}

\subsection{Mass and angular momentum accretion}
\label{ch:massAccr}

In theory it is expected that in case the local wind velocity is much larger than the orbital velocity, $v_{\rm w} \gg v_{\rm{orb}}$, the Bondi-Hoyle-Lyttleton (BHL) wind accretion scenario \citep{BHL1,BHL2} provides a fairly good description of the mass accretion \citep{TheunsII}. If $v_{\rm w} \lesssim v_{\rm{orb}}$ this is not the case, since the relatively low kinetic energy of the wind increases the impact of the gravitational interaction of the companion on the wind, resulting in a more complex accretion scenario.
The theoretical instantaneous BHL mass accretion efficiency is given by 
\begin{equation}
    \label{Eq:BHLMaccrEcc}
    \beta_{\rm BHL} = \frac{\dot{M}_\mathrm{acc}}{\dot{M}_\mathrm{wind}} = \alpha_{\rm{BHL}}
    \left( \frac{G M_{\rm{s}}}{r v_{\rm{w}}^{2}} \right)^{2} 
    \left[ 1 + \left(\frac{v_{\rm{orb}}}{v_{\rm{w}}}\right)^{2}\right]^{-3/2}
\end{equation}
with $\alpha_{\rm{BHL}} \approx 0.75$ a constant, $G$ the gravitational constant, $r$ the instantaneous orbital separation, $v_{\rm{w}}$ the estimated local wind velocity at the semi-major axis ($r =a$) in a single-star model, and 
 $v_{\rm{orb}} = \sqrt{G (M_{\rm s} +M_{\rm p}) ( 2/r - 1/a ) }$ 
the instantaneous relative orbital velocity of the stars with respect to each other. 
In our models, the estimated local wind velocities are $v_{\rm{w}} \approx 12.7,\ 15.0,\ 22.8 \, {\rm{km\,s^{-1}}}$ for models with initial velocity $v_{\rm{ini}} = 5,\ 10,\ 20 \, {\rm{km\,s^{-1}}}$ respectively (see Table~\ref{ta:impactMuOnVel}). The relative orbital velocity is $v_{\rm{orb}} \approx 19 \, {\rm{km\,s^{-1}}}$ in the circular orbits, and varies between $11$--$33 \, \rm{km \, s^{-1}}$ in the eccentric orbits. We do not expect our models to follow the BHL approximation, but it is still valuable to compare our results against these approximations.

The mass accretion efficiency in our models is calculated as the ratio of the mass accreted by the companion sink particle to the mass lost by the AGB star. 
Table~\ref{ta:BHLaverageMaccrEff} gives the time-averaged BHL mass accretion efficiency calculated over the last 2 orbital periods, when the wind structures have reached a steady state configuration, from the simulations  $\langle\beta_{\rm sim}\rangle$ and according to the theory 
$\langle\beta_{\rm BHL}\rangle$.
As expected from the relatively low wind velocity compared to the relative orbital velocity, the simulated accretion rates are up to a factor of 2 higher than the theoretical BHL values. The accretion efficiencies  vary between $\sim 0.04$--$0.21$, and material is accreted at a higher rate when the wind velocity is lower. Eccentricity is found to increase the average accretion rate, especially in case of a low wind velocity. 

To investigate the mass accretion in the eccentric cases in more detail, Fig.~\ref{fig:MassAccrEff} displays the phase-dependent calculated and BHL mass accretion efficiency for models v05e50 (top), v10e50 (middle) and v20e50 (bottom).
In all models, there is a peak in the accretion efficiency around periastron passage, as the companion passes closer to the AGB star in a higher density region. This accretion peak is sharp because of the high orbital velocity that makes periastron passage short.
All models show a secondary peak that is caused by the interaction of the accretion disk  with high-density shocks it may encounter every orbital period, or by the interaction of the disk with its surrounding structure that alters between a 2-edged flow and a bow shock. These peaks do not occur in the  theoretical BHL case, and are inherent to the complex interaction of the companion with the wind structures in case $v_{\rm w} \lesssim v_{\rm{orb}}$.

\begin{table}[H]
    \caption{Mean mass accretion efficiency}
    \begin{center}
    \begin{tabular}{lcc}
    \hline
    \hline
    \centering
       & $\langle\beta_{\rm BHL}\rangle$ & $\langle\beta_{\rm sim}\rangle$ \\
    \hline

      v05e00 & $0.106$ & $0.125$ \\
      v05e50 & $0.111$ & $0.206$ \\
      v10e00 & $0.075$ & $0.099$ \\
      v10e50 & $0.077$ & $0.123$ \\
      v20e00 & $0.027$ & $0.040$ \\
      v20e50 & $0.027$ & $0.043$ \\
      \hline
    \end{tabular}
    \end{center}
    {\footnotesize{\textbf{Notes.} Time-averaged BHL mass accretion efficiency from the theory $\langle\beta_{\rm BHL}\rangle$ and  simulations $\langle\beta_{\rm sim}\rangle$ over the last 2 orbital periods. }}
    \label{ta:BHLaverageMaccrEff}
\end{table}

\begin{figure}
    \centering
    \includegraphics[width = 0.49\textwidth]{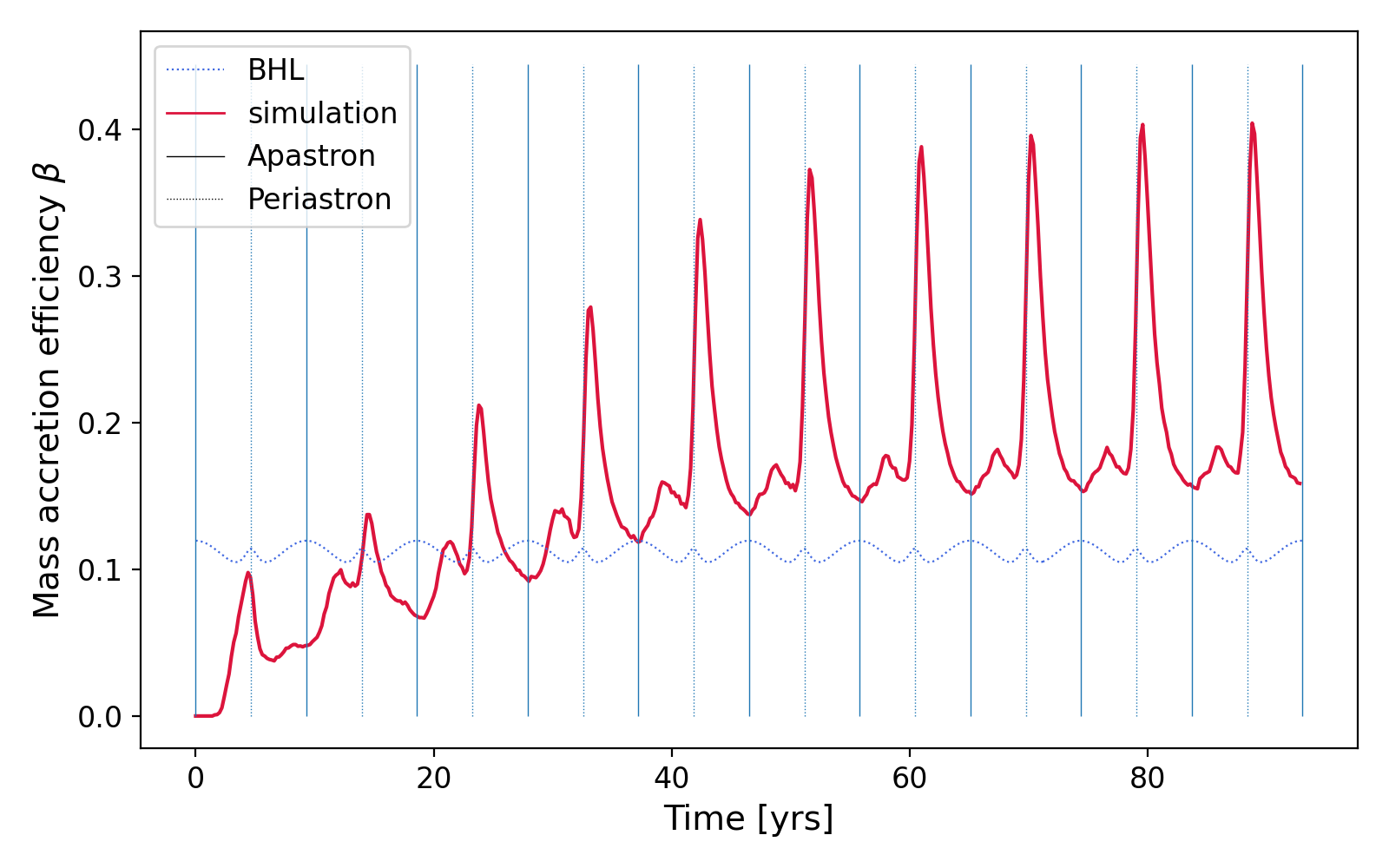}
    \includegraphics[width = 0.49\textwidth]{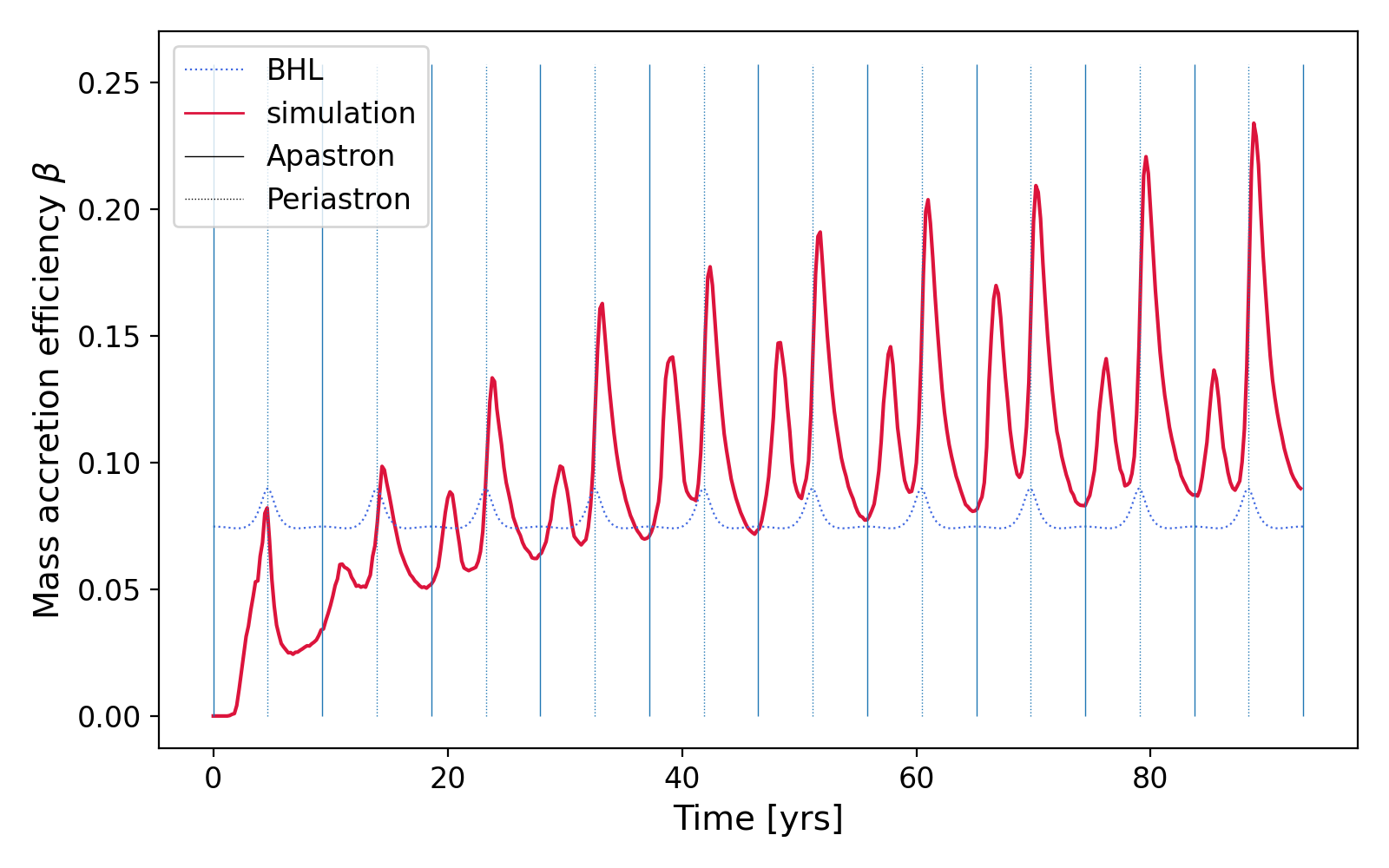}
    \includegraphics[width = 0.49\textwidth]{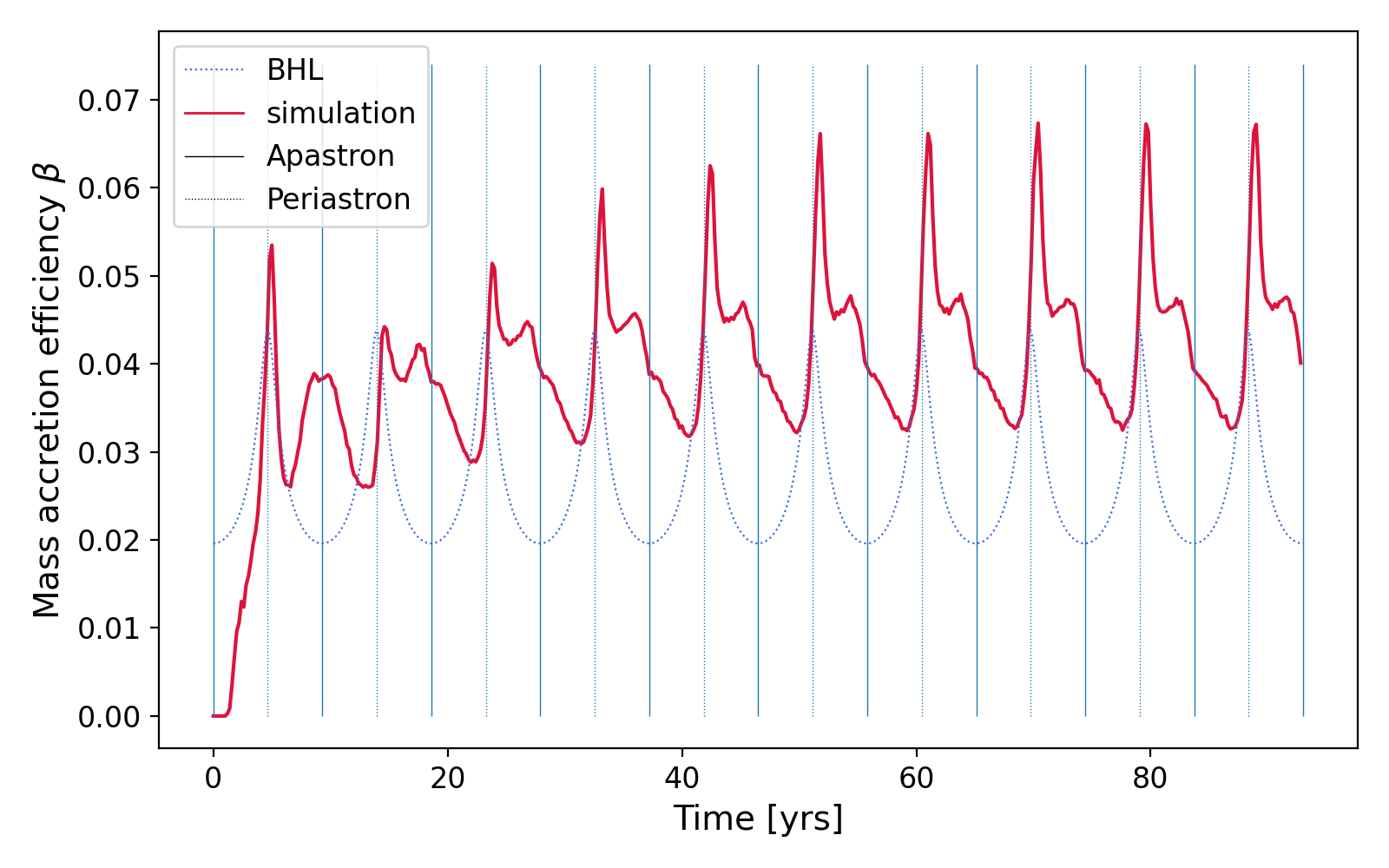}
    \caption{Mass accretion efficiency $\beta$ in models v05e50 (top), v10e50 (middle) and v20e50 (bottom), as calculated from the simulations (red), and compared to the BHL estimates (Eq.~\ref{Eq:BHLMaccrEcc}). The apastron and periastron orbital phases are indicated by vertical lines.
    }
    \label{fig:MassAccrEff}
\end{figure}

The mass accreted by the companion carries angular momentum and produces a torque on the stellar companion. The evolution of the torque on the companion star ${\rm d}J_{\rm s} / {\rm d}t$ is displayed in Fig.~\ref{fig:Jacc} and shows very similar patterns as the mass accretion rate (Fig.~\ref{fig:MassAccrEff}). In a circular orbit, after convergence of the mass accretion rate,  the torque on the companion is almost constant, while in eccentric systems, a main peak appears at apastron followed by a secondary weaker peak when the companion hits over-densities in the flow. In fact, the torque exerted by the accreted matter on the companion can be expressed as 
\begin{equation}
\frac{{\rm d}J_{\rm s}}{{\rm d}t} = \dot{M}_\mathrm{acc} j_\mathrm{acc}
\label{eq:jacc}
\end{equation}
where $j_\mathrm{acc}$ is the specific angular momentum of the accreted material, and $\dot{M}_\mathrm{acc}$ the accretion rate onto the companion. Since the companion is surrounded by an accretion disk, one would expect $j_\mathrm{acc} = ({\rm d}J_{\rm s} / {\rm d}t) / \dot{M}_\mathrm{acc}$ to be close to the Keplerian specific angular momentum $ j_{\rm K,s} =  (G M_{\rm s} R_{\rm s})^{1/2}$. Here $R_{\rm s}$ is the radius from where orbiting material is being accreted onto the secondary.
According to the accretion algorithm \cite[sect 2.8.2 of][]{phantom}, this radius lies between $0.8\, R_{\rm s, acc}$ and $R_{\rm s, acc}$.
In terms of specific angular momentum, this translates into $0.85 \lesssim j_\mathrm{acc}/j_{\rm K,s} \lesssim 0.95$, indicating that matter is accreted with a high specific angular momentum, close to the Keplerian value.
The deviation from the Keplerian value is due to the facts that ($i$) the accretion radius varies for each particle, and ($ii$) the disk is not perfectly Keplerian, as attested by the tangential velocity profiles (Fig.~\ref{fig:vt_vr_vKepl}) and the presence of spiral arms (Fig.~\ref{fig:vr_e00}).
Furthermore, we note that the specific angular momentum $j_\mathrm{acc}$ slightly increases with increasing wind velocity in the circular systems. This is likely due to the more stable flow in the vicinity of the companion (absence of bow shocks) and higher compactness of the accretion disk (smaller size). 
As shown by \cite{Packet1981}, the accretion of only a few percent of the original companion mass via the disc may spin up the mass accreting star to critical spin velocities, which may impact the global flow structure around the companion. For our simulations, this will be reached after $\sim 10^4$--$10^{5} \, {\rm yrs}$.

\begin{figure}
    \centering
    \includegraphics[width = 0.49\textwidth]{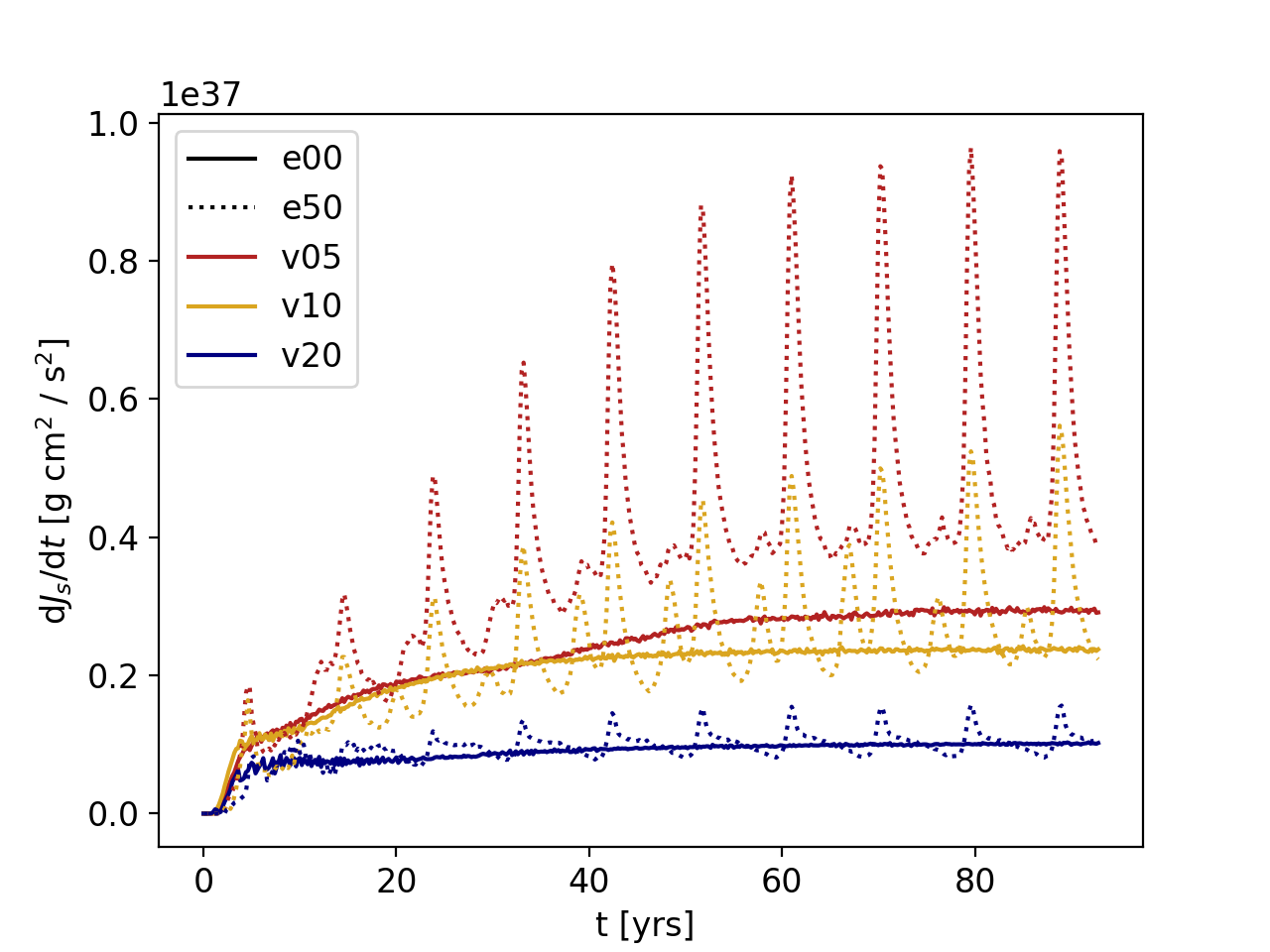}
    \caption{Torque ${\rm d}J_\mathrm{s} / {\rm d}t$ on the companion for the circular (solid line) and eccentric (dotted line) systems.}
    \label{fig:Jacc}
\end{figure}

\section{Further discussion}

\subsection{The possibilities to resolve or detect accretion disks in observations}
{The accretion disks in our simulations are small ($r_{\rm disk}<1 \, {\rm au}$) and thereby hard to resolve observationally. Additionally, they are relatively hot, with typical temperatures of $6\times 10^3$--$10^4 \, {\rm K}$, as can be seen from the temperature distribution in and around the accretion disk of model v05e00 (Fig.~\ref{fig:v05e00ADTemp}). Because of the high temperature, the disks are not observable in the submillimetre wavelength range, where AGB outflows are generally observed with facilities such as ALMA.
If the disk temperatures in our models are not overestimated, they could be detected through their ultraviolet (UV) excess, which is a wavelength range at which the luminous but cool AGB star is very faint. This is similar to symbiotic systems, in which there is UV emission from a hot white dwarf companion that is accreting matter from the stellar wind of a red giant. 
This was first suggested by \cite{Sahai2008}, who discovered significant far-UV excesses for $9$ out of $25$ AGB stars observed with GALEX, and suggested that this emission is linked to the presence of a binary companion that is either very hot, as in symbiotic systems, or that is surrounded by a hot accretion disk.}

{Accretion disks have been detected around post-AGB star companions, through thermal emission in near-infrared interferometry by \cite{Hillen2016} and \cite{Anugu2023} (with data from VLTI/PIONIER and CHARA, respectively). \cite{Hillen2016} estimated the temperature of the companion's accretion disk of the post-AGB binary IRAS 08544-4431 to $T_{\rm disk} = 4000 \pm 2000\, {\rm K}$, which is less than the disk temperature in our simulations ($T\approx 6000 - 10,000 \, K$, see Fig.~\ref{fig:v05e00ADTemp}).
However, \cite{DePrins2024} show that the disk temperature estimate from the near-infrared observations is underestimated, so the disk emission might peak in the ultraviolet (UV), with inner disk regions as hot as $10^4 \, {\rm K}$ or more, which is similar to the disks in our simulations. To reliably estimate the temperature distribution of the post-AGB companion's accretion disk, multi-wavelength interferometric observations are needed.}

\subsection{Transition to the post-AGB phase}
{It is very difficult to model the transition from AGB to the post-AGB and PNe phase, and to investigate how the asymmetries and structures that form in the AGB wind will shape the morphologies in later stages. 
In the post-AGB and (pre-) PN phase, a jet is frequently observed that is launched from an accretion disk around the companion, likely driven by the disk magnetic field \citep{DePrins2024}. 
In a recent study by \cite{Planquart2024a}, an accretion disk from which a jet is launched was also discovered around the companion of the AGB binary V Hydrae.
The accretion disk around a post-AGB companion may be the descendant of the accretion disk around the AGB companion, that we find in our simulations. 
However, the accretion disk and jet in post-AGB binaries are believed to by mainly fed by mass streaming from a circumbinary disk \citep{Bollen2022}, instead of from the AGB wind. 
It is unclear if the accretion disk that forms in AGB binaries does survive the transition to the post-AGB phase, when the feeding mechanism of the companion-surrounding accretion disk changes.
Further, it is not entirely clear how the observed post-AGB circumbinary disks form, but they are believed to arise by strong binary interaction before the post-AGB phase \citep{VanWinckel2009}.
To model the potential formation of circumbinary structures in AGB binaries that might be the progenitors of these post-AGB circumbinary disks, a better treatment of cooling and radiative transfer is needed \citep{Chen2020}.}

\subsection{Impact of mass-loss rate on wind structures}
\label{ImpactMLR}
{In the \textsc{Phantom} wind model, the wind mass-loss rate is an input parameter, which is set to $\dot{M} = 10^{-7}\,{\rm M_{\odot} \, {\rm yr}^{-1}}$ in our simulations. In Sect.~\ref{ch:SetupMethod} we noted that systems with wind velocities of $\sim 20 \, {\rm km \, s^{-1}}$ are expected to have higher mass-loss rates. Without the inclusion of H\,{\sc i} cooling, the hydrodynamic equations are invariant to a scaling in the mass-loss rate. 
For completeness, we here shortly investigate the effect of increasing the mass-loss rate in model v20e00 to a more realistic value of $\dot{M} = 10^{-5}\,{\rm M_{\odot} \, {\rm yr}^{-1}}$.
For the models analysed in Sect.~\ref{ch:impactCooling}, the increase in mass-loss rate does not affect the wind structures significantly, but leads to an overall increase of the CSE density.
In the higher-resolution model with small accretion radius (used to study the accretion disk in Sect.~\ref{ch:accrDisks}), the higher mass-loss rate effects the accretion disk and the surrounding wind structures. 
Fig.~\ref{fig:v20e00_M1e-5_rho} displays the density distribution around the companion in the orbital plane slice for the high-resolution model v20e00, with $\dot{M} = 10^{-5}\,{\rm M_{\odot} \, {\rm yr}^{-1}}$.
Compared to the same model with a lower mass-loss rate (Fig.~\ref{fig:ADzoomv20e00}), the disk size and density have increased, and instead of a 2-edged spiral, a structure resembling a bow shock forms. The temperature around the companion increased by a factor of $\sim 2$, such that the typical temperature within the disk becomes $1.2\times 10^4 - 2\times 10^4 \, {\rm K}$.
For this specific model, an increase in the mass-loss rate by a factor $100$ increases the disk mass by approximately a factor $\sim 50$, the disk radius by a factor $\sim 1.5$ (to $0.6 \, {\rm au}$), and the scale height with a factor $\sim 2.3$. The mass accretion efficiency slightly decreased from $0.040$ to $0.035$, such that the effective mass accretion rate increases to $3.5 \times 10^{-7}\,{\rm M_{\odot} \, {\rm yr}^{-1}}$.}


\section{Conclusion}
\label{ch:conclusion}
{In this work we use SPH simulations, performed with the \textsc{Phantom} code, to study the effects of atomic H\,{\sc i} cooling on the structures that form in the winds of evolved low- and intermediate-mass stars through the interaction with a binary companion. Further, we study the wind-capture accretion discs that form around the companion.} 

We consider a mass-losing AGB star with a solar-mass companion at an orbital separation of $6 \, {\rm au}$.
Compared to previous simulations done without HI cooling \citep{Malfait2021}, we find that the region surrounding the companion star is efficiently cooled down, with maximum temperature being an order of magnitude lower. This removes the irregularities present in \cite{Malfait2021} and \cite{Maes2021}, that were related to a `high-energy' region around the companion and that were causing irregular or square morphologies and bipolar-like outflows.
However, it is still possible to form very asymmetric and complex wind structures, in case of eccentric binary systems in which the kinetic energy of the wind is relatively small compared to the gravitational energy density around the companion.
Further, we found that changing the mean molecular weight $\mu$ in our model impacts the wind velocity profile, which is an important parameter determining which structures form around the companion. This indicates that it is important to self-consistently calculate $\mu$ in future models.

Our simulations show that wind material is able to accumulate closely around the companion, forming a high-density accretion disk. The disks consist of material that is orbiting with Keplerian to sub-Keplerian velocities, while slowly moving radially inward towards the companion {in a slightly eccentric orbit}.
The disks are flaring and have radial sizes of $\sim 0.4$--$0.9 \, {\rm au}$, and disk masses of $\sim 3 \times 10^{-8}$--$2 \times 10^{-7} \, {\rm M_\odot}$, with increasing size and radius for decreasing wind velocity.
Within the disk, two spiral-like features are identified, having lower radial velocities than the surrounding particles within the disk. In case of a relatively low wind velocity compared to the orbital velocity, these features also have a relatively high density, making them visible within the disk density distribution. The spiral with the largest density enhancement originates in the wake behind the companion, the second spiral in the bow shock in front of the companion, and they both spiral-in clockwise towards the companion sink particle.
In eccentric systems, the disk shape changes throughout an orbital period, due to the phase-dependent wind-companion interaction, and the interaction with other high-density structures present in the inner wind. 

The average mass accretion efficiencies in our simulations are up to a factor of $\sim$ two higher than the theoretical BHL predictions, and increase with decreasing velocity and higher eccentricity. The mass accretion rate ranges between $0.04$ and $0.21$ of the mass loss rate. 
In the eccentric orbits, there is a mass accretion peak around the close periastron passage, and an additional peak closer around apastron passage. 
Finally, we found that the companion accretes matter with a high specific angular momentum, close to the Keplerian value, which can result in a significant spin-up of the star. 

{The small spatial extent of the accretion disks cannot easily be resolved observationally. Further, the disks can not be observed in the submillimeter wavelength range where the AGB outflows are typically observed, due to their relatively high temperatures (between $\sim 6000 - 10^4 \, {\rm K}$). If the actual disk temperatures are indeed this high, the disks could be detected through UV flux excess, taking into account that such UV excess can have a different origin as well, such as accretion onto a hot companion.}


\begin{acknowledgements}
{We thank Toon De Prins for insightful discussions on accretion disks in post-AGB stars}, and we thank Daniel Price for his help concerning the development of \textsc{Phantom}. The animations and figures in this paper are made using the visualisation tools \textsc{Plons}\footnote{\url{https://github.com/Ensor-code/plons}} and \textsc{Splash} \citep{splash}.
JM, ME, and LD acknowledge support from the KU Leuven C1 excellence grant C16/17/007 MAESTRO and the FWO research project G099720N. L.S. is a senior F.N.R.S research associate.
FDC is a postdoctoral research fellow of the Research Foundation - Flanders (FWO), grant 1253223N.
LD acknowledges support from the KU Leuven C1 excellence grant BRAVE C16/23/009 and KU Leuven Methusalem grant SOUL METH/24/012.
\end{acknowledgements}


\bibliographystyle{aa}
\bibliography{references}	

\appendix

\section{Effect of H\,{\sc i} cooling}

{This section contains additional figures used to study the effect of H\,{\sc i} cooling in the \textsc{phantom} model on the wind.}

\begin{figure*}[h!]
	\centering
	\includegraphics[width = \textwidth]{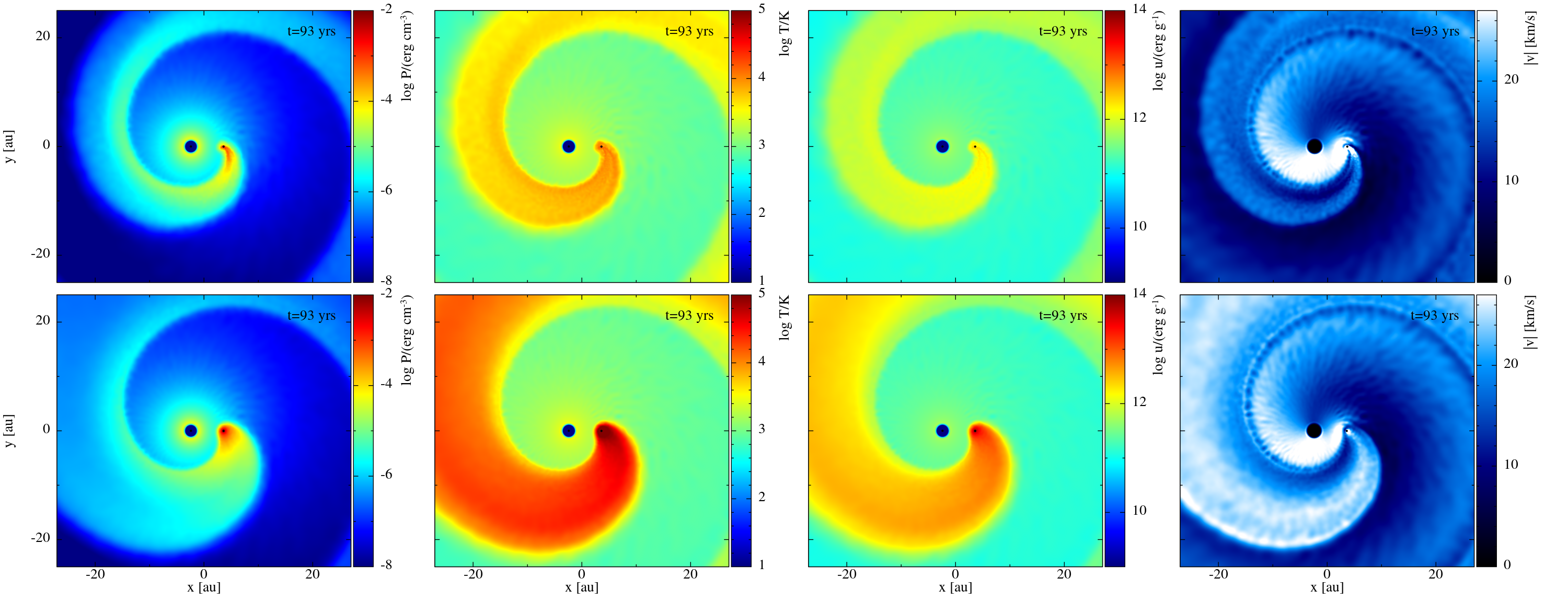}
	\caption{Comparison of pressure $P$, temperature $T$, internal energy $u$ and velocity $|v|$ profile in slices through the orbital plane of the binary model v20e00 with (top row) and without (bottom row) H\,{\sc i} cooling. The AGB and companion star are annotated as the bottom and upper dot, respectively, not to scale. The model without cooling is as described in \cite{Malfait2021}.}
	\label{Fig:v20e00TPu}
\end{figure*}

\begin{figure*}[h!]
	\centering
	\includegraphics[width = \textwidth]{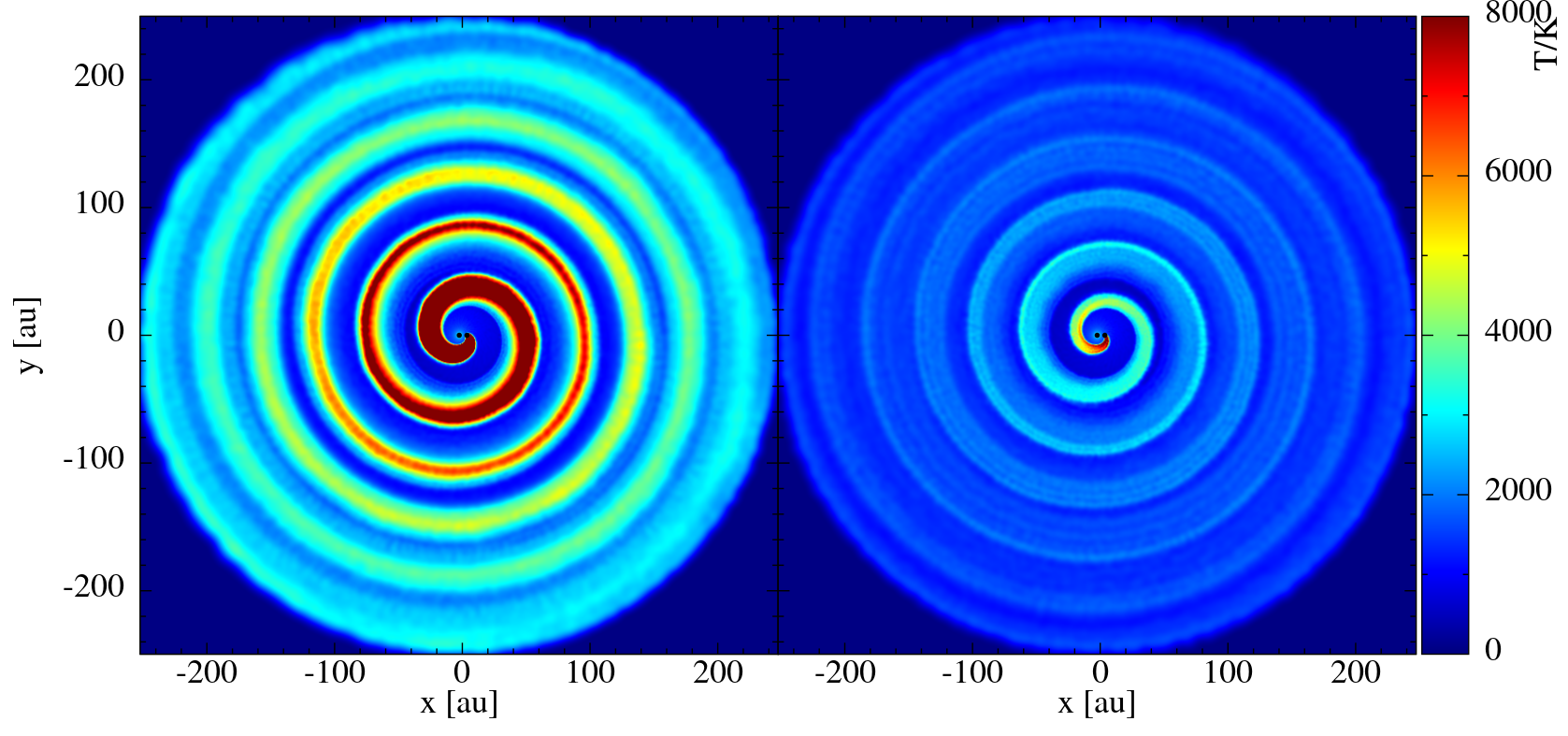}
	\caption{Comparison of temperature $T$ in slices through the orbital plane of the binary model v20e00 without (left) and with H\,{\sc i} cooling (right). The model without cooling is as described in \cite{Malfait2021}.}
	\label{Fig:v20e00Tfull}
\end{figure*}

\begin{figure}[h!]
    \centering
    \includegraphics[width = 0.49 \textwidth]{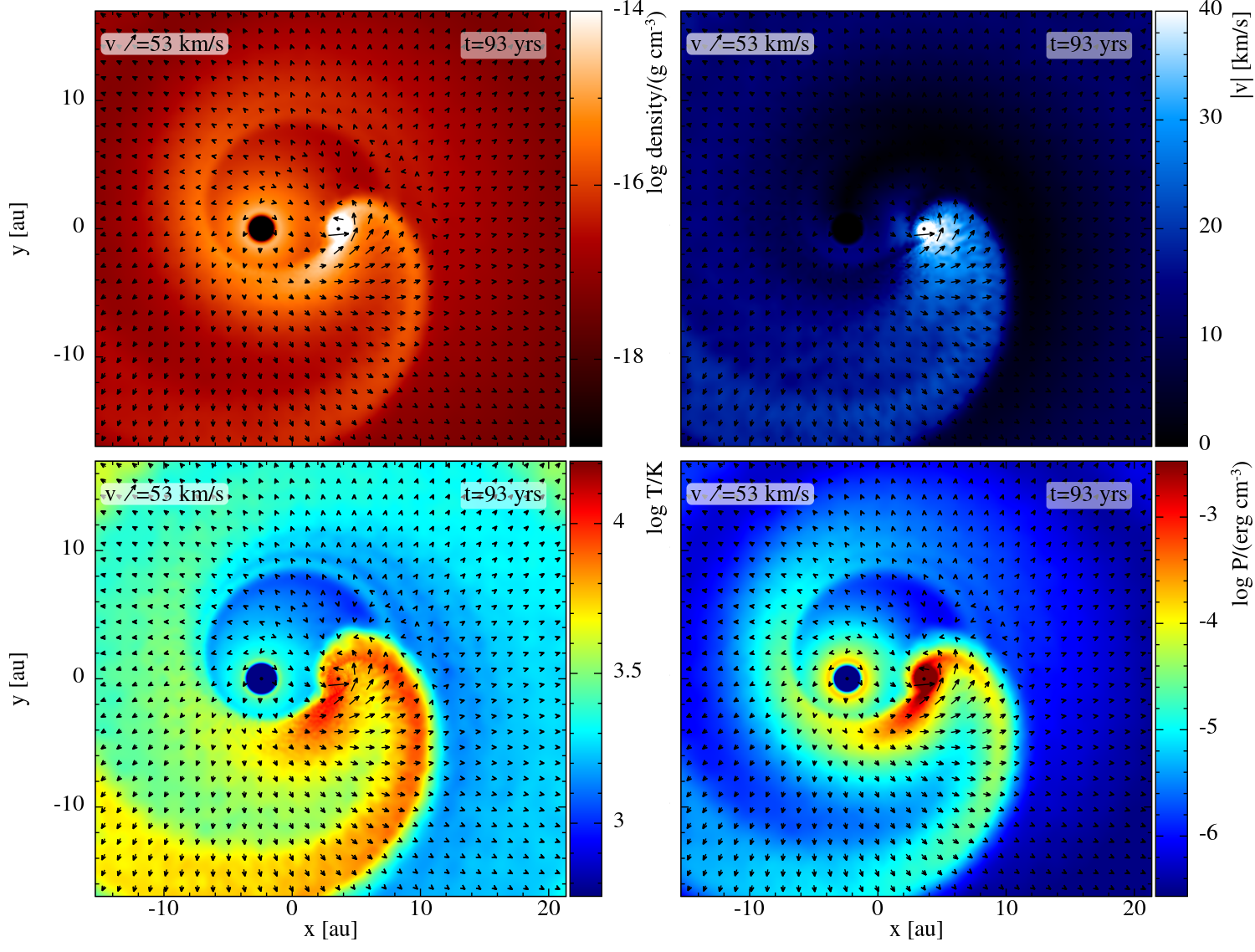}
    \caption{Same as Fig.~\ref{fig:v10e00zoom}, but for model v05e00. {A version of the density and temperature plot without H\,{\sc i} cooling is presented in Fig.~5 in Paper I, where the wind structures are unstable \citep{Malfait2021}.}}
    \label{v05e00zoom}
\end{figure}

\begin{figure}[h!]
	\centering
	\includegraphics[width =0.49 \textwidth]{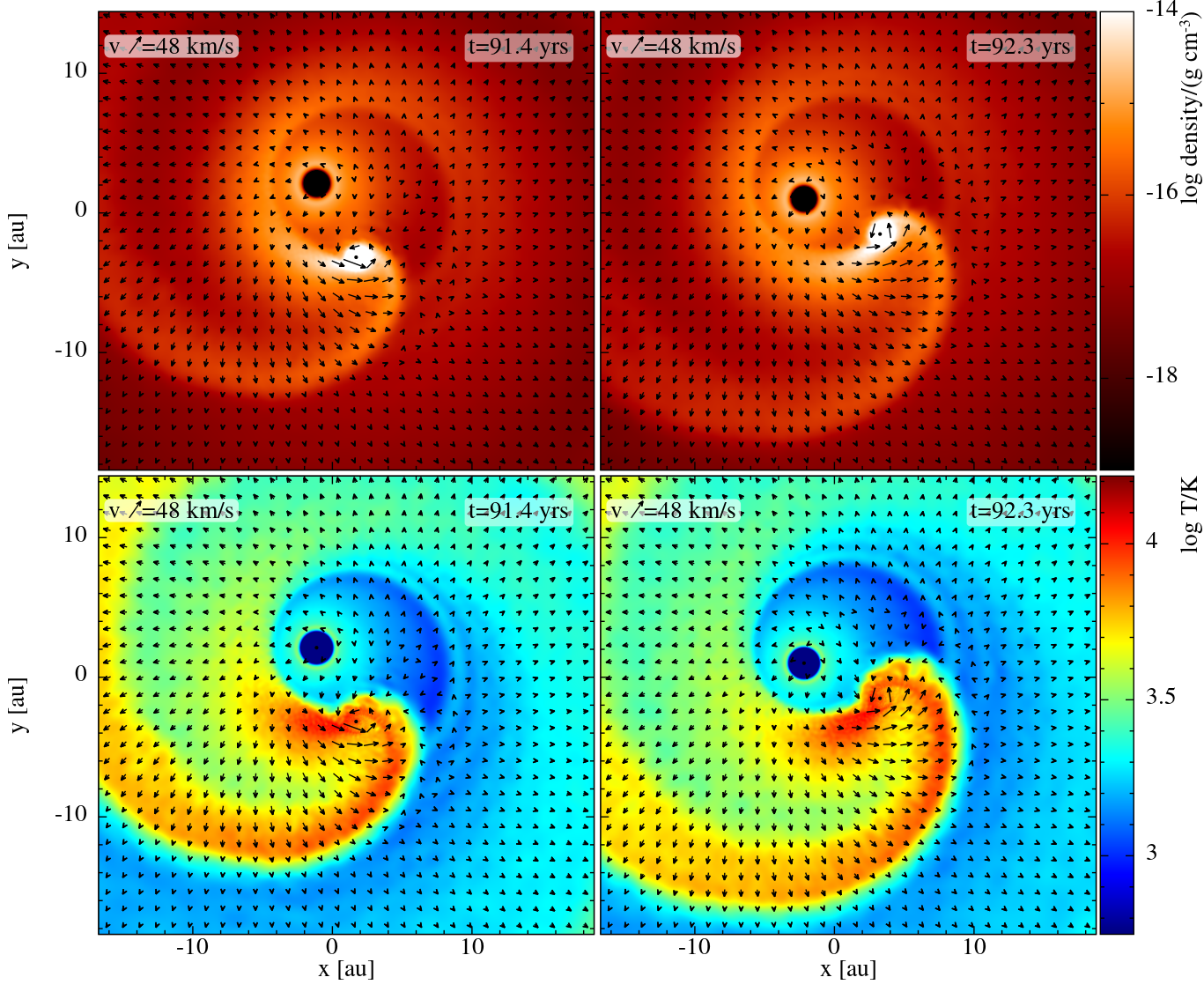}
	\caption{Density and temperature profile in the orbital plane of model v05e00 at two different timesteps, superimposed with the velocity vector profile which is shown as black arrows. {The inclusion of H\,{\sc i} cooling keeps the region around the companion stable. A version without H\,{\sc i} cooling is presented in Fig.~5 in Paper I.}}
	\label{v05e00densT2ts}
\end{figure}

\begin{figure*}[h!]
	\centering
	\includegraphics[width =  \textwidth]{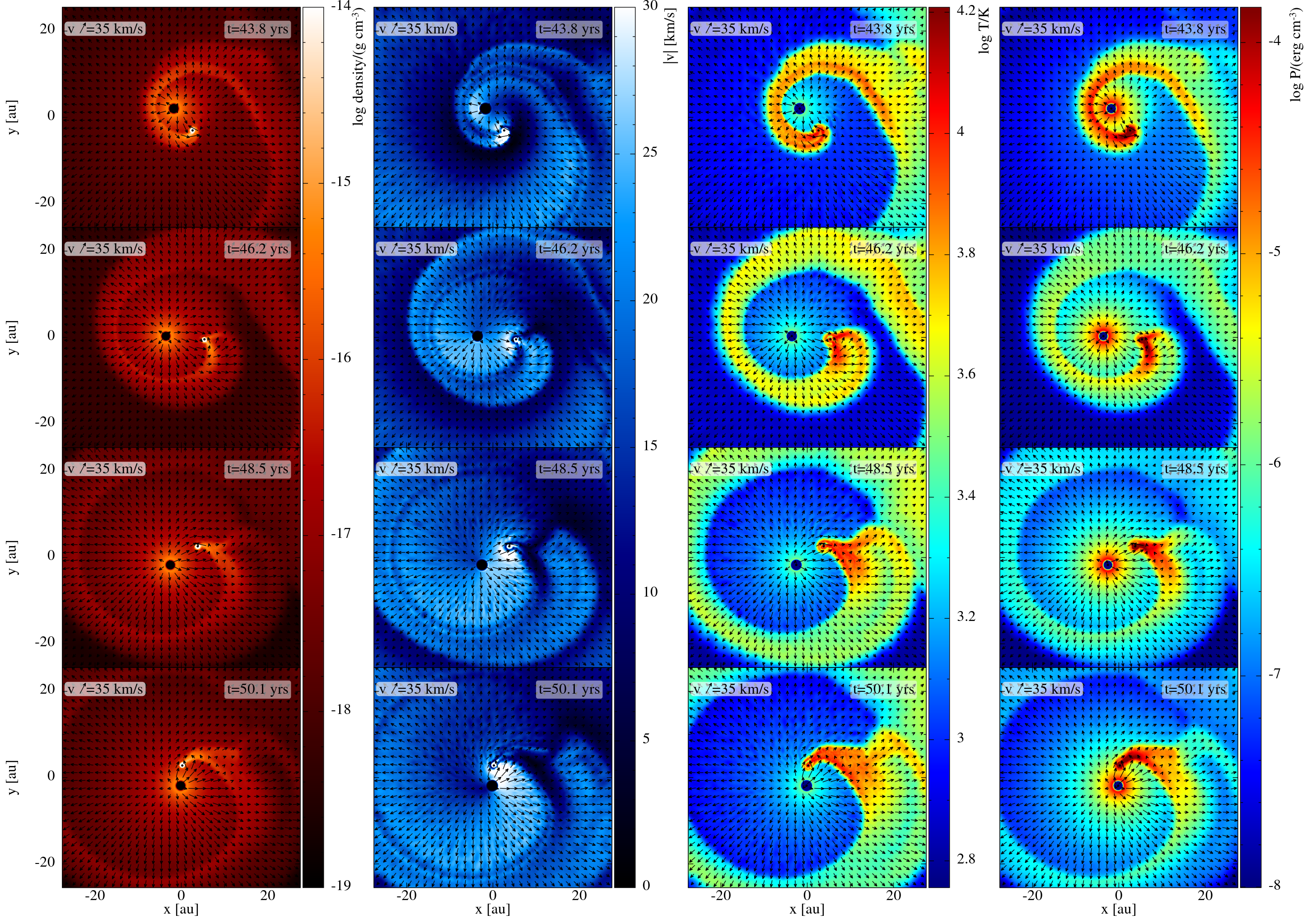}
	\caption{From left to right: density, velocity, temperature, and pressure distribution in the orbital plane of model v20e50 at four consecutive orbital phases $\phi = 1.73 \, \pi, 1.97 \, \pi, 0.19 \, \pi, 0.47 \, \pi$ (corresponding to $t = 43.8, 46.2, 48.5, 50.1 \, {\rm yrs}$ in the figure).  {A version without H\,{\sc i} cooling is presented in Fig.~9 in Paper I.}}
	\label{v20e50zoom}
\end{figure*}

\begin{figure*}[h!]
    \centering
    \includegraphics[width =  0.7\textwidth]{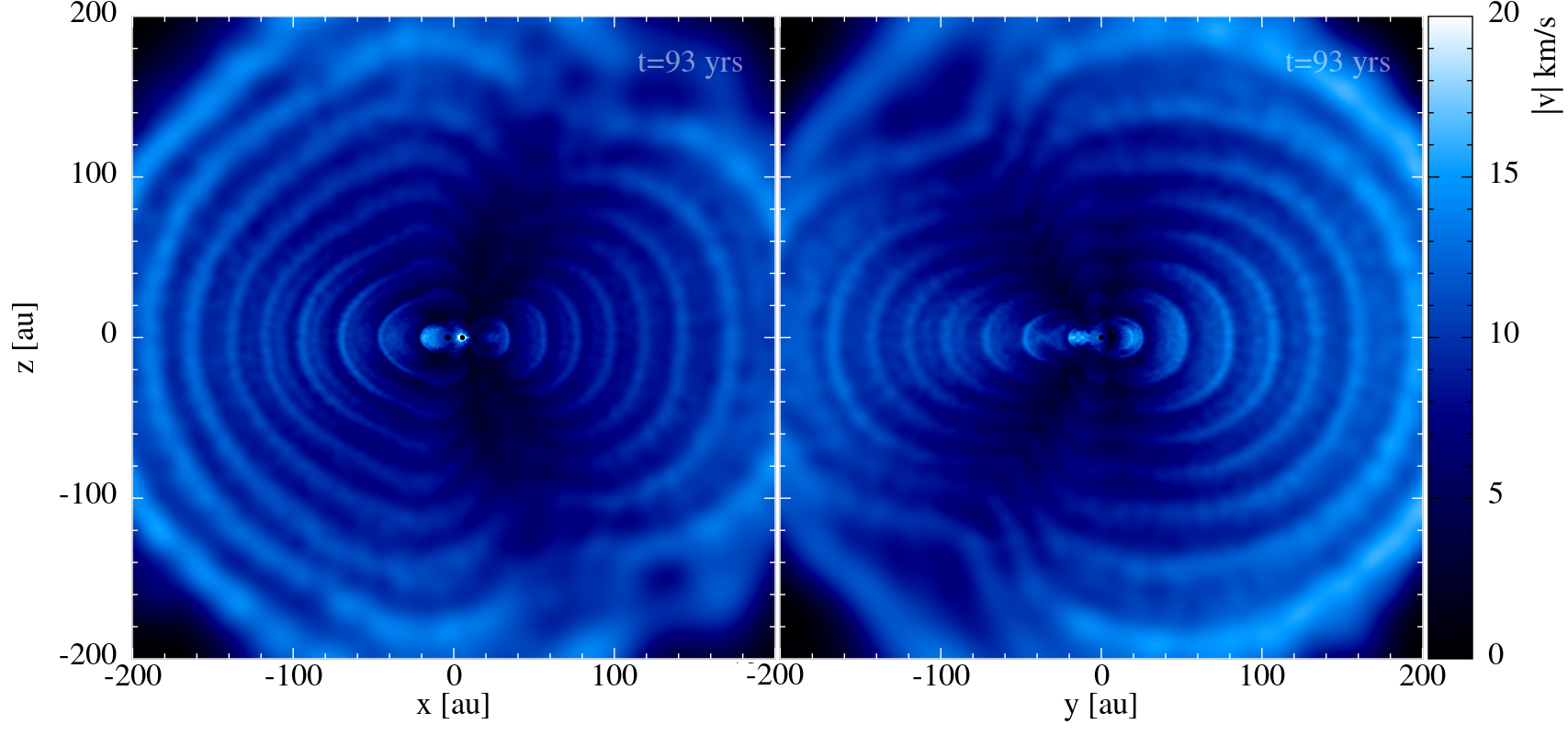}
    \caption{Velocity distribution in two perpendicular meridional plane slices (left, $x-z$ plane, and right, $y-z$ plane) of model v05e50. These plots illustrate that the inclusion of H\,{\sc i} cooling inhibits the formation of the bipolar outflows present in paper I (Fig.~13). }
    \label{Fig:v05e50vel_edgeOn}
\end{figure*}

\clearpage

\section{Resolution of accretion disks}
{This section contains an additional table and figures used to study the effect of the resolution of the \textsc{phantom} model on the formation of, and ability to resolve, accretion disks.}

\begin{figure*}[h!]
    \centering
    \includegraphics[width = \textwidth]{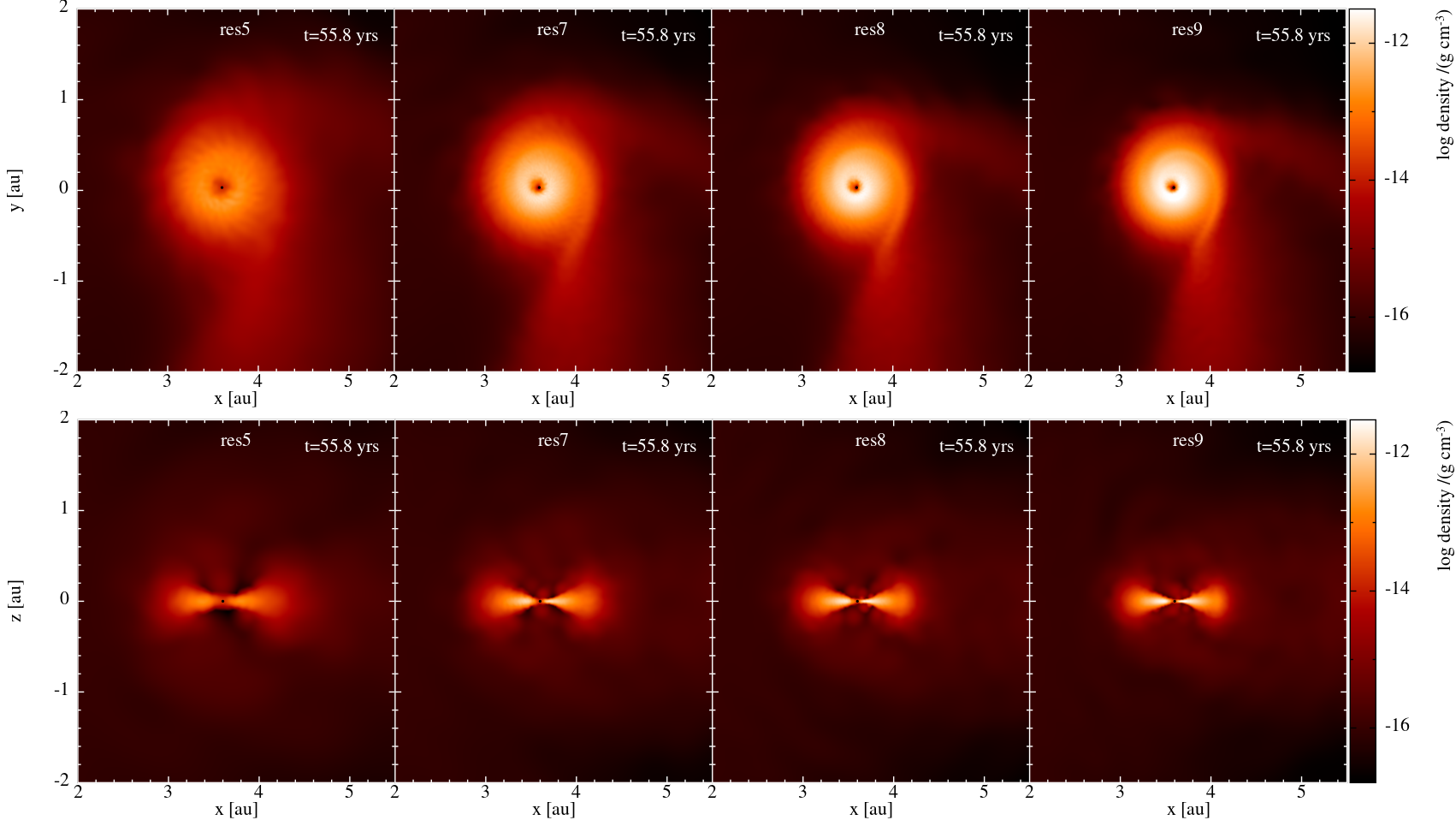}
    \caption{Density distribution in the orbital (top) and meridional (bottom) planes for model v10e00 with \texttt{iwind\_res{}}$ = 5, 7, 8, 9$ and $R_{\rm{s,accr}} = 0.04 \, \rm{au}$. }
    \label{fig:v10racc04}
\end{figure*}

\begin{figure*}[h!]
    \centering
    \includegraphics[width =0.7\textwidth]{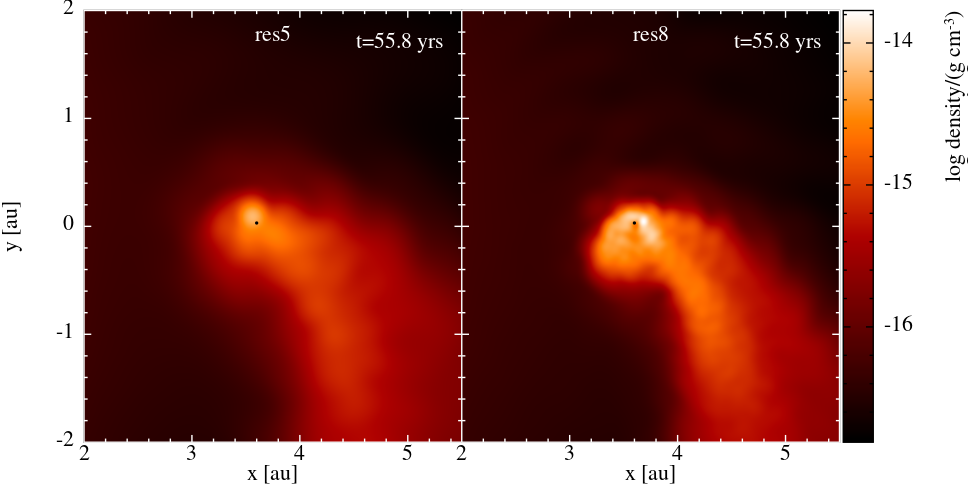}
    \caption{Density distribution in the orbital plane of model v20e00 with low and high resolution (\texttt{iwind\_res} $= 5, 8$ respectively) and $R_{\rm{s,accr}} = 0.04 \, \rm{au}$.}
    \label{v20e00ra04}
\end{figure*}

\begin{figure*}[h!]
    \centering
    \includegraphics[width = \textwidth]{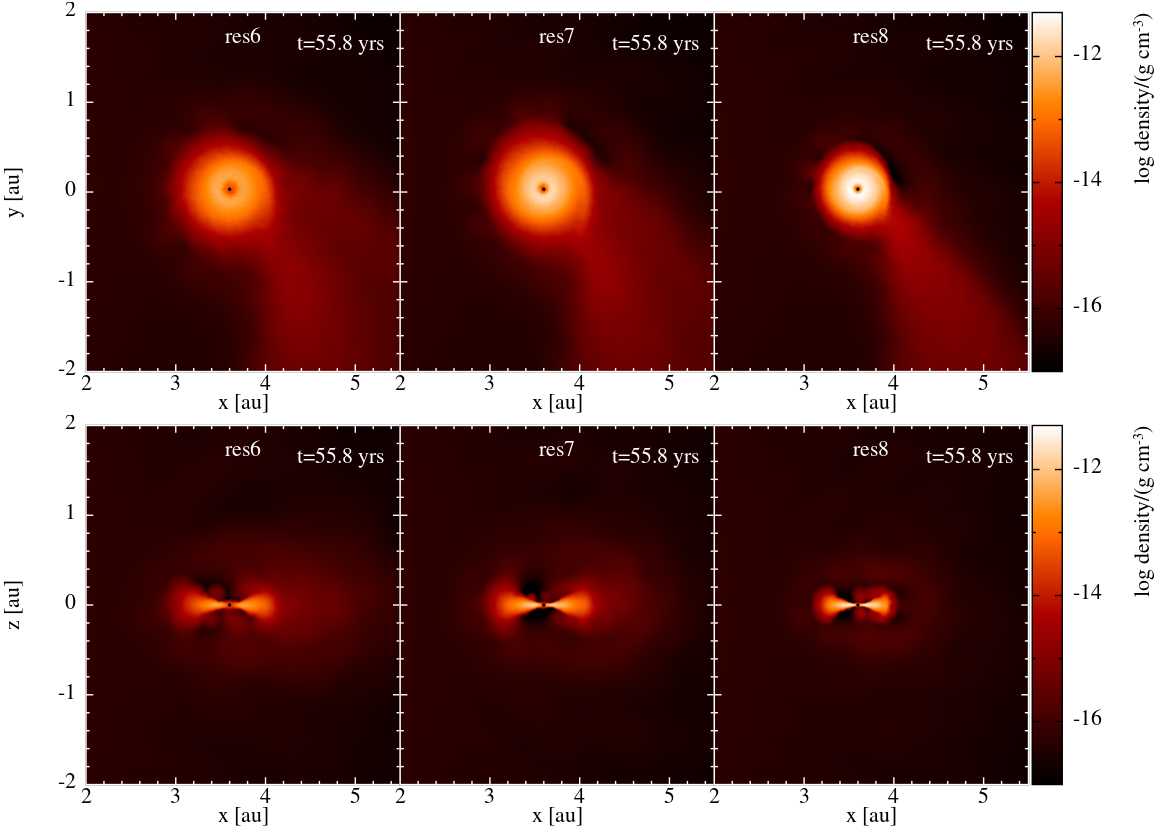}
    \caption{Density distribution in the orbital plane (top) and meridional plane slice (bottom) of model v20e00 with $R_{\rm{s,accr}} = 0.01 \, \rm{au} = 2.15\, R_\odot$ and \texttt{iwind\_res} $= 6, 7, 8$.}
    \label{fig:v20racc01}
\end{figure*}

\begin{table*}[h!]
	\caption{Model resolution options}
	\begin{center}
		\begin{tabular}{lccccccccccc}
  			\hline
			\hline
		    \texttt{iwind\_{res}}    & 1 & 2 & 3 &  4 &5 & 6 &7 &  8 &9 & 10 & 15 \\
                \texttt{nparticles}   &12 &92 &252 & 492 & 812 & 1212 & 1692 & 2252 & 2892 & 3612 & 8412\\
                particle mass [$10^{21}$ g] & $1.07 \times 10^{4}$ & $4.67\times 10^{2}$ &$1.02\times 10^{2}$&$37.4$&$17.6$ &$9.67$ & $5.86$& $3.82$&  $2.62$& $1.88$& $0.528$\\
                \hline
		\end{tabular}
	\end{center}
	{\textbf{Notes.} \footnotesize{For a given value of \texttt{iwind\_{res}},  \texttt{nparticles} represents the number of particles that will be launched from a sphere surrounding the AGB sink particle, with \texttt{nparticles}=10\,(2$\times$\texttt{iwind\_{res}}$-1)^{2} +2$. The last row indicates the SPH particle mass for a wind mass-loss rate of $\dot{M} = 10^{-7} \, {\rm{M_\odot\,yr^{-1}}}$ \cite[for details, see][]{Siess2022}.}
	}
	\label{resolutionTable}
\end{table*}

\clearpage

\section{Structure of accretion disks}

{This section contains an additional table and figures used to study the structure of accretion disks in our models.}

\begin{figure}[h!]
    \centering
    \includegraphics[width = 0.49\textwidth]{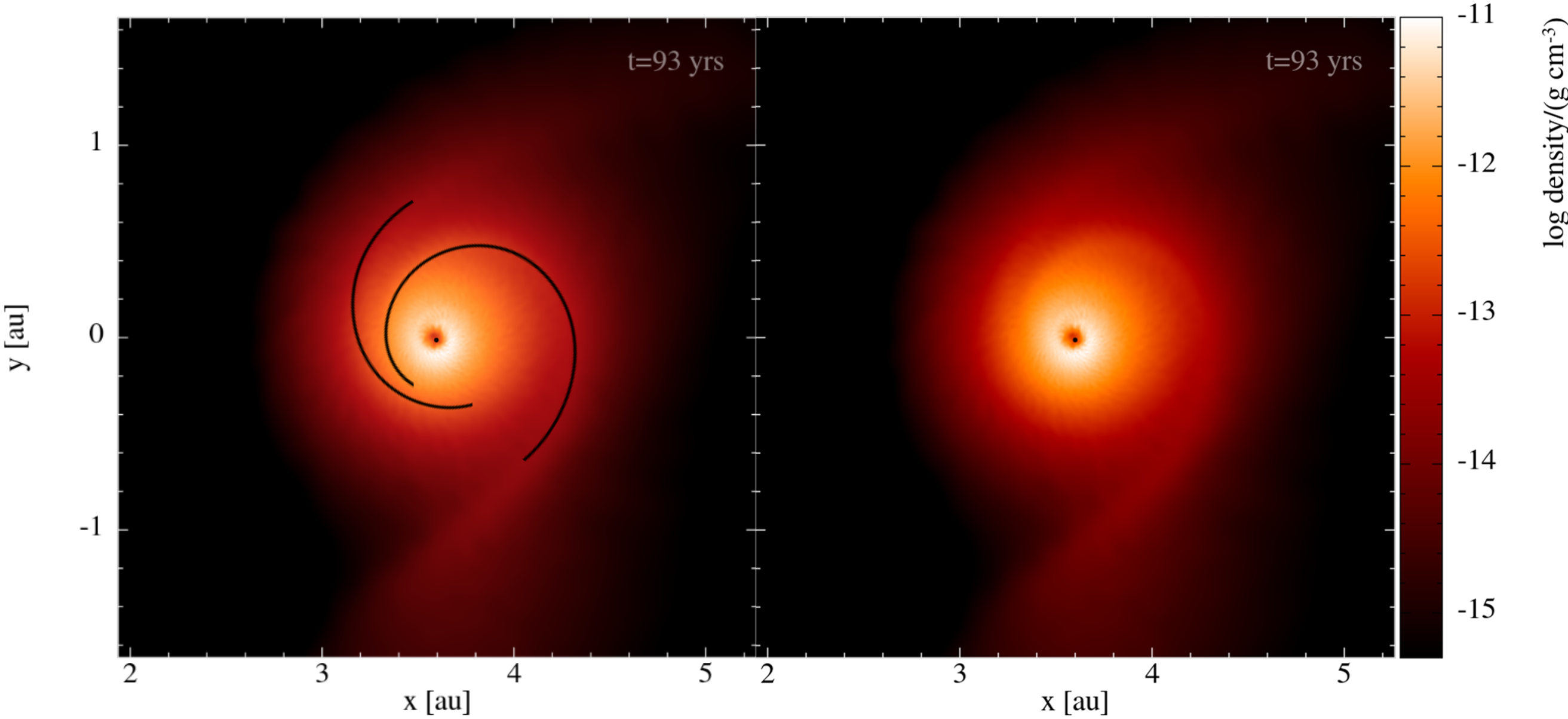}
    \caption{Accretion disk in model v05e00 with annotated spiral features on top of the density distribution in the orbital plane.}
    \label{fig:spiralsv05e00}
\end{figure}

\begin{figure*}[h!]
    \centering
    \includegraphics[width = \textwidth]{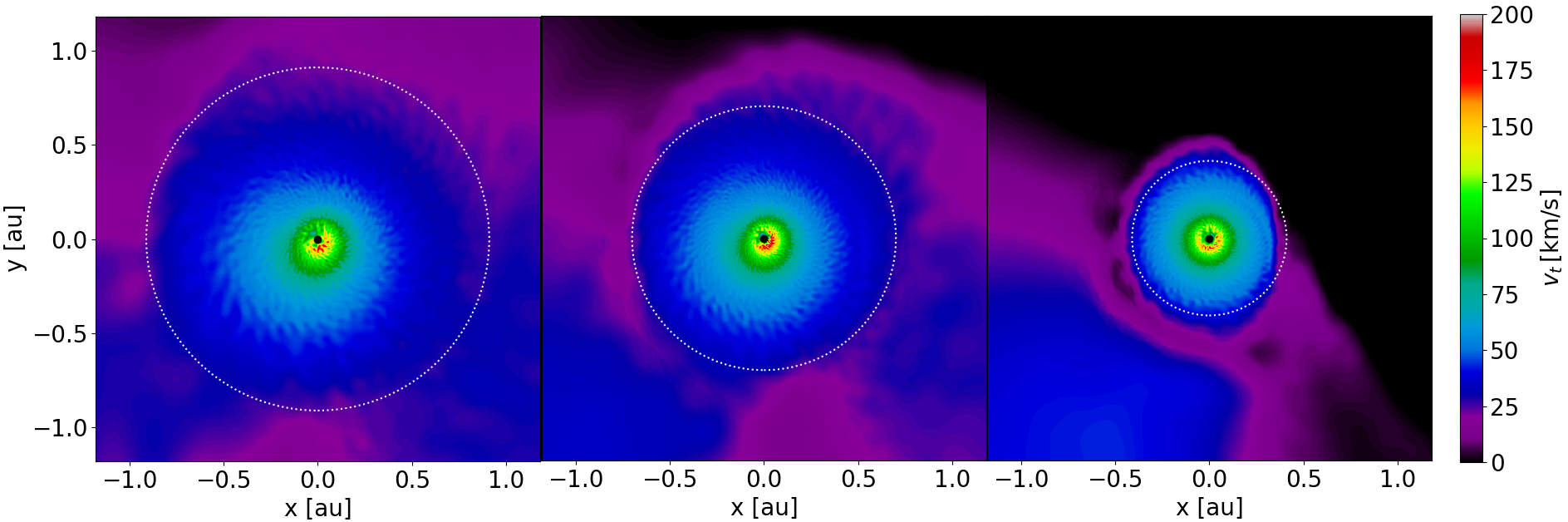}
    \caption{Tangential velocity distribution ($v_t$) in the accretion disks of models v05e00 (left), v10e00 (middle) and v20e00 (right), in the reference frame of the companion. The AGB star is located on the left side on the $y=0$ axis.  The estimated radii of the circumstellar disks as calculated in Sect.~\ref{ch:quantitativeAnalysisDisks} are annotated as a dotted circle.}
    \label{fig:vt_e00}
\end{figure*}

\begin{figure}[h!]
    \centering
    \includegraphics[width = 0.49 \textwidth]{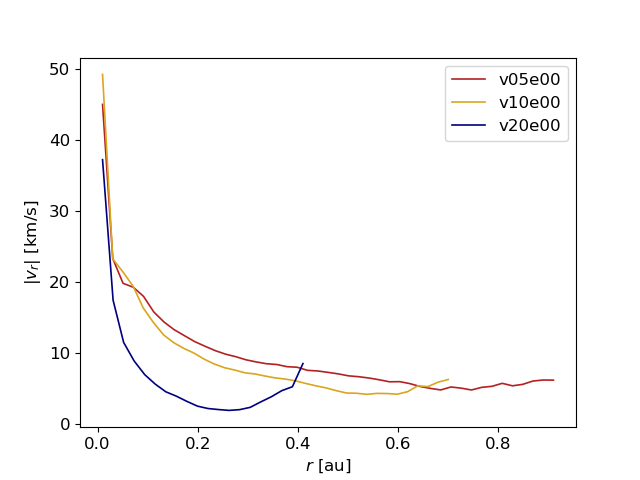}
    \caption{The absolute value of the radial velocity as a function of the distance $r$ from the companion, averaged over all angles $\theta$, within the disks of models v05e00, v10e00 and v20e00.}
    \label{fig:mean_abs_vr}
\end{figure}

\begin{figure*}[h!]
    \centering
    \includegraphics[width = 0.49\textwidth]{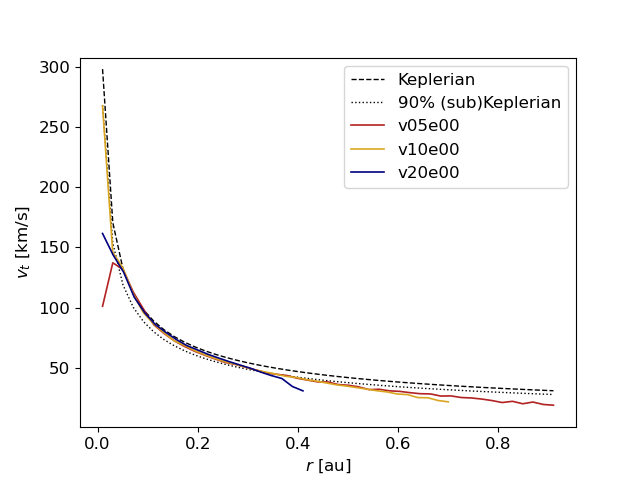}
    \includegraphics[width = 0.49 \textwidth]{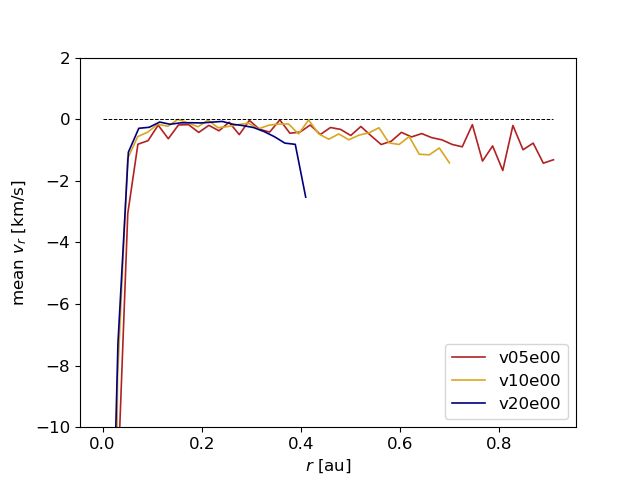}
    \caption{Left: tangential velocity distribution as a function of radius for the circular models, compared to a Keplerian and $90\%$ subKeplerian velocity distribution. Right: radial velocity at each radius $r$, averaged over all angles $\theta$, for the circular models, showing that in general material tends to spiral in towards the companion.}
    \label{fig:vt_vr_vKepl}
\end{figure*}

\begin{figure*}[h!]
    \centering
    \includegraphics[width = 0.7\textwidth]{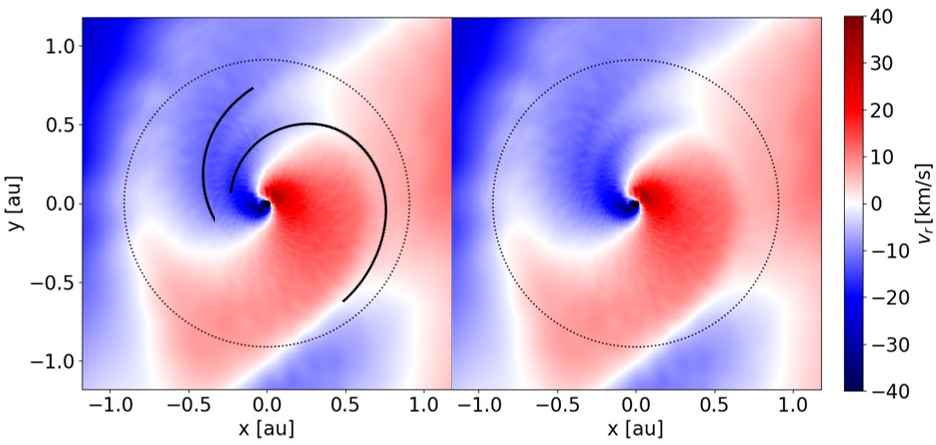}
    \caption{Radial velocity distribution in the orbital plane of the accretion disk in model v05e00, with annotated spiral features.}
    \label{fig:spiralsv05e00_vr}
\end{figure*}

\begin{figure}[h!]
    \centering
    \includegraphics[width = 0.49 \textwidth]{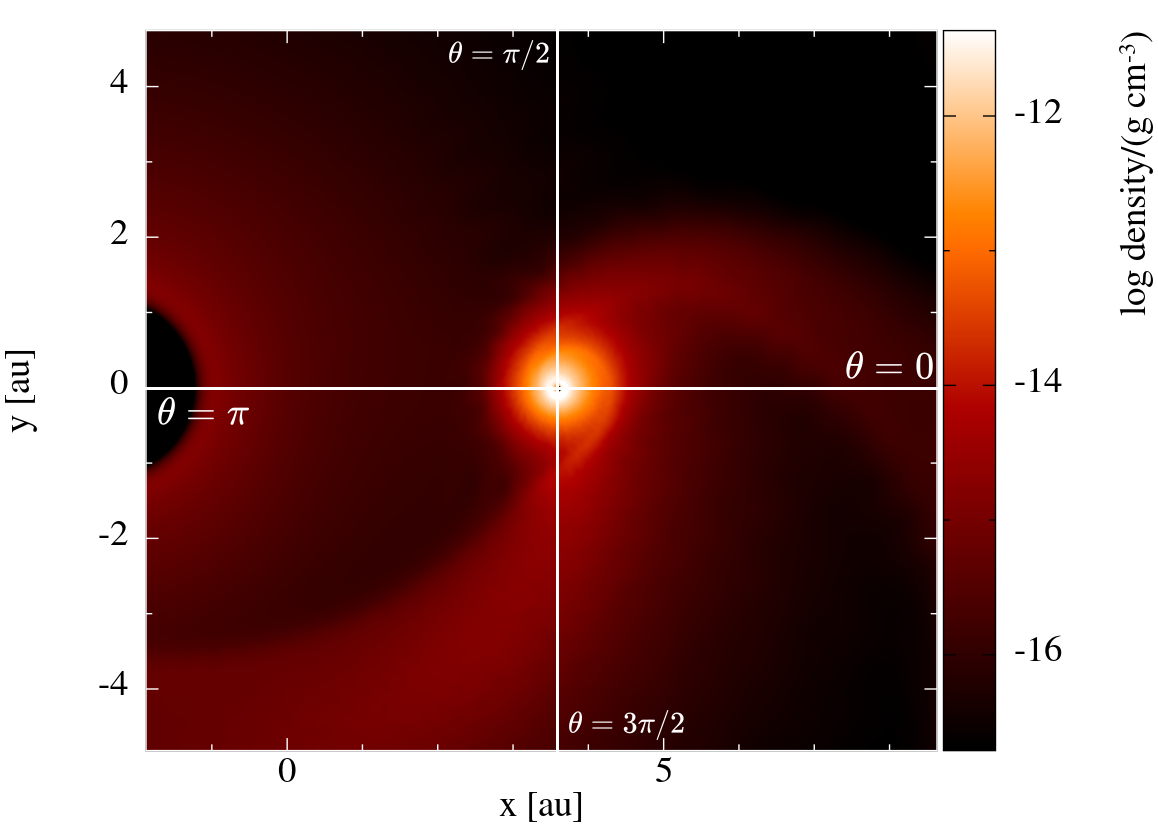}
    \caption{Coordinate system for analysis of accretion disks, showing the density distribution in the orbital plane of model v05e00. The AGB primary is located in the $\theta = \pi $ direction.}
    \label{fig:thetaDir}
\end{figure}
\begin{figure}[h!]
    \centering
    \includegraphics[width = 0.49 \textwidth]{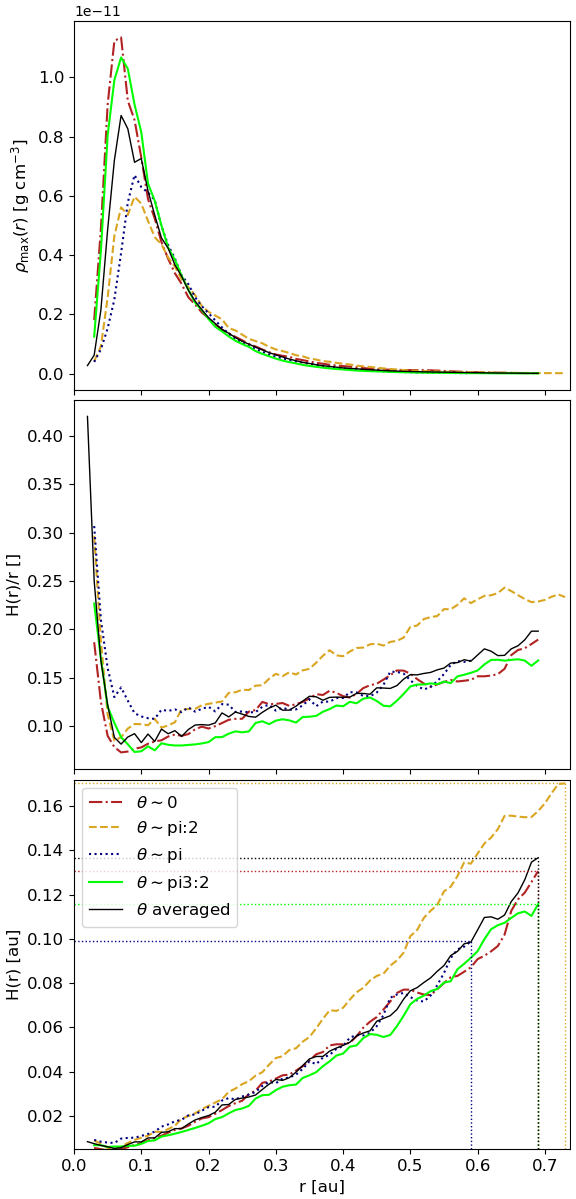}
    \caption{Same as Fig.~\ref{fig:v05e00_diffTh_SH_ar}, but for model v10e00.}. 
    \label{fig:v10e00_diffTh_SH_ar}
\end{figure}
\begin{figure}[h!]
    \centering
    \includegraphics[width = 0.49 \textwidth]{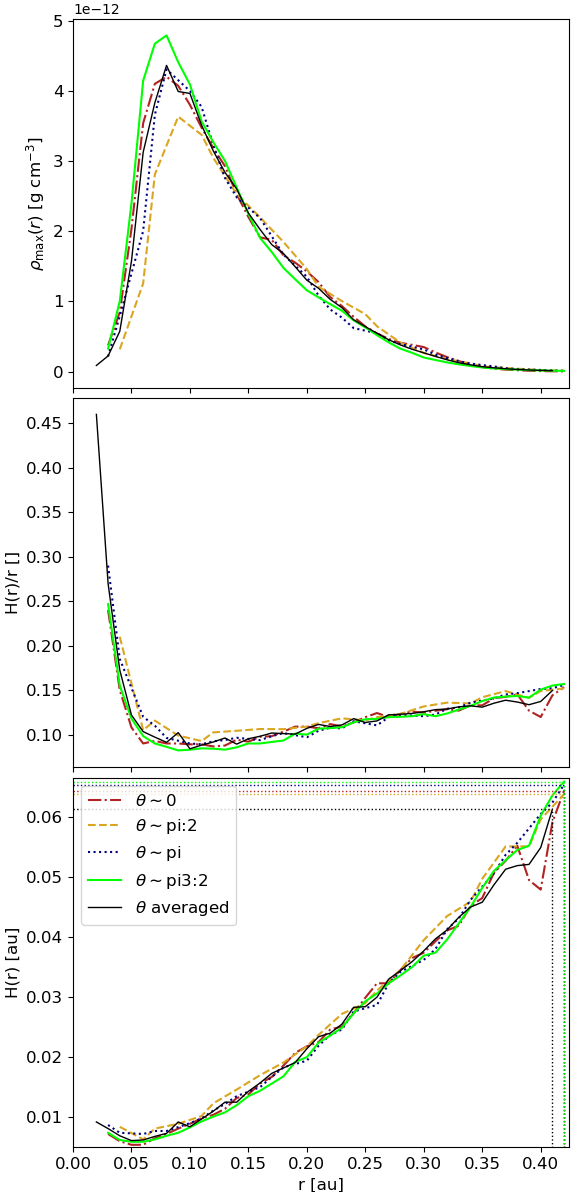}
    \caption{Same as Fig.~\ref{fig:v05e00_diffTh_SH_ar}, but for model v20e00.}
    \label{fig:v20e00_diffTh_SH_ar}
\end{figure}

\begin{figure*}[h!]
    \centering
    \includegraphics[width = \textwidth]{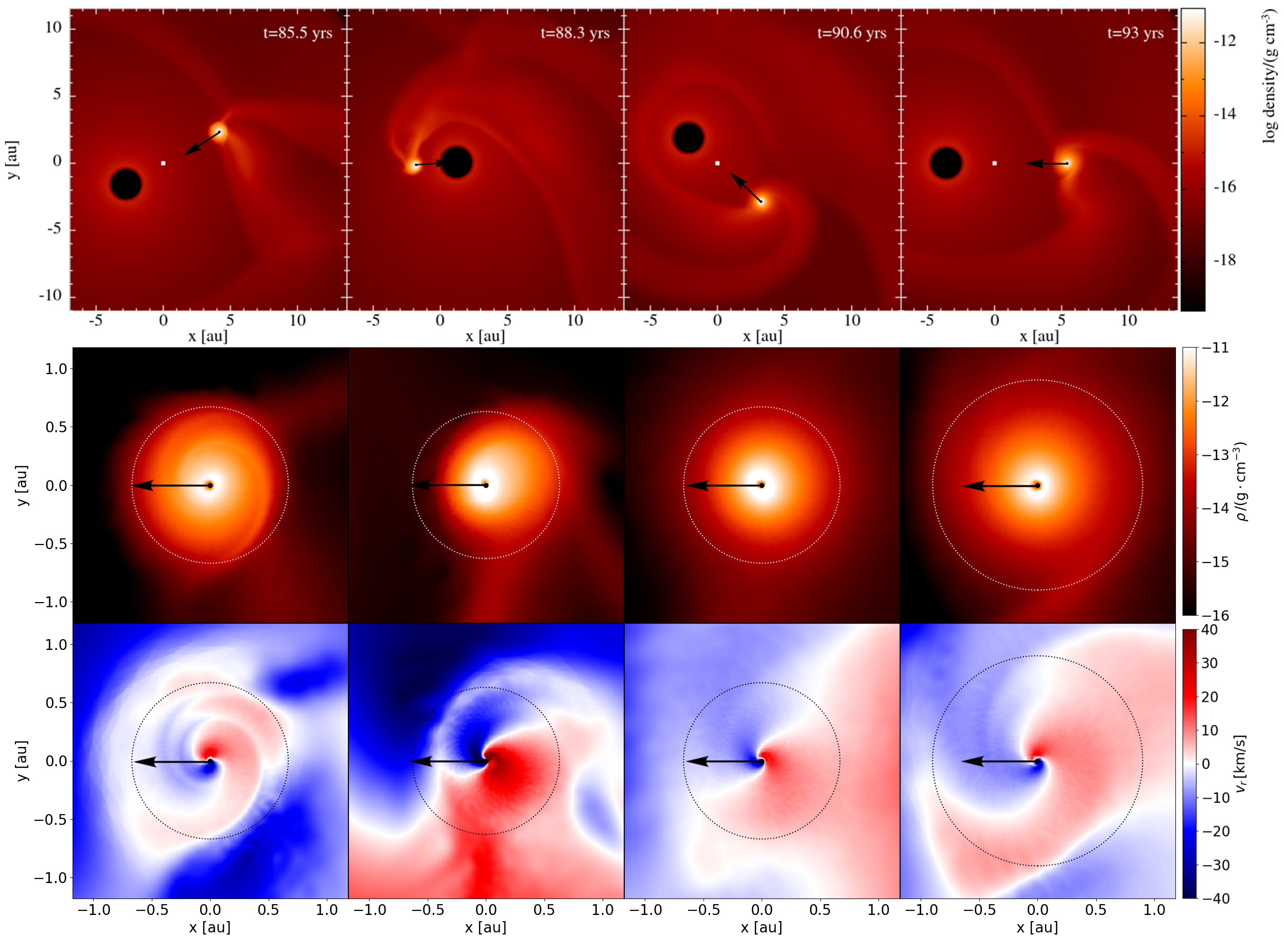}
    \caption{Same as Fig.~\ref{fig:v05e50}, but for model v10e50.}
    \label{fig:v10e50}
\end{figure*}

\begin{table}
    \caption{Disk properties of model v10e50 at different orbital phases}
    \begin{center}
    \begin{tabular}{lccccc}
    \hline
    \hline
    \centering
       $t$ [yrs] & $\phi$ & $r$ [au] & $M \, [M_\odot]$ & $H_{\rm{max}}$ [au] & $2H_{\rm{max}}$ [au] \\
       \hline
      85.5  & $0.16\, \pi$  & $0.67$ & $1.22 \times 10^{-7}$  & $0.11$ & $0.26$\\
      88.3  & $1.02\, \pi$  & $0.63$ & $1.23 \times 10^{-7}$  & $0.13$ & $0.26$\\
      90.6  & $1.77\, \pi$  & $0.67$ & $1.04 \times 10^{-7}$  & $0.12$ & $0.31$\\
      93.0  & $2.00\, \pi$  & $0.90$ & $1.05 \times 10^{-7}$  & $0.19$ &$0.46$ \\
       
    \end{tabular}
    \end{center}
    {\footnotesize{\textbf{Notes.} Same as Table~\ref{ta:diskPropv05e50} for model v10e50 (corresponding to the plots in Fig.~\ref{fig:v10e50}).}}
    \label{ta:diskPropv10e50}
\end{table}

\begin{figure*}[h!]
    \centering
    \includegraphics[width = 0.49 \textwidth]{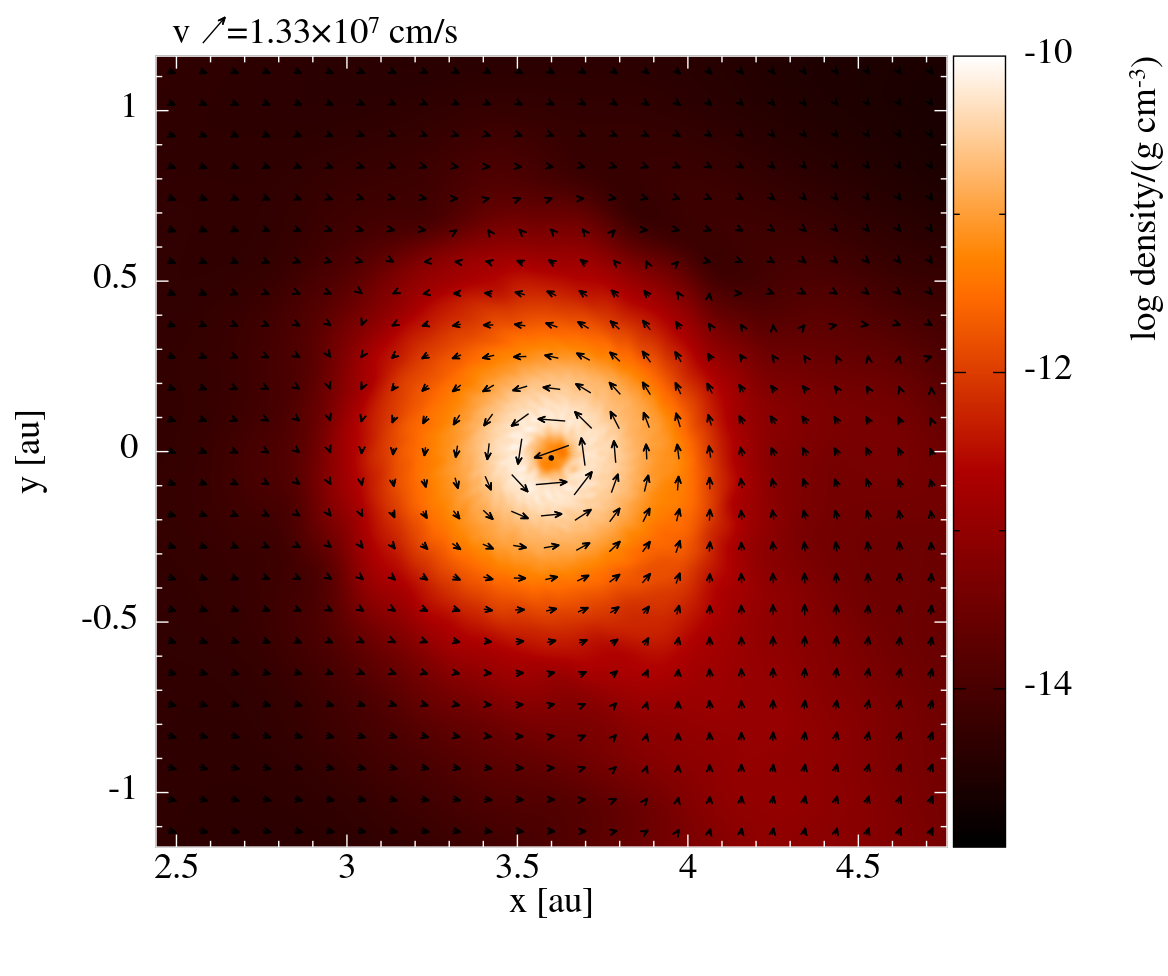}
    \includegraphics[width = 0.50 \textwidth]{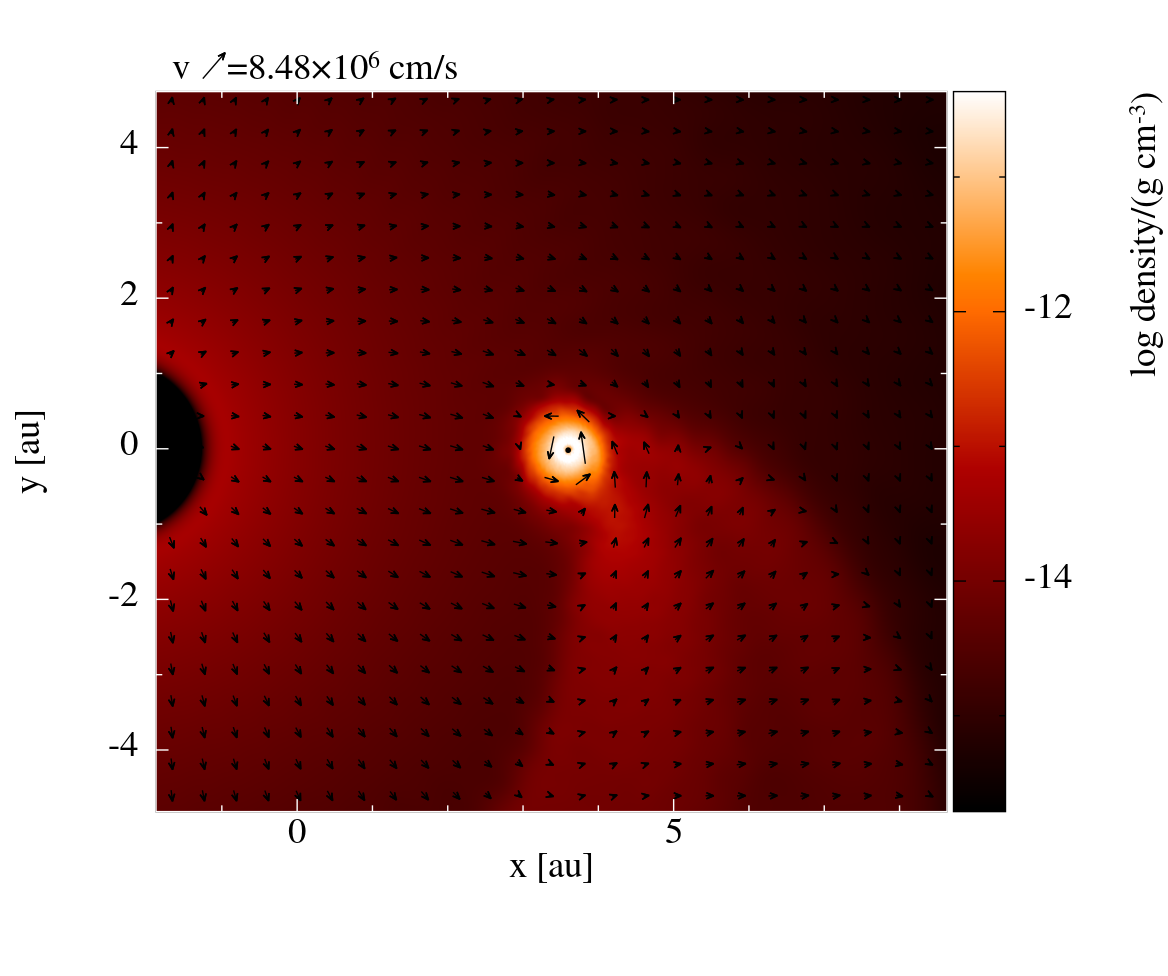}
    \caption{{Density distribution in a slice through the orbital plane of model v20e00, but with a mass-loss rate of $\dot{M} = 10^{-5}\,{\rm M_{\odot} \, {\rm yr}^{-1}}$ ($100$ times larger than the standard value used in this work), zoomed in on the accretion disk (left), and with density structure surrounding the accretion disk (right).  The colorbars are scaled with a factor $10$ compared to Figs.~\ref{fig:ADz10} and~\ref{fig:ADzoomv20e00}.}}
    \label{fig:v20e00_M1e-5_rho}
\end{figure*}

\clearpage

\section{Electron number density}
\label{AP:elNumbDens}
Following \cite{Palla_etal_1983}, we consider the following set of chemical reactions to calculate the electron density
\begin{align*} 
  &\ce{H^+ + e^- ->[k_1]H + h\nu} \\
  &\ce{H + e^- ->[k_2]H^- + h\nu} \\
  &\ce{H^+ H^- ->[k_3]H_2 + e^-} \\
  &\ce{H + H + H <=>[k_4][k_5]H_2 + H} \\ 
  &\ce{H + H + H_2 <=>[k_6][k_7]H_2 + H_2}  \\
  &\ce{H + e^- ->[k_8]H^+ + e^- + e^-} \\
  &\ce{H + H ->[k_9]H^+ + e^- + H} \\
\end{align*}
Assuming that all components $X= \{\Hy, \Hp, \Hy_2, \Hm, e^-\}$ are at equilibrium ($\partial_t X = 0$), we obtain the system of equations for the number density of $\Hp$, $\Hm$, and electrons
\begin{eqnarray}
    n_\Hp & = & \frac{k_9 n_\Hy^2 + k_8 n_e n_\Hy}{k_1 n_e} \\
    n_\Hm & = & \frac{k_2}{k_3} n_e \\
    \frac{\dd n_e}{\dd t} & = & \frac{\dd (n_\Hp -n_\Hm)}{\dd t}
    \label{eq:dnedt}
\end{eqnarray}
Integration of Eq.~\ref{eq:dnedt} leads to $n_\Hp = n_e + n_\Hm$. After substitutions and  resolution of a second degree equation, we obtain the following expression for the electron density
\begin{equation}
n_e = \frac{\sqrt{4 k_1 k_3 k_9 (k_2 + k_3) + k_3^2 k_8^2} + k_3 k_8}{2 k_1 (k_2 + k_3)} n_\Hy.
\end{equation}
The rate coefficients $k_i$ are temperature dependent, except for $k_3$ and are given in \cite{Palla_etal_1983}. The number of H atoms per unit volume is given by 
$$n_\Hy = \frac{\rho}{m_\Hy(1+4\epsilon_\He+ ..)} \approx \frac{\rho}{1.4\,m_\Hy}$$ 
where $\epsilon_\He$ is the number of He atom per H atom ($\epsilon_\He \approx 0.1$ for solar composition).

\section{Online material}
\label{movies}

{The input files and final output dumps of the Phantom simulations of this paper, and several movies to illustrate the complex flows within the binary systems, are available online at \url{https://zenodo.org/records/13224218}.
The movies can be accessed as well at \url{http://www.astro.ulb.ac.be/~siess/Phantom/HIcooling}.}

Movie 1 shows the evolution of the density distribution in the corotating orbital plane of model v10e50. It shows how the structures around the companion in this eccentric binary vary over each orbital period, and how the bow shock interacts and collides with other high density structures. This movie is linked to Fig.~\ref{fig:v10e50zoom}.\\

Movie 2 shows the evolution of the density distribution in the corotating orbital plane of model v05e50 during the final 4 orbital periods of the simulation. This movie focuses on the accretion disk around the companion, and reveals how the accretion disk shape varies throughout one orbital period due to the complex interplay between the phase-dependent wind-companion interaction strength and the interaction of the accretion disk with surrounding high density structures. This movie is linked to Fig.~\ref{fig:v05e50}.\\

Movie 3 shows the evolution of the density distribution in the corotating orbital plane of model v20e50 during the final 4 orbital periods of the simulation. This movie focuses on the accretion disk around the companion, and shows how the surrounding structure varies from a 2-edged spiral to bow shock within each orbital period. This movie appends to Fig.~\ref{fig:v20e50}.

\end{document}